\documentclass[a4paper,11pt]{article}
\usepackage{jcappub} 
\usepackage{graphicx}
\usepackage[varg]{txfonts}
\usepackage{graphicx}
\usepackage{psfrag}
\usepackage{hyperref}
\usepackage{gensymb}
\usepackage{xcolor}
\usepackage{tabularx}
\usepackage{lipsum}
\usepackage[normalem]{ulem}
\usepackage{amsmath}
\usepackage[labelfont=bf, labelsep=period]{caption}
\usepackage{subcaption}
\usepackage{mathtools}
\usepackage{balance}
\usepackage{adjustbox}
\usepackage{float} 
\usepackage{placeins}
\usepackage{caption}
\usepackage{stfloats}
\usepackage{pifont}
\usepackage{xcolor}
\usepackage{dsfont}
\usepackage{multirow}
\usepackage{changepage}
\usepackage{makecell}
\usepackage{soul}

\usepackage[colorinlistoftodos]{todonotes}
\usepackage{easyReview}

\usepackage{enumitem}

\usepackage{siunitx} 
\usepackage[switch,displaymath,mathlines]{lineno}

\title{Optimization of foreground moment deprojection for semi-blind CMB polarization reconstruction}

\def\reff@jnl#1{{\rm#1\/}}

\def\aj{\reff@jnl{AJ}}                  
\def\araa{\reff@jnl{ARA\&A}}            
\def\apj{\reff@jnl{ApJ}}                
\def\apjl{\reff@jnl{ApJ}}               
\def\apjs{\reff@jnl{ApJS}}              
\def\ao{\reff@jnl{Appl.Optics}}         
\def\apss{\reff@jnl{Ap\&SS}}            
\def\aap{\reff@jnl{A\&A}}               
\def\aapr{\reff@jnl{A\&A~Rev.}}         
\def\aaps{\reff@jnl{A\&AS}}             
\def\azh{\reff@jnl{AZh}}                        
\def\baas{\reff@jnl{BAAS}}              
\def\jcap{\reff@jnl{JCAP}}              
\def\jrasc{\reff@jnl{JRASC}}            
\def\memras{\reff@jnl{MmRAS}}           
\def\mnras{\reff@jnl{MNRAS}}            
\def\pra{\reff@jnl{Phys.Rev.A}}         
\def\prb{\reff@jnl{Phys.Rev.B}}         
\def\prc{\reff@jnl{Phys.Rev.C}}         
\def\prd{\reff@jnl{Phys.Rev.D}}         
\def\prl{\reff@jnl{Phys.Rev.Lett}}      
\def\physrep{\reff@jnl{Phys.Rep.}}      
\def\pasp{\reff@jnl{PASP}}              
\def\pasj{\reff@jnl{PASJ}}              
\def\qjras{\reff@jnl{QJRAS}}            
\def\skytel{\reff@jnl{S\&T}}            
\def\solphys{\reff@jnl{Solar~Phys.}}    
\def\sovast{\reff@jnl{Soviet~Ast.}}     
 \def\ssr{\reff@jnl{Space~Sci.Rev.}}    
\def\zap{\reff@jnl{ZAp}}                
\def\nat{\reff@jnl{Nature}}             

\author[a,b]{A. Carones,}
\author[c]{M. Remazeilles}
\affiliation[a]{Dipartimento di Fisica, Università di Roma ``Tor~Vergata'', via della Ricerca Scientifica 1, I-00133, Roma, Italy}
\affiliation[b]{Sezione INFN Roma~2, via della Ricerca Scientifica 1, I-00133, Roma, Italy}
\affiliation[c]{Instituto de F\'isica de Cantabria (CSIC-UC), Avda. los Castros s/n, 39005 Santander, Spain}

\emailAdd{alessandro.carones@roma2.infn.it, remazeilles@ifca.unican.es}

\abstract{Upcoming Cosmic Microwave Background (CMB) experiments, aimed at measuring primordial CMB polarization $B$-modes, require exquisite control of instrumental systematics and Galactic foreground contamination. Blind minimum-variance techniques, like the Needlet Internal Linear Combination (NILC), have proven effective in reconstructing the CMB polarization signal and mitigating foregrounds and systematics across diverse sky models without suffering from foreground mismodelling errors. Still, residual foreground contamination from NILC may bias the recovered CMB polarization at large angular scales when confronted with the most complex foreground scenarios. 
By adding constraints to NILC to deproject statistical moments of the Galactic emission, the Constrained Moment ILC (cMILC) method has been demonstrated to further enhance foreground subtraction, albeit with an associated increase in overall noise variance. Faced with this trade-off between foreground bias reduction and overall variance minimization, there is still no recipe on which moments to deproject and which are better suited for blind variance minimization. To address this, we introduce the optimized cMILC (ocMILC) pipeline, which performs full automated optimization of the required number and set of foreground moments to deproject, pivot parameter values, and deprojection coefficients across the sky and angular scales, depending on the actual sky complexity, available frequency coverage, and experiment sensitivity. The optimal number of moments for deprojection, before paying significant noise penalty, is determined through a data diagnosis inspired by the Generalized NILC (GNILC) method. 
Validated on $B$-mode simulations of the \textit{PICO} space mission concept with four challenging foreground models, ocMILC exhibits lower Galactic foreground contamination compared to NILC and cMILC at all angular scales, with limited noise penalty. This multi-layer optimization enables the ocMILC pipeline to achieve unbiased posteriors of the tensor-to-scalar ratio, regardless of foreground complexity.}


\begin{document}
\maketitle
\flushbottom

\section{Introduction}

Accurate measurements of the temperature anisotropies of the Cosmic Microwave Background (CMB) have yielded stringent constraints on viable cosmological models \citep{Planck_cosmopars,PhysRevD.107.023510}. Future CMB experiments, featuring enhanced sensitivity, will provide additional insights into CMB polarization anisotropies \citep{SO_2019,CMBS4,PTEP,PICO_inst}. On the one hand, an accurate reconstruction of the CMB $E$-mode polarization anisotropies will lead to a cosmic-variance limited measurement of the optical depth to last scattering surface $\tau$ \citep{PTEP,PICO_inst}, providing new constraints on the cosmic reionization history \cite{2020MNRAS.499..550Q,Planck_reio} and the sum of neutrino masses \citep{2015PhRvD..92l3535A,2016PhRvD..94h3522G,SO_2019}. On the other hand, a detection of the primordial CMB $B$-mode polarization  may open a new window on the physics of the early Universe by setting constraints on the tensor-to-scalar ratio $r$, thereby shedding light on the cosmic inflation scenario \citep{1997PhRvL..78.2058K}. 

To achieve these goals, future CMB experiments require exquisite control of the contamination from Galactic foregrounds and instrumental systematics. In recent years, several component separation pipelines,  both parametric \citep{Commander, FGBuster, Bsecret,Azzoni2021,Vacher2022,Commander3} and blind \citep{ILC,SMICA,SEVEM,NILC,Delta-map,cMILC,MCNILC}, have been developed or customized for this purpose. Given the huge amplitude discrepancy between Galactic foregrounds and CMB $B$-modes, these pipelines need to demonstrate robustness against the complex and unknown properties of Galactic foreground emissions, including spatial variations in their spectral energy distribution (SED) along and across the lines of sight \citep{Tassis2015}, as well as resulting foreground spectral distortions \citep{moments} and spectral decorrelation \citep{Pelgrims2021}.

Parametric component separation methods are engineered to yield minimal uncertainty in the retrieval of CMB $B$-modes, as long as the assumed parametric foreground model aligns with the actual foregrounds in the data. However, given the aforementioned intricacies in the foreground properties, parametric methods are also prone to potential biases in the recovery of the CMB $B$-mode signal when confronted with incorrect or incomplete modelling of poorly understood foregrounds \citep{Remazeilles2016}.
In contrast, blind minimum-variance approaches to component separation, like the Needlet Internal Linear Combination (NILC) method \citep{NILC}, offer the possibility of extracting the CMB polarization $E$- and $B$-mode fields from multifrequency data without making any assumption about the foregrounds, thus avoiding mismodelling errors. However, if the polarized Galactic foregrounds turn out to be highly complex, as in the case of  multi-layer contributions to thermal dust emission along the line-of-sight and spectral decorrelation \citep{d12}, minimum-variance methods like NILC may return CMB polarization maps that are still significantly contaminated by Galactic foreground residuals, as highlighted in \cite{cMILC, MCNILC, PICO}.

To overcome these issues, intermediate approaches at the intersection of parametric and blind methods can be pursued, using either partial information on Galactic foregrounds or some effective modelling. For instance, information on the spatial variations of the foregrounds across the sky can be inferred from the data and exploited to  optimize the domains over which the NILC variance is minimized, as implemented in the Multi-Clustering NILC method \citep{MCNILC}. As an alternative semi-blind approach, the Constrained Moment ILC (cMILC) method \cite{cMILC} relies on effective foreground modelling through statistical moment expansion around fixed pivot spectral parameters to project out some of these moments  through nulling constraints added to NILC, while leaving higher-order moments and unmodelled foregrounds for blind variance minimization. Through nulling constraints for moment deprojection, the cMILC method facilitates the handling of foreground moments and SED distortions resulting from line-of-sight and beam averaging  of spatial variations in foreground spectral properties by CMB instruments. In addition, it allows for imperfect parameterization of the foregrounds by leaving any higher-order correction to the moment expansion or any omitted foreground to blind variance minimization. This semi-blind technique has proven more effective than NILC in reducing Galactic contamination in the recovered CMB $B$-mode map when applied to complex foreground simulations of future CMB satellite experiments such as \textit{LiteBIRD} and \textit{PICO} \cite{cMILC,Fuskeland2023}.

However, the reduction of the residual foreground contamination bias in the cMILC CMB solution is counterbalanced by an increase of the overall variance associated with instrumental noise and unconstrained foreground moments, since each additional nulling constraint in cMILC effectively reduces the number of frequency channels available for variance minimization. Faced with this trade-off between bias reduction and variance increase, certain foreground moments may be more conducive to variance minimization than deprojection, particularly when they exert a dominant influence on the overall variance in the data, while other subdominant moments may be better addressed through deprojection.
However, no recipe for determining which moments to deproject and which to retain for variance minimization has been provided thus far.
In the baseline cMILC approach, the deprojection of zeroth-order moments is prioritized, anticipating their significant contribution to the bulk of Galactic foreground emission. Nonetheless, the optimal solution might favour the deprojection of higher-order components, relegating dominant moments to blind variance minimization. Moreover, if the actual  SED of polarized thermal dust and synchrotron emissions significantly deviates from the prevalent parameterization of a single modified blackbody (MBB) and a power-law, respectively, then the deprojection of SED moments with fixed values of pivot spectral parameters across the sky and angular scales may prove less effective.

In this work, we introduce the \emph{optimized cMILC} (ocMILC) pipeline, where the choice of moments to deproject, the values of pivot parameters, and the amount of deprojection of each moment is optimized through a fully data-driven approach. Specifically, the optimal number of moments to deproject before paying significant noise penalty is determined depending on the experimental configuration by applying the strategy of the Generalized Needlet ILC (GNILC) method \citep{GNILC}. This method provides a blind diagnosis of the actual complexity of the Galactic foreground emission across the sky and angular scales, yielding maps of the local number of independent foreground degrees of freedom as byproducts. 
Such a diagnostic tool is valuable for informing the foreground model parameterization and tuning component separation pipelines according to the observed sky complexity and experimental configuration.
The new ocMILC framework is tested and validated on a set of complex foreground scenarios using comprehensive simulations of the Probe of Inflation and Cosmic Origins (\textit{PICO}) mission concept \citep{PICO_inst}. We choose this experimental configuration as a proof of concept, given its high sensitivity and broad frequency coverage, allowing us to thoroughly assess the robustness of the proposed approach against various thermal dust and synchrotron models. We note that the validation of the ocMILC pipeline is performed under the assumption of perfectly reconstructed frequency maps from time-ordered data. However, as shown by the latest analyses of the \textit{Planck} data (NPIPE \cite{NPIPE}, SRoll \cite{sroll,sroll2}, Cosmoglobe \cite{cosmoglobe}), a parametric model of the Galactic emission may still be needed for the low-level data processing steps, as calibration, bandpass corrections, and multi-component destriping, involved in producing high-fidelity frequency maps.

The paper is organized as follows: In Section~\ref{sec:sims}, we provide a description of the \textit{PICO} sky simulations,
considering four foreground scenarios with different dust models. Section~\ref{sec:methods} outlines our component separation methodology, first reviewing the basics of the cMILC approach (Section~\ref{subsec:baseline}), then describing the proposed ocMILC pipeline (Section \ref{sec:ocMILC}), which involves a data-driven diagnosis of foreground complexity inspired by GNILC (Section~\ref{sec:GNILC}) and subsequent optimization of foreground moment deprojection (Section~\ref{subsec:optimization}). Section~\ref{sec:results} presents our results for the different foreground scenarios.
Our conclusions are summarized in Section~\ref{sec:concl}.

\section{Sky simulations}
\label{sec:sims}

As a proof of concept, we perform simulations of multifrequency sky maps based on the experimental setup of the \textit{PICO} mission. \textit{PICO}, a NASA probe-scale space mission concept, is anticipated to be the most sensitive among the proposed next-generation CMB space missions, targeting a sensitivity on the tensor-to-scalar ratio of $r \leq 0.0002$ at a $95\%$ confidence level (CL) and aiming for a cosmic-variance limited measurement of the optical depth to last scattering, $\sigma(\tau)=0.002$ \citep{PICO_inst}. To meet these objectives, \textit{PICO} will conduct full-sky observations in 21 frequency channels spanning from $21$ to $799$\,GHz with high signal-to-noise to disentangle Galactic foreground emission from the CMB polarization signal. Given its extensive frequency coverage and high sensitivity, \textit{PICO} serves as an ideal experimental framework to assess the robustness and flexibility of the proposed optimization of moment deprojection across various thermal dust and synchrotron models.

The \textit{PICO} sky maps are generated by co-adding simulated maps of CMB, Galactic foregrounds, and instrumental noise across the 21 frequency bands, according to the instrument specifications reported in \cite{PICO_inst}.

CMB polarization anisotropies are generated using the HEALPix Python package\footnote{\url{https://github.com/healpy/healpy}} \citep{healpix,healpy}, based on the CMB angular power spectra of the \textit{Planck} 2018 best-fit  parameters \citep{2020A&A...641A...6P} with a tensor-to-scalar ratio $r=0$, and include the contribution from the gravitational lensing effect.

The instrumental noise is simulated as white and isotropic Gaussian realizations, with standard deviations in each pixel corresponding to the polarization sensitivities of the baseline \textit{PICO} configuration as reported in \citep{PICO_inst}. The combined noise level for this configuration is equal to $0.87$ $\mu K\cdot\text{arcmin}$. In \citep{PICO_inst,PICO}, it has been demonstrated that the targeted sensitivity $r< 0.0002$ ($95\%$ CL) can be achieved for most foreground scenarios with the so-called Current Best Estimate (CBE) \textit{PICO} configuration, featuring a combined noise level of $0.61$ $\mu K\cdot\text{arcmin}$. However, for the purpose of our study relying on the more conservative baseline \textit{PICO} configuration, with a combined noise level of $0.87$ $\mu K\cdot\text{arcmin}$, we estimate a target sensitivity of about $r < (0.87)^2/(0.61)^2 \times 0.0002 = 0.0004$ ($95\%$ CL). Here, the noise levels are squared, as the tensor-to-scalar ratio parameter is associated with power spectrum rather than map noise. In Section \ref{subsec:ocMILC_res}, we show that, by adopting the CBE \textit{PICO} configuration, we can indeed achieve $r< 0.0002$ ($95\%$ CL) with the ocMILC method.

Maps of polarized Galactic emission are generated using the \texttt{PySM} Python package \citep{pysm,pysm3}. Given the current lack of knowledge about the polarized Galactic emission at the targeted sensitivity, we explore a range of foreground sky models that remain consistent with existing observations. Four distinct models of Galactic thermal dust emission are examined in this study and are described below, using the terminology of the PySM:
\begin{itemize}

    \item \texttt{d1}:  The thermal dust emission at each line-of-sight $\hat{n}$ and each frequency $\nu$ is characterized by a modified blackbody (MBB) SED in intensity units
\begin{equation}
	X_{\textrm{d}}(\hat{n},\nu) = X_{\textrm{d}}(\hat{n},\nu_{\textrm{d}}) \,  \left(\frac{\nu}{\nu_{\textrm{d}}} \right)^{\beta_{\textrm{d}}(\hat{n})} \frac{B_{\nu}\left(T_{\textrm{d}}(\hat{n})\right)}{B_{\nu_{\textrm{d}}}(T_{\textrm{d}}\left(\hat{n})\right)}\,,
	\label{eq:dust}
\end{equation}
where $X=\{\textit{Q},\textit{U} \}$ are Stokes parameters, $B_{\nu}(T)$ is the blackbody spectrum, $\beta_{\textrm{d}}$ is the dust spectral index, $T_{\textrm{d}}$ is the dust temperature, and $X_{d}(\hat{n},\nu_{\textrm{d}})$ represents the dust template at a reference frequency $\nu_{\textrm{d}}$. Both $\beta_{\textrm{d}}$ and $T_{\textrm{d}}$ vary across the sky, as estimated from the \textit{Planck} data analysis using the Commander algorithm \citep{2016A&A...594A..10P}. The thermal dust template is given by the Commander dust $Q$ and $U$ maps at $353$ GHz of the \textit{Planck} $2015$ data release, smoothed with a Gaussian beam of full width at half maximum (FWHM) $2.6^\circ$, with stochastic small-scale fluctuations added via the procedure described in \cite{pysm}. 

\item \texttt{d4}: In this model, the thermal dust emission is composed of two distinct MBB components, as fitted to \textit{Planck} and DIRBE/IRAS data in \cite{d4}. The temperatures of these two components vary across the sky, while the spectral indices remain constant with values $\beta^{(1)}_{\textrm{d}}=1.63$ and $\beta^{(2)}_{\textrm{d}}=2.82$. The $Q$ and $U$ templates at the reference frequency are obtained by scaling the dust intensity templates of \cite{d4} to $353$ GHz and applying the polarization fraction and polarization angles of the \texttt{d1} model. Gaussian small-scale fluctuations are generated similarly to the \texttt{d1} model.

\item \texttt{d7}: Simulations of thermal dust emission are based on a physical model of dust grains with a distribution of sizes, composition, and temperatures. Additionally, an environmental radiation strength field is assumed across the sky based on the dust temperature map obtained in \cite{2016A&A...594A..10P}. The model also accounts for magnetic dipole radiation from iron grains \citep{d7_magnet}, resulting in a SED that deviates from a simple MBB spectrum.

\item \texttt{d12}: In this model, the dust emission is composed of six distinct emitting layers \citep{d12}, so that the effective SED varies not only across the sky but also along the line-of-sight, leading to spectral decorrelation across frequencies. It represents the most complex dust model among the four considered in this study. Specifically, the maps of Stokes parameters at a frequency $\nu$ are given by the superposition of six templates scaled with a MBB law, each associated with a different map of spectral indices and temperatures:
\begin{equation}
    X_{\textrm{d}}(\hat{n},\nu) = \sum_{k=1}^{6}\,X_{\textrm{d}}^{(k)}(\hat{n},\nu_0)\, \left(\frac{\nu}{\nu_{0}} \right)^{\beta_{\textrm{d}}^{(k)}(\hat{n})} B_{\nu}\left(T_{\textrm{d}}^{(k)}(\hat{n})\right)\,,
\end{equation}
such that it matches the total dust emission of the \textit{Planck} GNILC intensity and polarization maps at $353$ GHz \citep{Planck2018_compsep}.

\end{itemize} 

Galactic synchrotron emission is simulated using the PySM \texttt{s1} model, which assumes a power-law SED in brightness temperature units
\citep{Rybicki}
\begin{equation}
	X_{\textrm{s}}(\hat{n},\nu) = X_{\textrm{s}}(\hat{n},\nu_{\textrm{s}})\,  \left(\frac{\nu}{\nu_{\textrm{s}}} \right)^{\beta_{\textrm{s}}(\hat{n})}\,,
	\label{eq:sync}
\end{equation}
where $\beta_{\textrm{s}}$ is the synchrotron spectral index, varying with the position $\hat{n}$ in the sky. $X_{\textrm{s}}(\hat{n},\nu_{\textrm{s}})$ represents the synchrotron template at a reference frequency $\nu_{\textrm{s}}$. The chosen templates are the \textit{WMAP} $9$-year $23$-GHz Q and U maps \citep{sync_temp}, smoothed with a Gaussian kernel of FWHM $5^\circ$, with small-scale fluctuations added using the procedure outlined in \cite{pysm}. The synchrotron spectral index map is derived through a spectral fit to the Haslam $408$-MHz data and the \textit{WMAP} $23$-GHz $7$-year data \citep{sync_index}.

In summary, four distinct \textit{PICO} data sets, hereafter referred to as \texttt{d1}, \texttt{d4}, \texttt{d7}, and \texttt{d12}, are created by combining the respective dust model maps with the PySM \texttt{s1} synchrotron maps and random realizations of CMB. At each frequency, these components are first smoothed using a Gaussian beam with the corresponding FWHM of that \textit{PICO} channel and then co-added to instrumental noise. Such resulting \textit{PICO} frequency maps are ultimately brought to a common resolution of $\textrm{FWHM}=38'$ and downgraded to a HEALPix resolution of $N_{\textrm{side}}=512$ before undergoing component separation analysis. For each data set, 50 simulations are generated, each featuring distinct realizations of CMB and instrumental noise. 

\section{Component separation methodology}
\label{sec:methods}

In polarization, we can assume that data at different frequencies, $d_{\nu}$, are the superposition of three distinct types of components: 
\begin{equation}
 d_{\nu}(\hat{n}) = c(\hat{n}) + f_{\nu}(\hat{n}) + n_{\nu}(\hat{n})\,,
 \label{eq:input_gnilc}
\end{equation}
where (assuming CMB temperature units) $c(\hat{n})$ denotes frequency-independent CMB anisotropies, $f_{\nu}(\hat{n})$ is the Galactic foreground emission, $n_{\nu}(\hat{n})$ the instrumental noise, $\hat{n}$ the sky position, and $\nu=1,\dotsc,N_{\nu}$, with $N_{\nu}$ the number of available frequency channels. 

An estimate $\hat{c}(\hat{n})$ of CMB anisotropies can be formed via a weighted linear combination of the data:
\begin{equation}
 \hat{c}(\hat{n}) = \boldsymbol{w}^T\cdot \boldsymbol{d} = \sum_{\nu} w_{\nu}(\hat{n})\, d_{\nu}(\hat{n})\,,
 \label{eq:wilc}
 \end{equation}
where the specific weights $w_{\nu}$ assigned to data $d_{\nu}$ across frequency channels are sought to minimize the variance due to foreground and noise contamination while preserving the CMB signal $c(\hat{n})$. This is the strategy of ILC methods such as NILC \cite{NILC}. In this section, we seek to optimize these weights to minimize foreground variance rather than global variance as a function of sky regions and angular scales with our proposed ocMILC (optimized cMILC) pipeline. We first review the basics of the cMILC formalism (Section~\ref{subsec:baseline}) which forms the skeleton of the proposed ocMILC pipeline, the specific properties of which are described afterward (Section~\ref{sec:ocMILC}).

\subsection{Baseline cMILC approach}
\label{subsec:baseline}

As demonstrated in \cite{cMILC}, although the blind NILC method provides the CMB solution with minimum overall variance, it does not achieve minimum foreground variance. This minimum foreground variance can be attained through the more constrained semi-blind cMILC method, which involves the deprojection of certain foreground components. Achieving minimum foreground variance is essential to alleviate systematic biases in the recovered tensor-to-scalar ratio, especially in cases where residual foreground contamination after component separation surpasses the CMB and noise power at large angular scales. Such unfavourable outcome was observed in the case of NILC for the multi-layer dust scenario with PICO \citep{PICO} and for less complex foreground skies in experimental configurations with narrower frequency coverage \citep{Fuskeland2023}.

Galactic emission exhibits three-dimensional characteristics \citep{Tassis2015}, extending along the line of sight, and its spectral parameters vary spatially both along and across the lines of sight. When integrated along the line of sight and within the instrumental beam by CMB instruments, the superposition of emitting elements with varying spectral parameters results in spectral distortions of the baseline SEDs of the Galactic components. As demonstrated in the seminal work of \cite{moments}, these foreground distortions can be accurately captured through a Taylor moment expansion of the baseline SEDs around pivot values of the spectral parameters. Foreground subtraction can then be enhanced by \emph{deprojecting} certain moments of the Galactic foreground emission from the output CMB solution, a technique introduced by the cMILC method \citep{cMILC}.

 Referring to Equation~\ref{eq:sync}, the Galactic synchrotron emission can be expanded around a baseline power-law SED up to second order in $\beta_{\textrm{s}}(\hat{n})$ as:
\begin{equation}
    \begin{aligned}
        X_{\textrm{s}}(\hat{n},\nu)=X_{\textrm{s}}(\hat{n},\nu_{\textrm{s}})\, &\Bigg[f_{\textrm{sync}}(\nu,\bar{\beta}_{\textrm{s}})+  \left(\beta_{\textrm{s}}(\hat{n})-\bar{\beta}_{\textrm{s}}\right) \partial_{\beta_{\textrm{s}}}f_{\textrm{sync}}(\nu,\bar{\beta}_{\textrm{s}})\\
        &+\frac{1}{2} \left(\beta_{\textrm{s}}(\hat{n})-\bar{\beta}_{\textrm{s}}\right)^2 \partial^{2}_{\beta_{\textrm{s}}}f_{\textrm{sync}}(\nu,\bar{\beta}_{\textrm{s}})+\mathcal{O}\left( \left(\beta_{\textrm{s}}(\hat{n})-\bar{\beta}_{\textrm{s}}\right)^3\right) \Bigg]\,,
    \end{aligned}
\label{eq:sync_exp}
\end{equation}
where 
\begin{equation}
    f_{\textrm{sync}}(\nu,\bar{\beta}_{\textrm{s}})=\left(\frac{\nu}{\nu_{\textrm{s}}} \right)^{\bar{\beta}_{\textrm{s}}}
\end{equation}
is a power-law with fixed pivot spectral index $\bar{\beta}_{\textrm{s}}$. Equation~\ref{eq:sync_exp} highlights moments of the synchrotron spectral index, $(\beta_{\textrm{s}}(\hat{n})-\bar{\beta}_{\textrm{s}})^k$, with uniform SEDs $\partial^{k}_{\beta_{\textrm{s}}}f_{\textrm{sync}}(\nu,\bar{\beta}_{\textrm{s}})$.

Similarly, from Equation~\ref{eq:dust}, the moment expansion of the thermal dust emission around a baseline MBB SED up to second order reads as:
\begin{equation}
    \begin{aligned}
        X_{\textrm{d}}(\hat{n},\nu)=X_{\textrm{d}}(\hat{n},\nu_{\textrm{d}})\,  &\Bigg[f_{\textrm{dust}}(\nu,\bar{\beta}_{\textrm{d}},\bar{T}_{\textrm{d}})+ \left(\beta_{\textrm{d}}(\hat{n}) - \bar{\beta}_{\textrm{d}}\right)  \partial_{\beta_{\textrm{d}}}f_{\textrm{dust}}(\nu,\bar{\beta}_{\textrm{d}},\bar{T}_{\textrm{d}})\\
        &+\left(T_{\textrm{d}}(\hat{n}) - \bar{T}_{\textrm{d}}\right)  \partial_{T_{\textrm{d}}}f_{\textrm{dust}}(\nu,\bar{\beta}_{\textrm{d}},\bar{T}_{\textrm{d}}) + \frac{1}{2}\left(\beta_{\textrm{d}}(\hat{n}) - \bar{\beta}_{\textrm{d}}\right)^2  \partial^{2}_{\beta_{\textrm{d}}}f_{\textrm{dust}}(\nu,\bar{\beta}_{\textrm{d}},\bar{T}_{\textrm{d}})\\
        &+  \left(\beta_{\textrm{d}}(\hat{n}) - \bar{\beta}_{\textrm{d}}\right) \left(T_{\textrm{d}}(\hat{n}) - \bar{T}_{\textrm{d}}\right)  \partial_{\beta_{\textrm{d}}}\partial_{T_{\textrm{d}}}f_{\textrm{dust}}(\nu,\bar{\beta}_{\textrm{d}},\bar{T}_{\textrm{d}})\\
        &+ \frac{1}{2} \left(T_{\textrm{d}}(\hat{n}) - \bar{T}_{\textrm{d}}\right)^2  \partial^{2}_{T_{\textrm{d}}}f_{\textrm{dust}}(\nu,\bar{\beta}_{\textrm{d}},\bar{T}_{\textrm{d}})+ \mathcal{O}\left(\left(\beta_{\textrm{d}}(\hat{n}) - \bar{\beta}_{\textrm{d}}\right)^3,\left(T_{\textrm{d}}(\hat{n}) - \bar{T}_{\textrm{d}}\right)^3\right)\Bigg]\,,
    \end{aligned}
\label{eq:dust_exp}
\end{equation}
where 
\begin{equation}
\label{eq:dust_baseline_krj}
         f_{\textrm{dust}}(\nu,\bar{\beta}_{\textrm{d}},\bar{T}_{\textrm{d}})=\left(\frac{\nu}{\nu_{\textrm{d}}} \right)^{\bar{\beta}_{\textrm{d}} +1}\frac{\exp{\left(h\nu_{\textrm{d}}/k\bar{T}_{\textrm{d}}(\hat{n})\right)}-1}{\exp{\left(h\nu/k\bar{T}_{\textrm{d}}(\hat{n})\right)}-1}
\end{equation}
is the MBB function for data in brightness temperature units, with fixed pivot values of the spectral index and temperature $\bar{\beta}_{\textrm{d}}$ and $\bar{T}_{\textrm{d}}$.

We assume the observed averages of spectral parameters across the sky as pivot values to ensure that most of the Galactic emission projects onto the zeroth-, first-, and second-order moments of the expansion: $\bar{\beta}_{\textrm{s}} \simeq -3$ \citep{page_2007,sync_index,krach_2018,Planck2018_compsep}, $\bar{\beta}_{\textrm{d}} \simeq 1.5$, and $\bar{T}_{\textrm{d}} \simeq 20\ \textrm{K}$ \citep{Planck2018_compsep}. However, further refinement of the most suitable pivot values for spectral parameters could be achieved by considering the observed average in the sky area outside the Galactic mask used in the analysis or by requiring the cancellation of first-order moments in the average \citep{Vacher2023}.

The cMILC weights $\boldsymbol{w}$ are built to deproject some of these foreground moments by adding to NILC nulling constraints against their homogeneous SEDs, while the remaining unconstrained contamination is left to blind variance minimization:
\begin{equation}
    \begin{cases}
    \min_{\substack{\boldsymbol{w}}}\left(\boldsymbol{w}^{T}C\,\boldsymbol{w}\right) \\
    \boldsymbol{w}^{T}\cdot \ \boldsymbol{a}_{\textrm{CMB}}=1 \\
    \boldsymbol{w}^{T}\cdot  \boldsymbol{f}_{\textrm{sync}}(\bar{\beta}_{\textrm{s}})=0 \\
    \boldsymbol{w}^{T}\cdot \boldsymbol{\partial}_{\beta_{\textrm{s}}}\boldsymbol{f}_{\textrm{sync}}(\bar{\beta}_{\textrm{s}})=0 \\
    \boldsymbol{w}^{T}\cdot \boldsymbol{f}_{\textrm{dust}}(\bar{\beta}_{\textrm{d}},\bar{T}_{\textrm{d}})=0 \\
    \boldsymbol{w}^{T}\cdot \boldsymbol{\partial}_{\beta_{\textrm{d}}}\boldsymbol{f}_{\textrm{dust}}(\bar{\beta}_{\textrm{d}},\bar{T}_{\textrm{d}})=0 \\
    \boldsymbol{w}^{T}\cdot \boldsymbol{\partial}_{T_{\textrm{d}}}\boldsymbol{f}_{\textrm{dust}}(\bar{\beta}_{\textrm{d}},\bar{T}_{\textrm{d}})=0 \\
    \vdots \\
    \end{cases}\,,
\label{eq:w_constraints}
\end{equation}
where $\boldsymbol{w} = \{w_\nu(\hat{n})\}$ 
represents the vector of frequency-dependent, pixel-dependent cMILC weights, 
$C=\{C_{\nu\nu'}(\hat{n})=\langle d_{\nu}(\hat{n})\, d_{\nu'}(\hat{n}) \rangle\}$  denotes the data covariance matrix with the same dependencies as the weights, $\boldsymbol{\partial^{k}}_{\beta_{\textrm{d}}} \boldsymbol{\partial^{n}}_{T_{\textrm{d}}} \boldsymbol{f}_{\textrm{dust}}(\bar{\beta}_{\textrm{d}},\bar{T}_{\textrm{d}}) = \{\partial^{k}_{\beta_{\textrm{d}}} \partial^{n}_{T_{\textrm{d}}} f_{\textrm{dust}}(\nu,\bar{\beta}_{\textrm{d}},\bar{T}_{\textrm{d}})\}$ and $\boldsymbol{\partial^{k}}_{\beta_{\textrm{s}}} \boldsymbol{f}_{\textrm{sync}}(\bar{\beta}_{\textrm{s}}) = \{\partial^{k}_{\beta_{\textrm{s}}} f_{\textrm{sync}}(\nu,\bar{\beta}_{\textrm{s}})\}$ are the frequency-dependent but pixel-independent dust and synchrotron moment SEDs, and ${\boldsymbol{a}_{\textrm{CMB}} = \{a_{\textrm{CMB}, \nu}\}}$ stands for the CMB SED, which is independent of frequency in CMB temperature units. The number of nulling constraints on moments in Equation~\ref{eq:w_constraints} is only limited by the number of available frequency channels and the sensitivity of the experiment.

By collecting the SED vectors of the CMB and selected foreground moments into the mixing matrix
\begin{equation*}
    A=\left(\boldsymbol{a}_{\textrm{CMB}}\quad \boldsymbol{f}_{\textrm{sync}}(\bar{\beta}_{\textrm{s}})\quad \boldsymbol{f}_{\textrm{dust}}(\bar{\beta}_{\textrm{d}},\bar{T}_{\textrm{d}})\quad \boldsymbol{\partial}_{\beta_{\textrm{d}}}\boldsymbol{f}_{\textrm{dust}}(\bar{\beta}_{\textrm{d}},\bar{T}_{\textrm{d}})\quad \ldots\right)
\end{equation*}
and the deprojection coefficients into the vector:
\begin{equation}
    \boldsymbol{e}=\left(1\quad 0\quad 0\quad 0\quad \ldots\right)^T\,,
    \label{eq:e}
\end{equation}
the cMILC weights can be expressed in the compact form:
\begin{equation}
\boldsymbol{w} = \boldsymbol{e}^{T}\left[A^{T}C^{-1}\,A\right]^{-1}A^{T}C^{-1}\,.
\label{eq:cMILC_weights} 
\end{equation}
The cMILC method, initially introduced in \cite{cMILC}, has since been applied in various subsequent works \citep{2021MNRAS.500..976R,Adak2021,2022JCAP...10..063G,Fuskeland2023,2023arXiv230908170Z,Dou2023}. In such a framework, NILC represents simply a subcase of cMILC where only the first element of $\boldsymbol{e}$ and first column of $A$ are employed in Equation \ref{eq:cMILC_weights}.

Adding constraints to null some of the foreground moments in Equation~\ref{eq:w_constraints} and, thereby, remove systematic biases generally results in an increase of the variance of the noise and unconstrained foreground components, such as non-included higher-order moments. There is thus a trade-off between deprojection and blind variance minimization, where components that significantly contribute to the overall variance in the data may be better suited for variance minimization rather than deprojection. However, in the baseline cMILC approach, there is no inherent procedure for selecting which moments should be prioritized for deprojection and which should be left for variance minimization, nor is there a method for assessing the optimal number of moments to deproject before incurring significant noise penalties. Moreover, if the effective SED of foreground emission turns out to highly deviate from the adopted baseline SEDs of Equations~\ref{eq:sync_exp} and \ref{eq:dust_exp}, moment expansion may not fully capture such distortions, in which case partial deprojection or variance minimization could be more effective. Therefore, there is room for optimization of moment deprojection in cMILC, as anticipated and preliminarily explored in \cite{cMILC}.

\subsection{Optimized cMILC (ocMILC) method}
\label{sec:ocMILC}

Our proposed ocMILC (optimized cMILC) method, an enhanced version of cMILC, aims to optimize moment deprojection and variance minimization in four ways: 
(i) allowing the number of deprojected moments to vary across sky regions and multipole ranges via a blind diagnosis of local foreground complexity; 
(ii) selecting moments for deprojection independently of the natural order of the Taylor expansion; 
(iii) allowing pivot values to vary across sky regions and multipole ranges to reduce the significance of higher-order moments in the moment expansion; 
(iv) partially filtering out the contribution of specific moments by modifying the deprojection vector in Equation~\ref{eq:e} as follows:
\begin{equation}
\boldsymbol{e}=\left(1\quad \epsilon_{1}\quad \epsilon_{2}\quad \epsilon_{3}\quad \ldots\right)^T\,,
\label{eq:e_new}
\end{equation}
with $\vert\epsilon_{i}\vert < 1$. This last adjustment aims to strike a balance between reducing bias and increasing noise variance, drawing inspiration from \cite{depro_paper}. The method is fully data-driven.

The first stage in the ocMILC pipeline aims at assessing the optimal number of moments to deproject across the sky and angular scales. This is achieved by leveraging information from byproducts of the Generalized Needlet ILC (GNILC) method \cite{GNILC_intro,GNILC}, which allows for a blind diagnosis of the effective foreground complexity in the data by determining how many independent modes of the Galactic emission exhibit greater power than noise across the sky and angular scales. The GNILC diagnosis tool which is implemented in ocMILC is detailed in Section \ref{sec:GNILC}, while subsequent optimization layers of the ocMILC pipeline are described in Section \ref{subsec:optimization}. 
 
\subsubsection{Diagnosis of foreground complexity}
\label{sec:GNILC}

The complex properties of the polarized Galactic foregrounds at microwave frequencies are not yet fully understood at the anticipated sensitivity levels of future CMB experiments. 
Furthermore, the integration of foreground emission along the line-of-sight and within the beam by CMB instruments leads to spatial averaging of each foreground spectral parameter, which results in spectral distortions of the expected foreground SEDs \citep{moments,Stolyarov2005} and spectral decorrelation of the foreground emission across frequencies \citep{Tassis2015,Pelgrims2021}. Additional spectral averaging occurs during data processing steps such as spherical harmonic transforms. All these processes effectively augment the list of expected foreground parameters, adding further complexity to the modelling of the foregrounds. 
To avoid making incorrect assumptions about the number of foregrounds and their spectral properties, it is useful to adopt a data-driven approach which can inform us on the actual complexity of foregrounds in sky observations. The GNILC method \cite{GNILC_intro,GNILC} offers an avenue, by providing maps of the effective number of \emph{independent} foreground components across the sky and different angular scales as byproducts, based on the experiment's sensitivity and frequency coverage. These GNILC byproduct maps, revealing the effective dimension of the foreground subspace, can in turn inform the cMILC method \citep{cMILC} about the optimal number of nulling constraints or moments to deproject without incurring significant noise penalty, depending on the sky area and the range of angular scales. 
Besides, parametric component separation methods may also benefit from these diagnostic maps to gauge the expected volume of the foreground parameter space depending on the sky regions.

Instead of making spectral assumptions, GNILC relies on statistical decorrelation to estimate the number of independent foreground degrees of freedom and perform component separation. It capitalizes on the fact that while the noise is largely uncorrelated across frequencies, the foreground emission,  $f_{\nu}(\hat{n})$, remains strongly correlated, such that it can be decomposed, at any frequency $\nu$, onto a common basis of a few independent (i.e. uncorrelated, not physical) templates $\boldsymbol{t}(\hat{n})=\{t_i(\hat{n})\}_{i\in [1,m]}$:
\begin{equation}
f_{\nu}(\hat{n}) = \sum_{i=1}^{m} \alpha_{\nu, i}(\hat{n})\, t_i(\hat{n})\,,
 \label{eq:templates}
\end{equation}
where the number $m$ of independent templates is typically smaller than the number of frequency channels. The focus of ocMILC is on mapping this number $m$ rather than the templates themselves to provide insight into the complexity of the foreground sky. 

From Equation~\ref{eq:input_gnilc}, the elements of the $N_{\nu} \times N_{\nu}$ data covariance matrix for frequency pairs $(\nu,\nu')$ read as
\begin{equation}
    C_{\nu\nu'}(\hat{n})=\langle d_{\nu}(\hat{n})\, d_{\nu'}(\hat{n}) \rangle = C^{\textrm{CMB}}_{\nu\nu'}(\hat{n}) + F_{\nu\nu'}(\hat{n}) + N_{\nu\nu'}(\hat{n})\,,
\label{eq:covariances}    
\end{equation}
where $C^{\textrm{CMB}}$, $F$, and $N$ denote respectively the CMB, foreground, and noise covariance matrices. In Equation~\ref{eq:covariances} we fairly assume that the three components are statistically uncorrelated. In practice, the ensemble average of Equation~\ref{eq:covariances} is substituted with a sample average:
\begin{equation}
C_{\nu\nu'}(\hat{n}) = \sum_{\hat{n}^{\prime}} W\left(\hat{n},\hat{n}^{\prime}\right)d_{\nu}\left(\hat{n}^{\prime}\right)d_{\nu'}\left(\hat{n}^{\prime}\right)\,,
\end{equation}
where $W\left(\hat{n},\hat{n}^{\prime}\right)$ are the weights associated to a Gaussian convolution kernel centred at $\hat{n}$, with a standard deviation large enough to sample an adequate number of modes. 

Considering that our focus is on Galactic foregrounds, we rewrite Equation~\ref{eq:covariances} as 
\begin{equation}
    C(\hat{n}) = F(\hat{n}) + C_{N}(\hat{n})\,,
\label{eq:covariances_1}    
\end{equation}
where $C_{N}$ is the \emph{nuisance} covariance matrix collecting the contribution from CMB and instrumental noise covariance matrices. 

Unlike the foreground covariance matrix, the nuisance covariance matrix can be estimated from simulations based on the knowledge of the instrument and the CMB statistics. The estimated nuisance covariance matrix, $\hat{C}_{N}(\hat{n})$, is then used as an input of the GNILC pipeline to compute the \emph{whitened} data covariance matrix:
\begin{equation}
    \tilde{C}(\hat{n}) =\left( \hat{C}_{N}^{-1/2} C \hat{C}_{N}^{-1/2}\right)(\hat{n}) \simeq \left(\hat{C}_{N}^{-1/2} F \hat{C}_{N}^{-1/2}\right)(\hat{n}) + \mathds{1}\,,
    \label{eq:whitened_covar}
\end{equation}
where the last equality follows from Equation~\ref{eq:covariances_1} and is exact if the estimated nuisance covariance matrix perfectly captures the statistics of the CMB and noise components, and $\mathds{1}$ denotes the identity matrix. 

Therefore, the eigendecomposition and diagonalization of the whitened data covariance matrix must have the following structure:
\begin{equation}
    \tilde{C}(\hat{n}) \simeq \begin{pmatrix}
        U_{F}(\hat{n})\ |\ U_{N}(\hat{n})
    \end{pmatrix}\left(\begin{array}{ccc|ccc}
        \lambda_{1}(\hat{n}) +1 & & & & &\\
        & \ddots & & & & \\
        & & \lambda_{m(\hat{n})}(\hat{n}) +1 & & &\\
        \hline
        & & & 1 & &\\
        & & & & \ddots &\\
        & & & & & 1
   \end{array}\right)
   \left(\begin{array}{c}
        U_{F}(\hat{n})^T \\
        \hline
        U_{N}(\hat{n})^T
    \end{array}\right)\,,
\label{eq:C_tilde}
\end{equation}
where all eigenvalues significantly greater than one correspond to the independent eigenmodes of the Galactic emission which locally dominate over the variance of the noise and CMB components. Therefore, in each point of the sky $\hat{n}$, we are able to estimate the number of independent Galactic foreground modes, $m(\hat{n})$, above the noise, given the sensitivity of the considered experiment, through this model-independent approach where we only need to assume the two-point statistics of the noise and CMB components in the polarization data. The $m$ independent foreground eigenmodes are simply collected in the $N_\nu\times m$ matrix $U_{F}$. 

In the language of principal component analysis (PCA), the template maps of Equation~\ref{eq:templates} are given by ${\boldsymbol{t}(\hat{n}) = U_{F}(\hat{n})^T\hat{C}_{N}^{-1/2}(\hat{n})\boldsymbol{d}(\hat{n})}$ and correspond to the $m$ principal components detected above the noise along the line-of-sight $\hat{n}$. They form uncorrelated, not physical, foreground templates,  as indicated by their diagonal covariance matrix. 
If the foreground emission was completely uncorrelated across frequencies, as instrumental noise is, the method would return $m=N_\nu$, i.e. as many independent templates as the number of available frequency channels. 
In contrast, the greater the correlation among physical foregrounds across frequencies and with each other, the lower the value of $m$ will be. In addition, when noise dominates the local data, the effective number $m$ of independent foreground components being detected is also reduced.

However, foreground emission models with same overall amplitude, i.e. equal 'foreground-to-noise' ratio, yet different levels of decorrelation, can lead to a different number of independent components, thus holding different degrees of complexity and producing different GNILC outcomes.

We note that, in the above treatment, we considered the CMB as a nuisance whose covariance matrix can be predicted. This may not be exactly the case at large angular scales for CMB $B$-modes since the value of the tensor-to-scalar ratio $r$ is unknown. However, this issue can be addressed in ocMILC by considering only the noise in the nuisance covariance matrix ($C_N = N$), in which case our method outputs the number of independent modes from both foregrounds and CMB that dominate over the noise:
\begin{equation}
    m(\hat{n}) = m_{\textrm{fgds}}(\hat{n}) + m_{\textrm{CMB}}(\hat{n})\,,
\end{equation}
with $m_{\textrm{CMB}}(\hat{n})=0$ or $1$, as the CMB is fully correlated between frequencies. Given the sensitivity of future CMB experiments, we expect the lensing CMB $B$-mode signal to dominate over instrumental noise across the range of observed angular scales (see e.g. figure 1 of \cite{PICO}). Therefore,
\begin{equation}
    m(\hat{n}) = m_{\textrm{fgds}}(\hat{n}) + 1\,.
\label{eq:m_fgds_used}
\end{equation}

The GNILC approach implemented in ocMILC thus allows one to perform a blind diagnosis of the complexity of the Galactic foreground emission in the observed sky and to possibly compare it with that expected from various foreground models. 
 
We expect the 'foregrounds-to-noise' ratio, as traced by the whitened covariance matrix $\tilde{C}(\hat{n})$ (Equations~\ref{eq:whitened_covar} and \ref{eq:C_tilde}), to vary not only across the sky but also across angular scales. Therefore, most likely, $m_{\textrm{fgds}}$ will be higher at large angular scales where diffuse Galactic emission is brighter than CMB and noise by orders of magnitude, while it will be lower at small angular scales where the noise dominates the covariance budget. 
This reasoning supports an implementation of our diagnostic tool in wavelet domain, using a specific frame of spherical wavelets known as needlets \citep{Narcowich2006,Marinucci2008}.  
In such a framework, the data $d_{\nu}(\hat{n})$ are convolved by a set of needlet bandpass filters in harmonic domain in order to form needlet coefficient maps, $\beta^{j}_{\nu}(\hat{n})$, including sky emission from specific ranges of angular scales:
\begin{equation}
\beta^{j}_{\nu}(\hat{n})
=\sum_{\substack{\ell,m}} \left(a_{\ell m, \nu}\, b^{j}_\ell\right)Y_{\ell m}(\hat{n})\,,
\label{eq:needlets_map}
\end{equation}
where $a_{\ell m, \nu}$ are the spherical harmonic coefficients of the frequency maps $d_{\nu}(\hat{n})$, $b^{j}_\ell$ is the $j^{\rm th}$ needlet bandpass, and $Y_{\ell m}(\hat{n})$ are spherical harmonics. 
By construction, the needlet bandpasses satisfy ${\sum_{j} (b^{j}_\ell)^2 = 1}$ for all multipoles, so that the inverse needlet transform 
\begin{equation}
d_{\nu}(\hat{n})
=\sum_{j}\sum_{\substack{\ell,m}} b^{j}_\ell \left(a_{\ell m, \nu}\, b^{j}_\ell\right) Y_{\ell m}(\hat{n})
\label{eq:needlets_map_back}
\end{equation}
retrieves the initial data. 

The methodology presented above is applied to the multifrequency needlet coefficient maps, $\beta^{j}_{\nu}(\hat{n})$, instead of the original sky maps, $d_{\nu}(\hat{n})$. The data and nuisance covariance matrices, $C^j(\hat{n})$ and $C_N^j(\hat{n})$ of Equation~\ref{eq:covariances_1}, are thus computed separately for each range of angular scales selected by the needlet bands. This implementation allows us to produce maps of the number of independent foreground modes, $m^j_{\textrm{fgds}}(\hat{n})$, across the sky and in different ranges of angular scales.

By diagnosing the number of independent modes of the Galactic foreground emission above the noise level, the blind approach outlined in this section can be leveraged to enhance the optimization of CMB component separation. In ocMILC, we exploit this information to optimize the selection of deprojected moments of the Galactic foreground emission within the framework of the semi-blind cMILC component separation method.

\subsubsection{Optimization of moment deprojection}
\label{subsec:optimization}

For each range of angular scales (needlet scale $j$), the GNILC diagnosis conducted in ocMILC (Section~\ref{sec:GNILC}) provides a partitioning of the sky into clusters of uniform dimension, $m^j_{\textrm{fgds}}$, of the expected foreground subspace. Each cluster of the partition, as illustrated later in Section~\ref{subsec:GNILC_results}, offers constraints on the optimal number of moments to deproject in that region of the sky and range of angular scales without incurring significant noise penalty. In ocMILC, this information is fully leveraged, and only $m_{\textrm{fgds}}^{j}(\hat{n})$ moments are filtered out at pixel $\hat{n}$ and needlet scale $j$. 

  Once the optimal number of moments to deproject is determined across the sky and needlet scales, ocMILC selects the optimal configuration of $m^j_{\textrm{fgds}}$ moments, pivot values, and deprojection coefficients among the following sets of
 \begin{enumerate}
\item moment SEDs:
\begin{equation}
    \begin{aligned}
   \Big[f_{\textrm{dust}},\ f_{\textrm{sync}},\ \partial_{\beta_{\textrm{d}}}f_{\textrm{dust}},\ \partial_{\beta_{\textrm{s}}}f_{\textrm{sync}},\ \partial_{T_{\textrm{d}}}f_{\textrm{dust}},\ \partial^{2}_{\beta_{\textrm{d}}}f_{\textrm{dust}},
    \partial_{\beta_{\textrm{d}}}\partial_{T_{\textrm{d}}}f_{\textrm{dust}},\ \partial^{2}_{\beta_{\textrm{s}}}f_{\textrm{sync}},\ \partial^{2}_{T_{\textrm{d}}}f_{\textrm{dust}} \Big]\,,
    \label{eq:list_moms}
\end{aligned}
\end{equation}
 \item dust pivot parameter values: 
 \begin{equation}
 \bar{\beta}_{\textrm{d}}\in [1.2,1.8]\ \text{with}\ \Delta \bar{\beta}_{\textrm{d}} = 0.05,\ \bar{T}_{\textrm{d}}\in [17\,\text{K},22\,\text{K}]\ \text{with}\ \Delta \bar{T}_{\textrm{d}} = 0.25,
\end{equation}
\item and deprojection values: 
\begin{equation}
 \epsilon_{k}\in [-0.05,-0.01,-0.005,-0.001,-0.0005,0.,0.0005,0.001,0.005,0.01,0.05].
\end{equation}
\end{enumerate}
A distinct deprojection coefficient is assigned to each selected moment.
The optimal combination of moments, pivots, and deprojection coefficients is determined so as to minimize the cMILC output variance corrected for residual noise variance in each GNILC cluster:
\begin{equation}
        \langle c_{\textrm{cMILC}}^{2}\rangle - \langle n_{\textrm{res}}^{2}\rangle = w^{T}_{\textrm{cMILC}} \left(C\, - N\right)\, w_{\textrm{cMILC}}\,,
\label{eq:min_ocMILC}
\end{equation}
where $C$ represents the data covariance matrix and $N$ denotes the instrumental noise covariance matrix.

The GNILC partitioning of the sky into clusters of pixels characterized by uniform foreground complexity is used to recalculate the local covariance matrix of the data and ocMILC weights in each cluster independently, drawing inspiration from the multiclustering strategy of \citep{MCNILC}. The covariance matrices in Equation~\ref{eq:min_ocMILC} and the optimization procedure are thus computed within each independent GNILC region, representing a connected set of pixels sharing the same value of $m^j_{\textrm{fgds}}$. 
Therefore, the ocMILC CMB map is the solution with minimum local foreground variance obtained by partially (or fully) deprojecting $m^j_{\textrm{fgds}}$ moments with varying pivot parameters across the sky and angular scales.  

\section{Data analysis and results}
\label{sec:results}

In this section, we analyse the simulated \textit{PICO} \texttt{d1}, \texttt{d4}, \texttt{d7}, and \texttt{d12} data sets described in Section~\ref{sec:sims}, representing four different foreground scenarios,  using the ocMILC pipeline.
The performance of ocMILC is compared against that of the baseline cMILC approach and the fully blind NILC approach. 

\begin{figure}
	\centering
	\includegraphics[width=\textwidth]{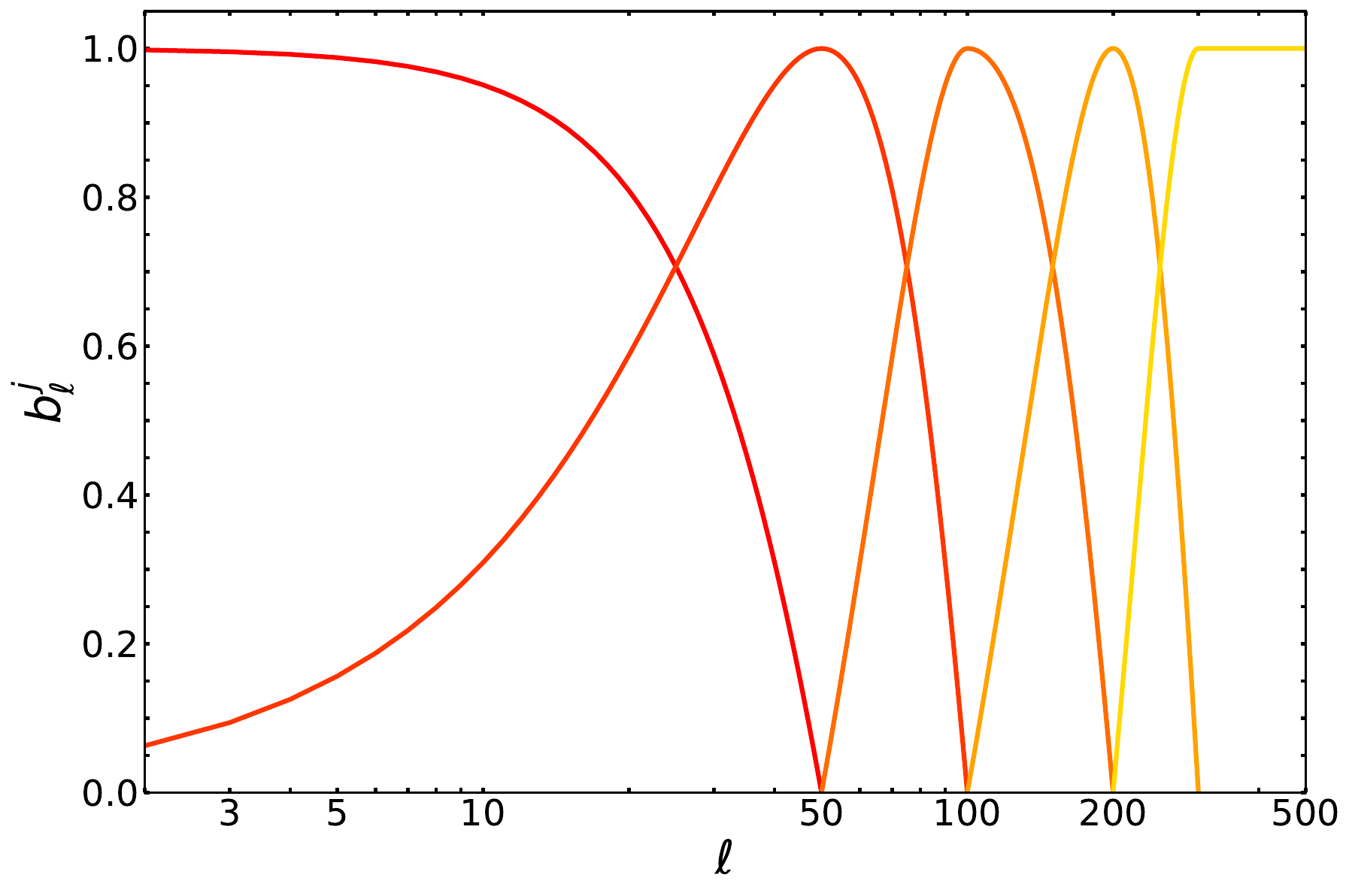}
	\caption{Cosine needlet bandpass windows in harmonic domain used in this work. }
	\label{fig:needlets}
\end{figure}

The data are decomposed into needlet coefficient maps (Equation~\ref{eq:needlets_map}) for analysis by the NILC, cMILC, and ocMILC pipelines. In this work, we employ cosine-shaped needlet bandpasses defined as
\begin{equation}
   b^{j}_\ell = \begin{cases}
       \cos\left( \frac{\ell_{\rm peak}^j\,  -\, \ell}{\ell_{\rm peak}^j\,  -\, \ell_{\rm peak}^{j-1}}\frac{\pi}{2}\right)   & \text{if}\ \ell_{\rm peak}^{j-1}\leq \ell < \ell_{\rm peak}^j \\
       \\
	\cos\left( \frac{\ell\, -\, \ell_{\rm peak}^j}{\ell_{\rm peak}^{j+1}\, -\, \ell_{\rm peak}^j}\frac{\pi}{2}\right)   & \text{if}\ \ell_{\rm peak}^j\leq \ell < \ell_{\rm peak}^{j+1}
   \end{cases}\,,
\end{equation}
where $\ell_{\rm peak} = 0,50,100,200,300$ for $j=1,\ldots,5$. These needlet bandpasses in harmonic domain are plotted in Figure~\ref{fig:needlets}, with each of them covering a specific range of multipoles.
Since the needlet bands have compact support in harmonic domain, the needlet coefficient maps, $\beta^{j}_{\nu}(\hat{n})$, are obtained at a HEALPix $N_{\textrm{side}}$ equal to the smallest power of $2$ larger than $\ell^{j}_{\textrm{max}}/2$, where $\ell^{j}_{\textrm{max}}$ is the maximum multipole of the $j^{\rm th}$ needlet band. 

The outcomes from the pipelines are first inspected by means of the angular power spectra of foreground and noise residuals, calculated by applying the weights derived from the specified data set to the foreground and noise inputs of each simulation. Such spectra are compared against the primordial CMB $B$-mode signal across a range of tensor-to-scalar ratio values, ${r\in [0.4\times 10^{-3},4\times 10^{-3}]}$, where $r=0.4\times 10^{-3}$ represents the actual upper bound at $95\%$ CL targeted by \textit{PICO} for its baseline instrumental configuration, as adopted in this work (refer to Section~\ref{sec:sims} for details on rescaling \textit{PICO} CBE to baseline sensitivity).
 Throughout the paper, to alleviate residual foreground contamination along the Galactic plane, the angular power spectra for \texttt{d1}, \texttt{d4}, and \texttt{d7} are computed over $f_{\rm sky}=70\%$ of the sky using the publicly available \textit{Planck} GAL70 mask applied to the maps. For  \texttt{d12}, a more conservative masking strategy is employed, by extending the \textit{Planck} GAL70 mask through smoothing the average \texttt{d12} foreground residuals with a $5^\circ$ beam and thresholding them to achieve a final sky fraction of $f_{\rm sky}=40\%$.  Although the  \texttt{d12} mask is constructed from foreground residuals that may not be accessible in real data analysis, it is employed solely for comparing outcomes from different pipelines within the same clean $40\%$ region of the sky. 
The angular power spectra are computed using the \emph{pymaster} Python package\footnote{\url{https://namaster.readthedocs.io/en/latest/}} of the MASTER estimator \citep{MASTER, MASTER2}, which facilitates proper deconvolution of masking and beam smoothing effects \citep{pymaster}. 

To further assess the quality of the CMB $B$-mode reconstruction, we examine the impact of residuals on the estimation of the tensor-to-scalar ratio, $r$. We estimate the tensor-to-scalar ratio from the recovered $B$-mode power spectra using the inverse-Wishart log-likelihood \citep{Hamimeche2008}:
\begin{equation}
    \log\mathcal{L}(r)=-f_{\textrm{sky}}\frac{2\ell +1}{2}\left[\frac{\hat{C}_{\ell}}{C_{\ell}(r)} + \log{C_{\ell}(r)} - \frac{2\ell -1}{2\ell +1}\log{\hat{C}_{\ell}}\right]\,.
\label{eq:like}
\end{equation}
Here $\hat{C}_{\ell}=A_{\textrm{lens}} C_{\ell}^{\textrm{lens}} + C_{\ell}^{\textrm{fgds}} + C_{\ell}^{\textrm{noi}}$ represents the angular power spectrum of the output CMB $B$-mode map, averaged over all realizations, with $C_{\ell}^{\textrm{lens}}$ the lensing power spectrum, $A_{\textrm{lens}}$ its residual amplitude after some expected level of delensing, $C_{\ell}^{\textrm{fgds}}$ and $C_{\ell}^{\textrm{noi}}$ the foreground and noise residual power, while $C_{\ell}(r) =  r\, C_{\ell}^{\,r=1} + A_{\textrm{lens}} C_{\ell}^{\textrm{lens}} + C_{\ell}^{\textrm{noi}}$ represents our theoretical power spectrum model, where $C_{\ell}^{\,r=1}$ is the primordial CMB $B$-mode power spectrum for tensor perturbations with $r=1$. We adopt $A_{\textrm{lens}}=0.27$ to align with the expected $73\%$ delensing capability of  \textit{PICO} \citep{PICO}. As our simulations do not include primordial tensor perturbations, the best-fit value of $r$ is only driven by residual foreground contamination, while its uncertainty is influenced by residual lensing, noise, and foregrounds.
While the likelihood function in Equation~\ref{eq:like} is not exact for a masked sky as it neglects correlations between different multipoles induced by masking, we ensure that these off-diagonal terms of the covariance matrix are negligible by binning the angular power spectrum with a constant binning scheme of $\Delta \ell = 10$. 

We begin by discussing some limitations in the performance of the baseline cMILC approach (Section~\ref{subsec:cmilc_res}), then present the results obtained using the ocMILC method for all foreground scenarios (Section~\ref{subsec:ocMILC_res}). This includes showcasing diagnostic maps illustrating the actual foreground complexity in each dataset (Section~\ref{subsec:GNILC_results}) and then presenting ocMILC outcomes related to CMB $B$-mode reconstruction and tensor-to-scalar ratio estimation (Section~\ref{subsec:optimization_res}).

\subsection{cMILC results and limitations}
\label{subsec:cmilc_res}

To illustrate some limitations of the baseline cMILC approach, we applied both NILC and cMILC pipelines to different \textit{PICO} data sets as discussed in Section~\ref{sec:sims}, using the needlet frame implementation presented above.
We implemented three versions of cMILC:
\begin{enumerate}
    \item cMILC-0, where all zeroth-order moments are deprojected.
    \item cMILC-01, where all zeroth- and first-order moments are deprojected.
    \item cMILC-012, where all zeroth-, first-, and second-order moments are deprojected.
\end{enumerate}
In all cases, the pivot values of the spectral parameters were held fixed to their average across the sky in the PySM \texttt{d1s1} model: $\bar{\beta}_{\textrm{s}} = -3$, $\bar{\beta}_{\textrm{d}} = 1.54$, and $\bar{T}_{\textrm{d}} = 19.6\ \textrm{K}$.

\begin{figure*}
	\centering
	\includegraphics[width=0.5\textwidth]{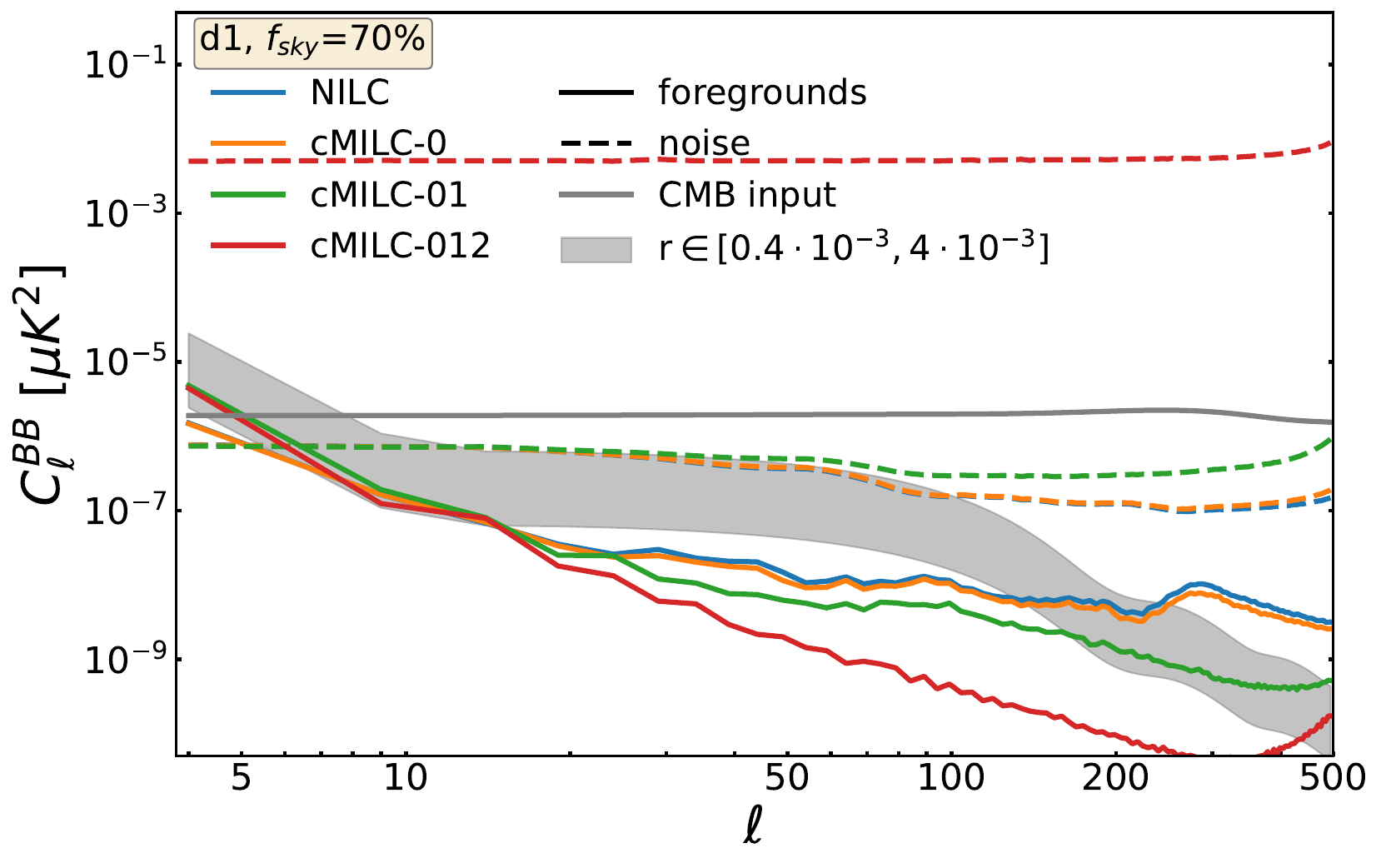}~
    \includegraphics[width=0.5\textwidth]{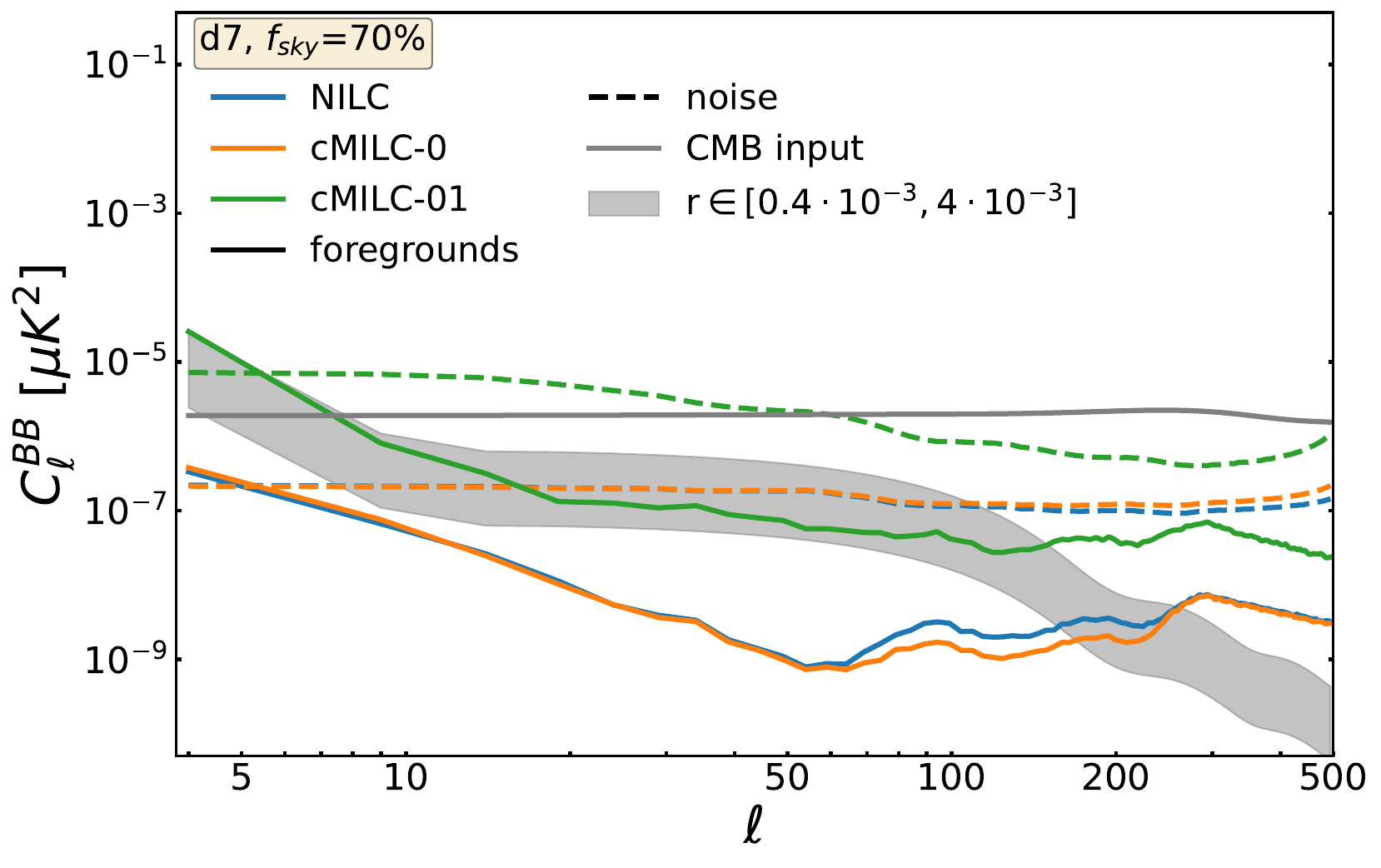}~
	\caption{
 Angular power spectra of foreground (solid) and noise (dashed) residuals in $B$-modes resulting from the application of NILC (blue), cMILC-0 (orange), and cMILC-01 (green) to \textit{PICO} \texttt{d1} (left) and \texttt{d7} (right) data sets. For the \texttt{d1} case, results of cMILC-012 (red) are also presented. All the power spectra are binned with $\Delta \ell=5$ and computed over $70\%$ of the sky. The grey shaded area denotes the range of primordial tensor CMB $B$-mode power spectra corresponding to tensor-to-scalar ratio values $r\in [0.0004,0.004]$.}
	\label{fig:cMILCs_diffmodels}
\end{figure*}

The results for \textit{PICO} data sets \texttt{d1} and \texttt{d7}  are reported in Figure~\ref{fig:cMILCs_diffmodels}, showing the angular power spectra of foreground and noise residuals. 
Residuals from the most constrained version cMILC-012 are exclusively shown for the \texttt{d1} data set.  The results from Figure~\ref{fig:cMILCs_diffmodels} lead to several conclusions. For the \textit{PICO} \texttt{d1} data set, the cMILC foreground residuals are lower than those of NILC at intermediate and small angular scales. However, on the largest angular scales, the averaging processes significantly distort the foreground SEDs, and the rigid cMILC deprojection fails to adequately suppress these low-$\ell$ distortions of the Galactic emission compared to simple variance minimization. As previously demonstrated in \cite{cMILC}, imposing the deprojection of more moments in Equation~\ref{eq:w_constraints} results in a penalty to the residual noise budget. This is particularly noticeable when examining the residuals from the application of cMILC-012 (where all zeroth-, first-, and second-order moments are deprojected) to the \textit{PICO} \texttt{d1} data set: the CMB solution, derived from the linear combination of 21 frequency channels with nine fully deprojected moments, suffers from a significant increase in noise contamination. Therefore, a compromise in the optimal number of moments to deproject must be found. 
Moreover, in this baseline cMILC approach, the moments have been projected out in the natural order of the expansion. However, there is no a priori reason to filter out lower-order moments first in the context of minimum-variance methodologies. It may be more effective to leave some major low-order contributions for blind variance minimization, as they are anyway prevailing in the budget of the observed data covariance matrix of Equation~\ref{eq:covariances}. When considering \textit{PICO} maps with different dust models, such as \texttt{d7} (right panel of Figure~\ref{fig:cMILCs_diffmodels}), the cMILC foreground residuals might even appear brighter compared to NILC in some cases (cMILC-01), indicating that the employed moment deprojection may not entirely capture the actual deviations of the dust model from the MBB SED. 
The application of ocMILC's optimized moment deprojection effectively overcomes all these limitations of the baseline cMILC approach, as demonstrated in the following sections.

\subsection{ocMILC results}
\label{subsec:ocMILC_res}

As described in Section \ref{sec:ocMILC}, the ocMILC pipeline consists of two main steps: a blind GNILC diagnosis of the effective foreground complexity to determine the optimal number of moment deprojection constraints across the sky and angular scales, and the subsequent optimized moment deprojection. The outcomes from these two steps are presented in Sections \ref{subsec:GNILC_results} and \ref{subsec:optimization_res}, respectively.

\subsubsection{Diagnostic maps of foreground complexity}
\label{subsec:GNILC_results}

We perform our GNILC diagnosis on the simulated \textit{PICO} \texttt{d1}, \texttt{d4}, \texttt{d7}, and \texttt{d12} data sets, each representing different foreground scenarios detailed in Section~\ref{sec:sims}. 
In this analysis, the CMB signal is considered part of the nuisance budget of GNILC, allowing us to estimate the number, $m_{\textrm{fgds}}$, of independent degrees of freedom associated with Galactic emission detectable by a telescope like \textit{PICO} for each foreground scenario. We perform this diagnosis for both $E$- and $B$-modes, presenting results for the baseline realization $n_{\textrm{sim}}=0$, although similar conclusions can be drawn for other realizations.

\begin{figure*}
	\centering
	\includegraphics[width=0.5\textwidth]{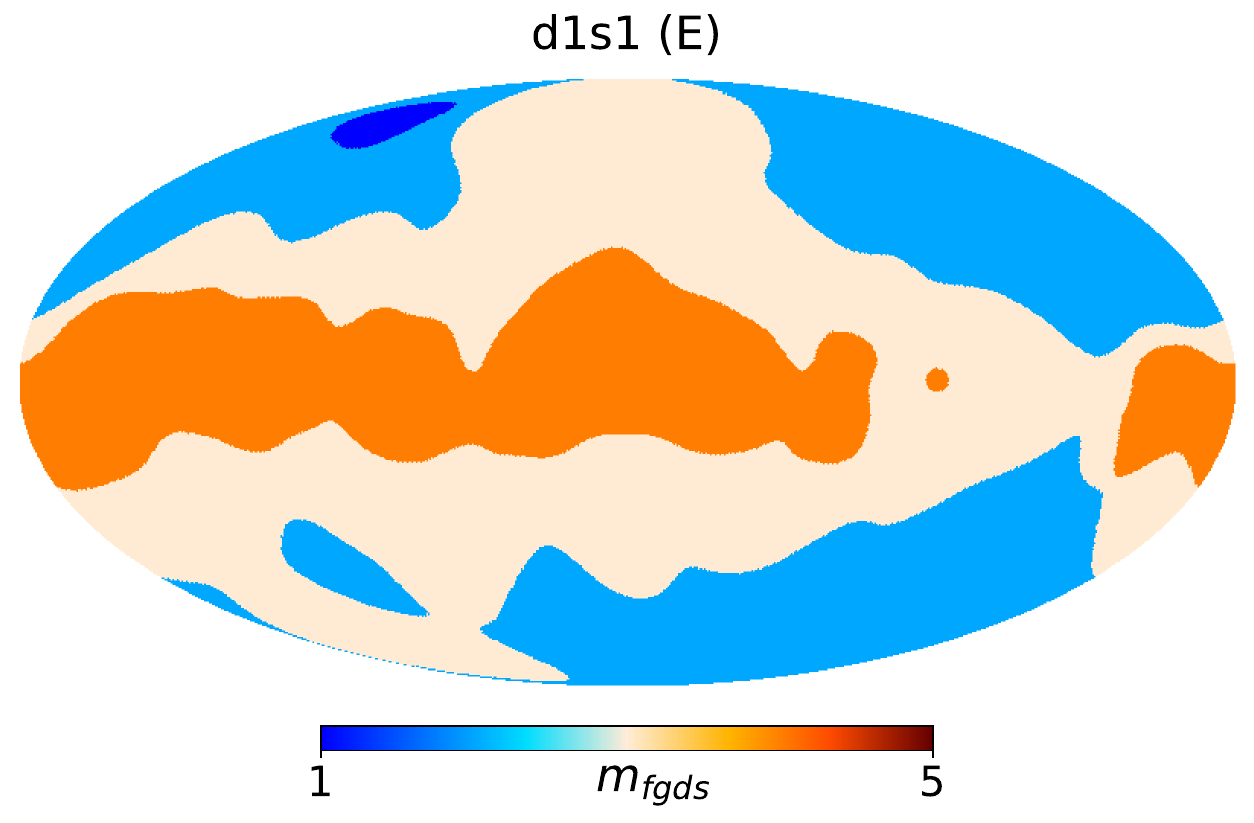}~
    \includegraphics[width=0.5\textwidth]{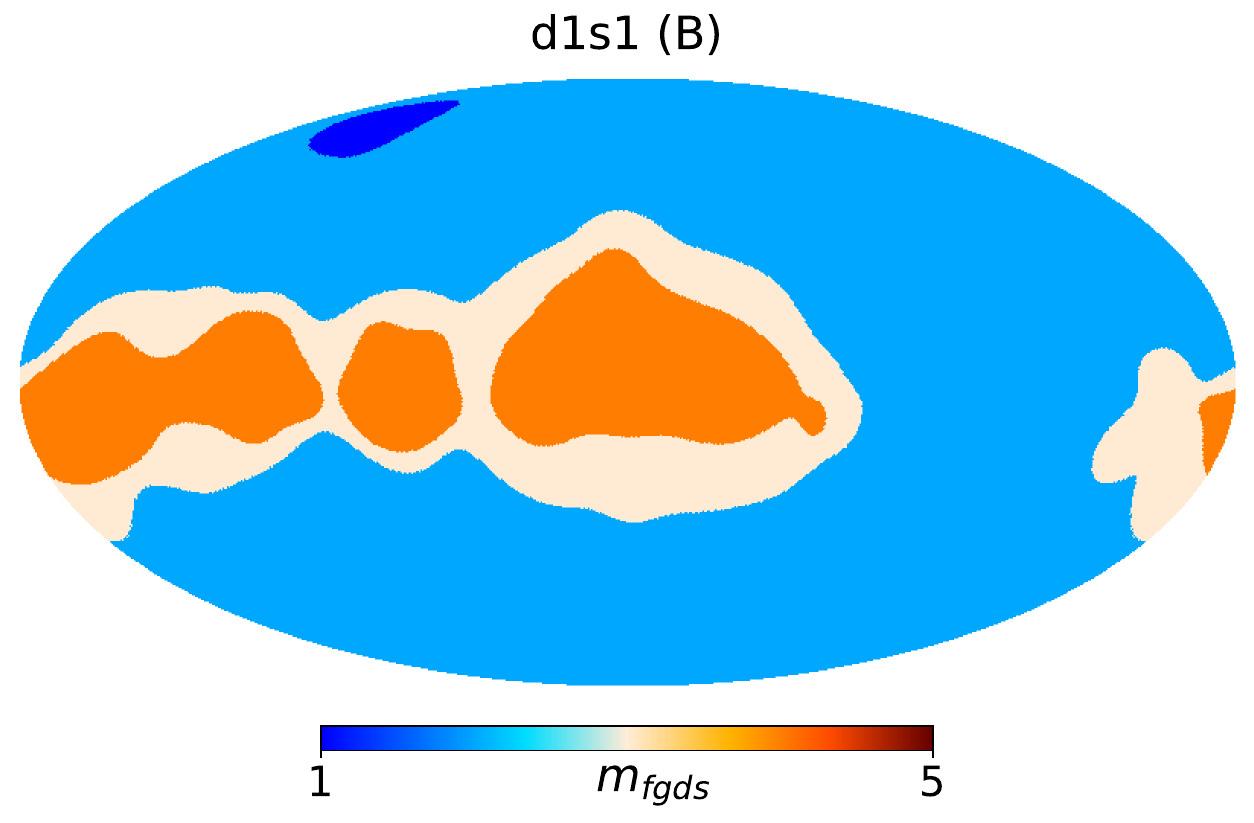} \\
	\includegraphics[width=0.5\textwidth]{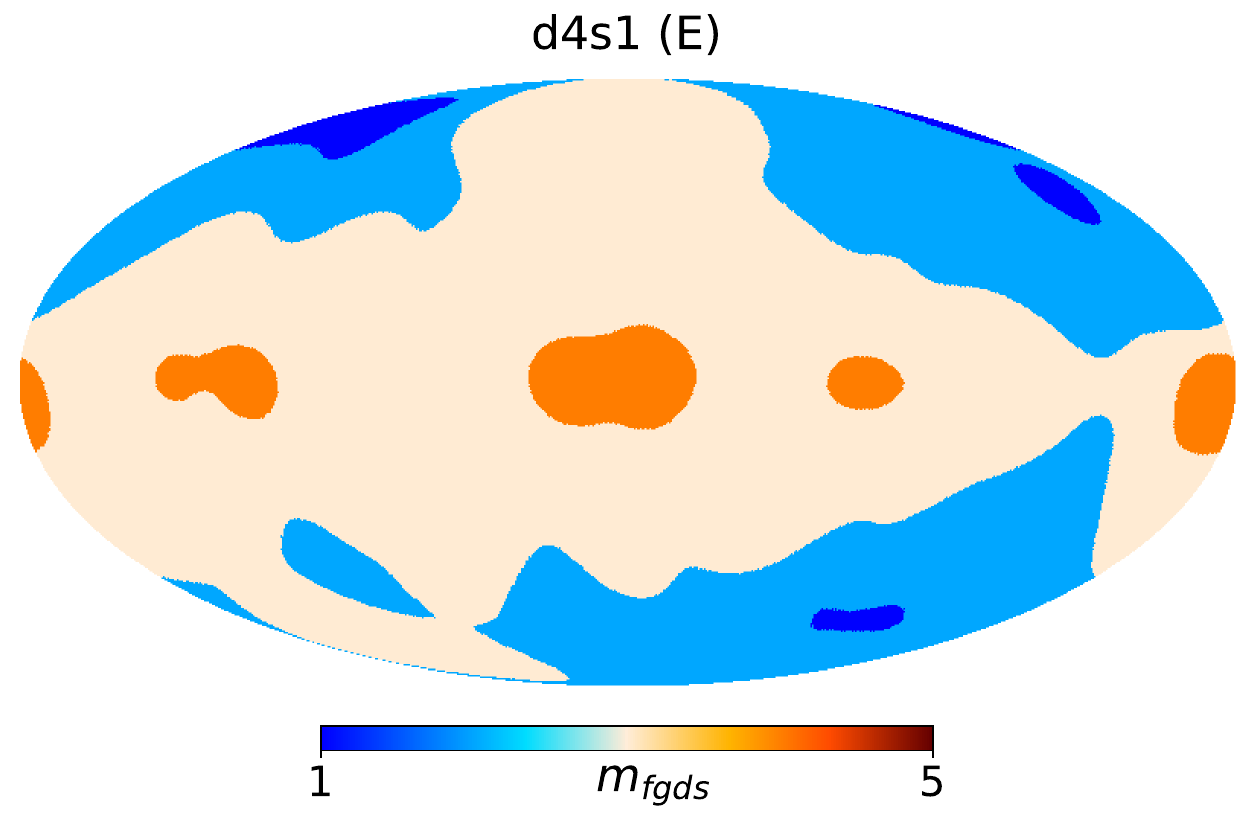}~
 \includegraphics[width=0.5\textwidth]{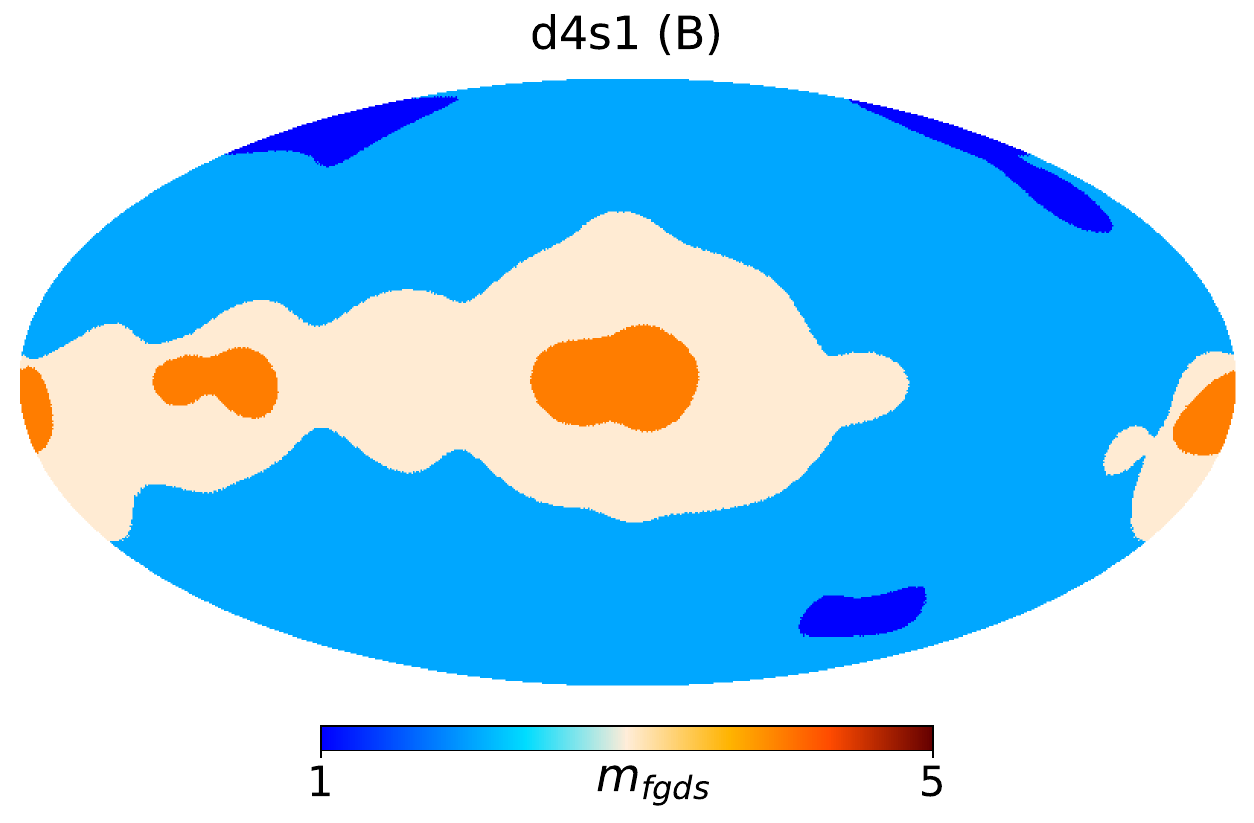} \\
	\includegraphics[width=0.5\textwidth]{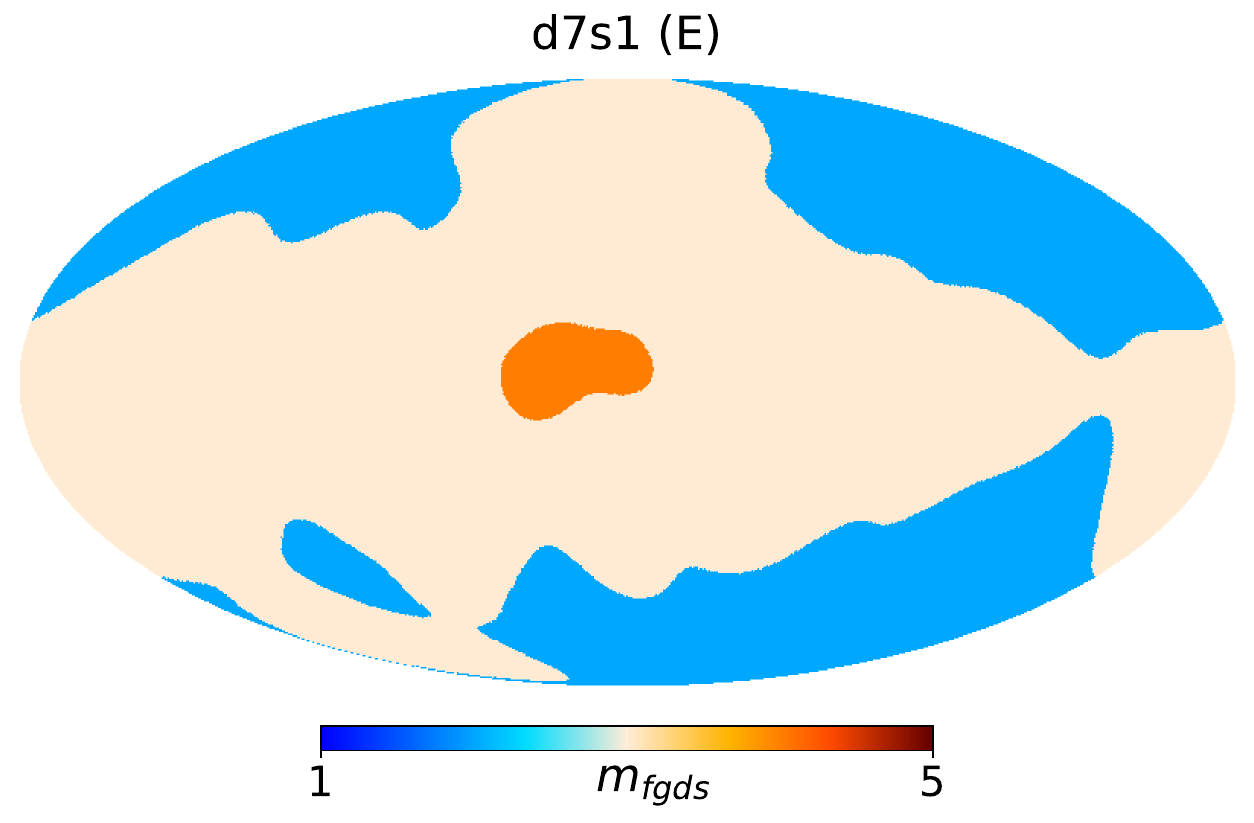}~ 
 \includegraphics[width=0.5\textwidth]{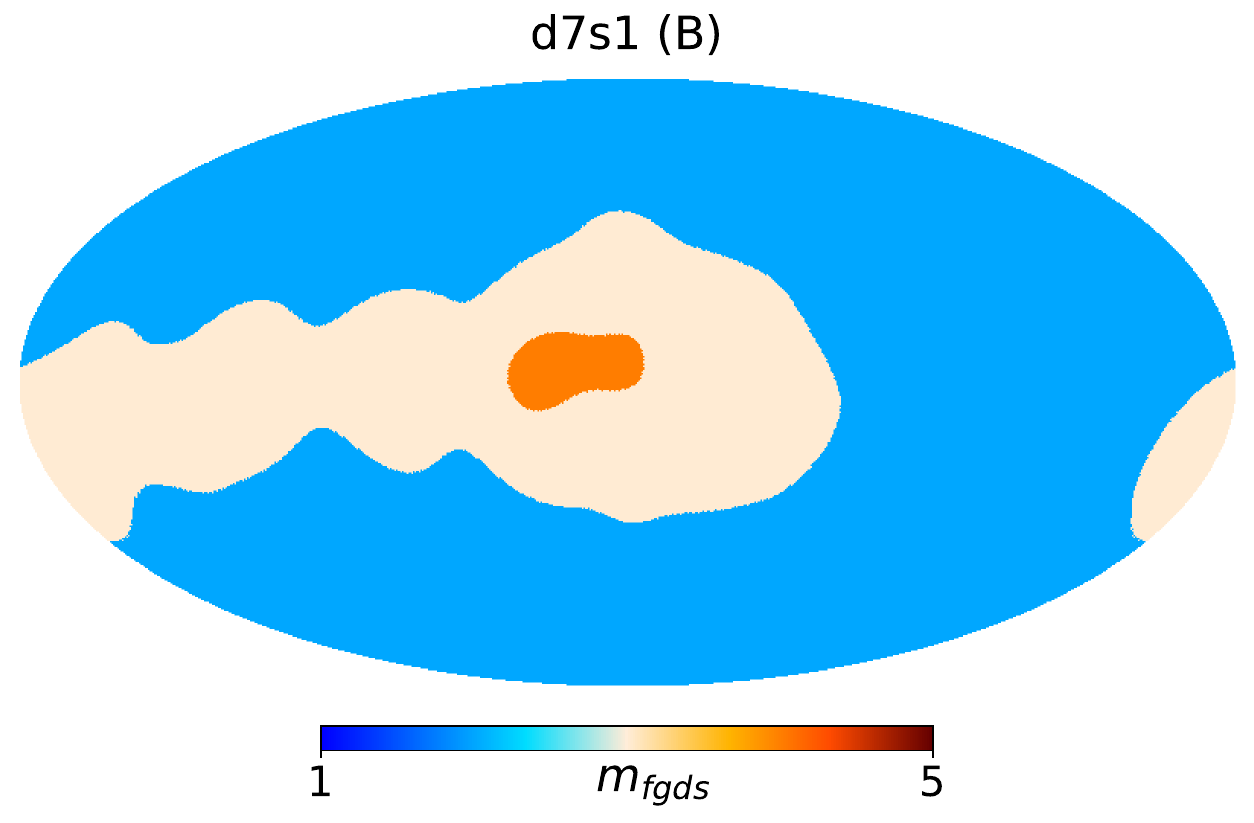}\\
	\includegraphics[width=0.5\textwidth]{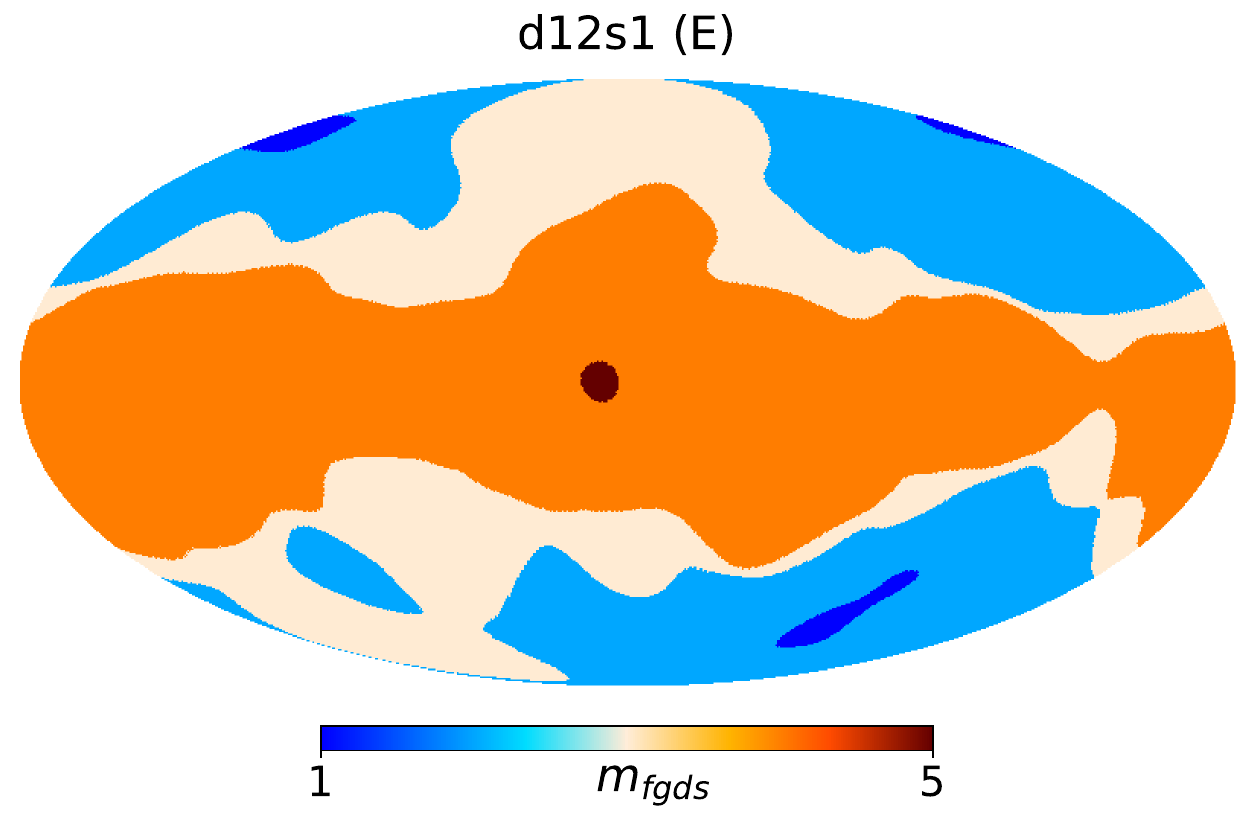}~
	\includegraphics[width=0.5\textwidth]{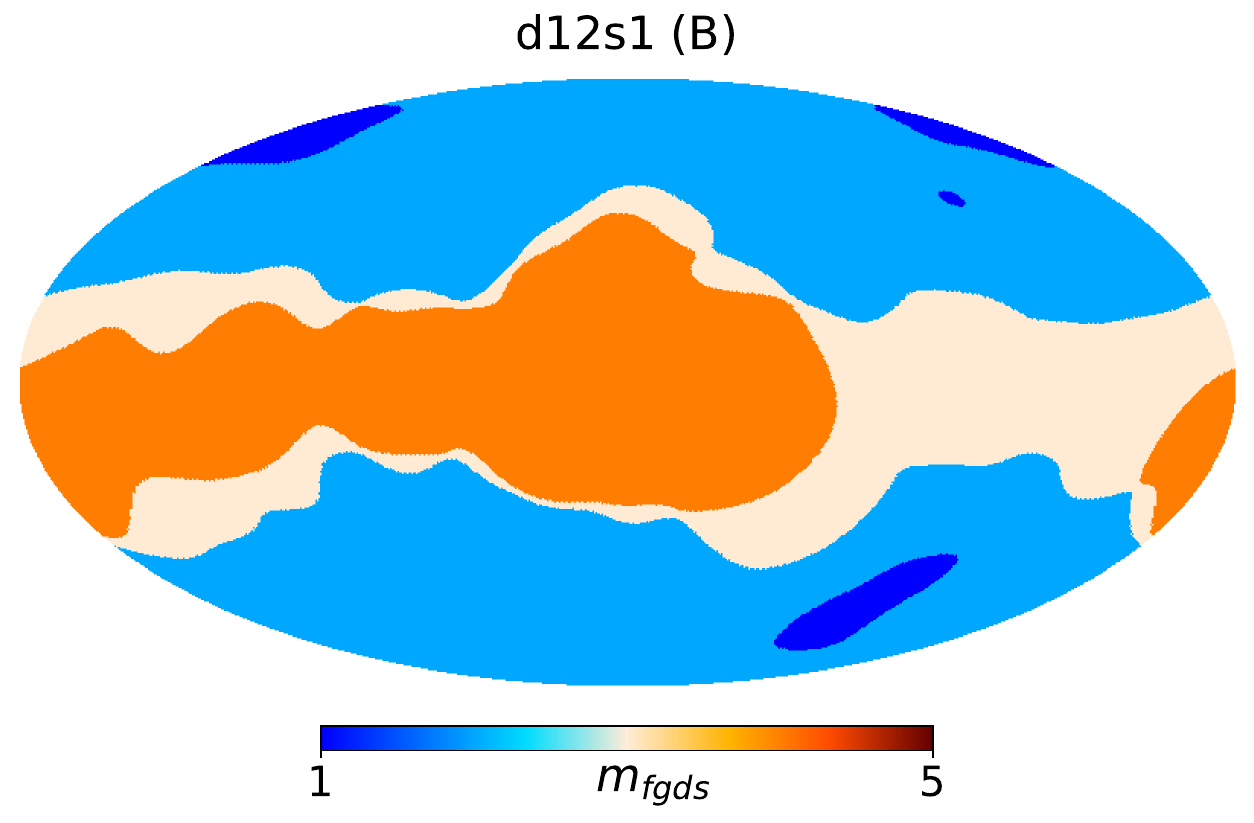}
	\caption{Maps of the detected number, $m^j_{\textrm{fgds}}$, of independent foreground $E$-mode components (left) and $B$-mode components (right) at needlet scale $j=5$. The results are shown for four distinct foreground scenarios, \texttt{d1} (first row), \texttt{d4} (second row), \texttt{d7} (third row), and \texttt{d12} (fourth row), and they reveal varying degrees of complexity. These diagnostic maps, which partition the sky into clusters of uniform foreground complexity, were generated using the GNILC methodology outlined in Section~\ref{sec:GNILC}, applied to full-sky frequency maps of \textit{PICO} $E$- and $B$-modes. }
	\label{fig:ms_PICO_diffmodels_nl4}
\end{figure*}

Maps of $m^j_{\textrm{fgds}}(\hat{n})$ for the fifth needlet scale ($j=5$) are shown in Figure~\ref{fig:ms_PICO_diffmodels_nl4} for the four distinct data sets.
They provide the effective number of independent  foreground components detected across the sky in both in $E$- and $B$-mode fields at this scale, thereby partitioning the sky into clusters of uniform foreground complexity. These diagnostic maps reveal different degrees of complexity across the sky depending on the underlying foreground scenario, even though all considered foreground models are consistent with \textit{Planck} observations. In particular, \texttt{d12} and \texttt{d1} models exhibit a larger number of independent foreground modes detected across more extensive sky regions compared to \texttt{d4} and \texttt{d7} models, with indications of higher complexity for \texttt{d12}. These results hold for both $E$- and $B$-modes.  
The effective lower complexity of \texttt{d4} with two MBB components for dust compared to the single-MBB \texttt{d1} model, is attributed to the absence of spatial variations of each MBB spectral index in \texttt{d4}.
Indeed, unlike the \texttt{d1} model, where spatial variations of the spectral index induce higher-order moments, the constant spectral indices of the two MBB components in \texttt{d4} results in the absence of moments beyond zeroth-order in Equation \ref{eq:dust_exp}, thereby reducing the list of effective components contributing to the overall foreground emission.
The highest level of complexity detected for \texttt{d12} can be attributed to the multi-layer structure of the dust emission in this model. 
Also in this case, not all the coadded seven physical components are detected by the GNILC diagnosis, as some may be correlated or below the noise level.
In summary, with high polarization sensitivity, as in the case of \textit{PICO}, our approach offers a reliable tool for discriminating between different foreground models. 
Moreover, the observed decrease in the effective number of independent foreground modes towards high Galactic latitudes aligns with the spatial extent of Galactic emission in the sky and the prevalence of noise in the data at these latitudes.
In $E$-modes, we also observe larger values of $m_{\textrm{fgds}}$ than in $B$-modes over larger regions of the sky. This discrepancy arises from the foreground $E/B$ ratio, which exceeds $2$. However, this trend is partially compensated by the even higher CMB $E/B$ ratio, mitigating the discrepancy between $E$- and $B$-modes concerning the number of independent foreground modes.

In Table~\ref{tab:ms_nl} we report the excursion set of $m_{\textrm{fgds}}$ observed in the four \textit{PICO} data sets at the five different needlet scales. For each value, the corresponding sky fraction is also reported in parentheses. As expected, the effective number of independent degrees of freedom of the Galactic emission decreases towards smaller angular scales, where the foreground power decreases while the CMB and noise have relatively larger power.  
Differences among the four data sets are noticeable. The \texttt{d1} and \texttt{d12} data sets exhibit larger values of $m_{\textrm{fgds}}$ compared to \texttt{d4} and \texttt{d7} at the largest angular scales and over larger portions of the sky at intermediate and small angular scales, both in $E$- and $B$-modes. Compared to other data sets, the \texttt{d12} data set, which incorporates multilayer dust emission, exhibits higher complexity in broader regions of the sky for all needlet scales. 
Again, in general, higher values of $m_{\textrm{fgds}}$ are found on larger portions of the sky for $E$-modes. At $j=2$, corresponding to angular scales of $\sim 3\degree$, we find no sensible differences among the first three foreground models, as we reach the resolution limit of original Planck reference templates used in these PySM models. Conversely, at smaller angular scales, artificial fluctuations have been added to these templates through different power spectra extrapolations \cite{pysm}, thus returning again visible differences across the sky among the three models as diagnosed by GNILC.

\begin{table*}
\caption{Number $m_{\textrm{fgds}}$ of independent modes characterizing Galactic foreground emission, as detected by GNILC within each needlet band, for the \textit{PICO} baseline configuration and four distinct foreground scenarios. 
The results are presented separately for $E$-modes (top table) and $B$-modes (bottom table). In cases where multiple values are observed, the corresponding sky fraction (in percentage) associated with each value of $m_{\textrm{fgds}}$ is provided in parentheses.}
\label{tab:ms_nl}
\centering
\begin{tabular}{c}
{$E$-modes}
\end{tabular} \\
\centering
\begin{adjustwidth}{-2.2cm}{}
\hspace{1.7cm}\resizebox{16.5cm}{!}{\begin{tabular}{|c|ccccc|}
\hline 
      \multirow{2}{*}{Model}
      &
      \multicolumn{5}{c|}{$m_{\textrm{fgds}}$\ ($f_{\textrm{sky}}(\%)$)} \\
      & $j=1$ & $j=2$ & $j=3$ & $j=4$ & $j=5$ \\
      \hline
  \texttt{d1s1} & $6$ & $5\ (97),6\ (3)$ & $3\ (3),4\ (81),5\ (16)$&  $3\ (36.8),4\ (63),5\ (0.2)$ & $1\ (0.5) ,2\ (30.3),3\ (46.2),4\ (23)$ \\ [0.2cm]
  \texttt{d4s1} & $5$  & $4\ (1.2),5\ (98.8)$ & $3\ (54),4\ (31),5\ (15)$ &  $3\ (82),4\ (18)$ & $1\ (2),2\ (30),3\ (63),4\ (5)$ \\ [0.2cm]
  \texttt{d7s1} & $5$  & $5$ & $3\ (13),4\ (71),5\ (16)$&  $3\ (62),4\ (37.9),5\ (0.1)$ & $2\ (31.3),3\ (67),4\ (1.7)$ \\ [0.2cm]
  \texttt{d12s1} & $6$ & $5\ (11),6\ (89)$ & $4\ (70),5\ (26),6\ (4)$ &  $3\ (18),4\ (80),5\ (3)$ & $1\ (1),2\ (27.8),3\ (26.5),4\ (44.5),5\ (0.2)$ \\
 \hline
\end{tabular} }\vspace{0.2 cm}\\
\end{adjustwidth}
\begin{tabular}{c}
{$B$-modes}
\end{tabular} \\
\begin{adjustwidth}{-2.2cm}{}
\hspace{1.7cm}\resizebox{16.5cm}{!}{\begin{tabular}{|c|ccccc|}
\hline 
      \multirow{2}{*}{Model}
      & \multicolumn{5}{c|}{$m_{\textrm{fgds}}$\ ($f_{\textrm{sky}}(\%)$)} \\
      & $j=1$ & $j=2$ & $j=3$ & $j=4$ & $j=5$ \\
      \hline
  \texttt{d1s1} & $6$ & $4\ (13),5\ (87)$ & $3\ (4),4\ (95),5\ (1)$ & $2\ (11.5),3\ (28),4\ (60.5)$ & $1\ (0.7),2\ (68),3\ (16.3),4\ (15)$ \\ [0.2cm]
  \texttt{d4s1} & $5$ & $4\ (16),5\ (84)$ & $3\ (59),4\ (40),5\ (1)$ & $2\ (12),3\ (72),4\ (16)$ & $1\ (3),2\ (67),3\ (25.5),4\ (4.5)$ \\ [0.2cm]
  \texttt{d7s1} & $5$ & $4\ (13),5\ (87)$ & $3\ (17),4\ (82),5\ (1)$ & $2\ (12),3\ (53),4\ (35)$ & $2\ (71.5),3\ (27),4\ (1.5)$ \\ [0.2cm]
  \texttt{d12s1} & $6\ (82),7\ (18)$ & $5\ (17),6\ (83)$ & $4\ (77),5\ (23)$ & $2\ (9),3\ (15.3),4\ (75),5\ (0.7)$ & $1\ (2),2\ (51),3\ (20.4),4\ (26.6)$ \\
 \hline
\end{tabular}}
\end{adjustwidth}
\end{table*}

\subsubsection{CMB $B$-mode reconstruction}
\label{subsec:optimization_res}

The clusters of pixels depicted in the diagnostic maps of Figure~\ref{fig:ms_PICO_diffmodels_nl4}, each featuring distinct dimensions of the foreground subspace, guide ocMILC in determining the appropriate number of constraints for moment deprojection in each cluster at each needlet scale. This constitutes the first layer of optimization in ocMILC. The next three layers of optimization in ocMILC consist in selecting, for each specified number of constraints, the optimal combination of moments, pivot values, and deprojection coefficients to minimize residual foreground contamination in CMB $B$-mode reconstruction, as detailed in Section~\ref{subsec:optimization}.

\begin{figure}
	\centering
    \includegraphics[width=\textwidth]{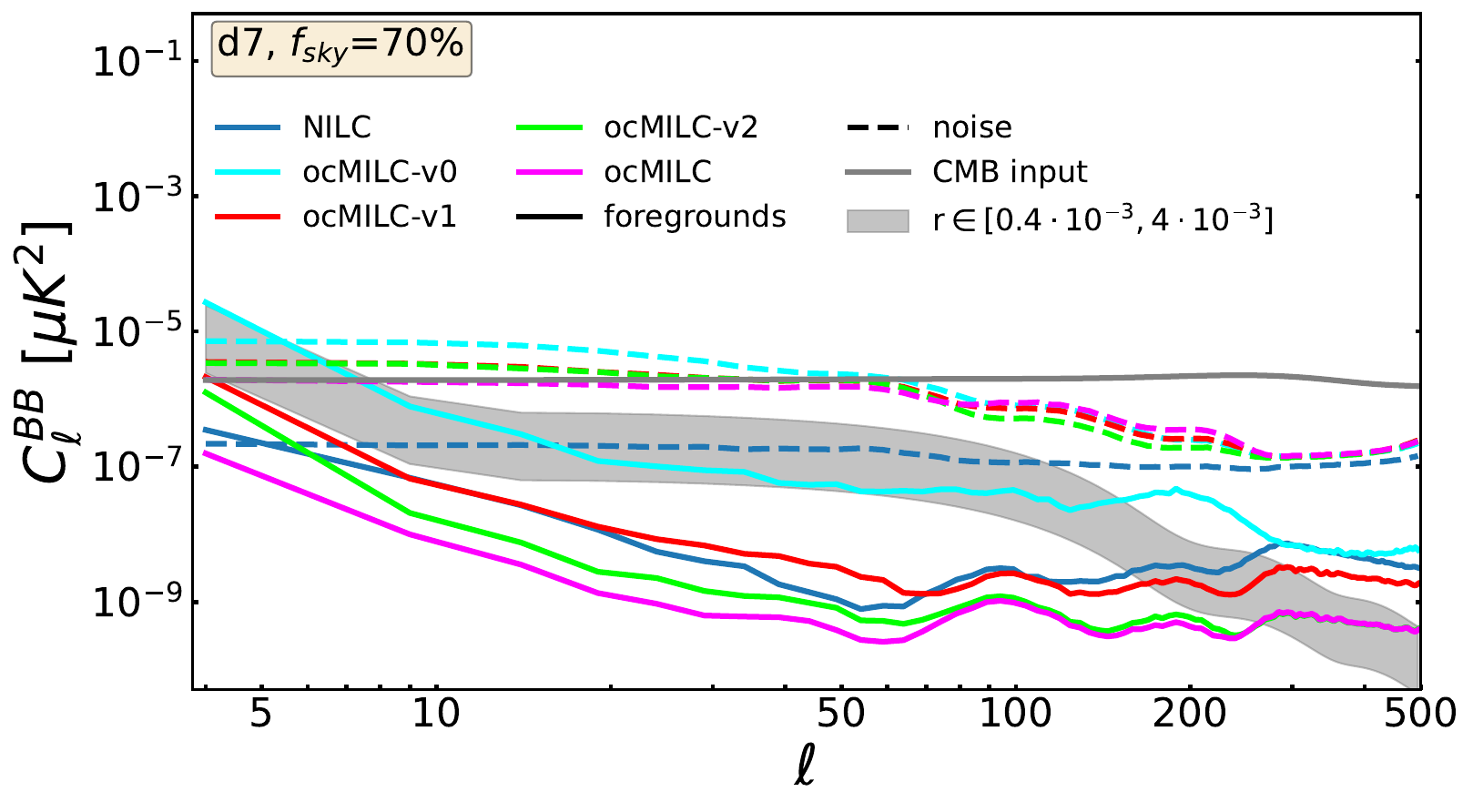}
	\caption{Angular power spectra of foreground (solid) and noise (dashed) residuals over $f_{\textrm{sky}}=70\%$ of the sky resulting from the application of NILC (blue), ocMILC-v0 (cyan), ocMILC-v1 (red), ocMILC-v2 (green), and ocMILC (magenta) to the \textit{PICO} \texttt{d7} $B$-mode data set. This illustrates the progressive improvement in residuals achieved by combining several layers of optimization in ocMILC.}
	\label{fig:ocMILC_diffversions}
\end{figure}

The improvement induced by combining the various layers of optimization in ocMILC is demonstrated by comparing the power spectrum of ocMILC residuals with those obtained when optimization is exclusively applied to specific layers. This comparison is performed using the example of the \textit{PICO}  \texttt{d7} data set (similar results are obtained for other models) and considering the following intermediate ocMILC pipelines:
\begin{itemize}
\item ocMILC-v0, where optimization focuses on the number of moments, based on information from GNILC diagnostic maps, deprojecting the first $m_{\textrm{fgds}}$ moments in the natural order of the Taylor expansion (Equation~\ref{eq:list_moms}); 
\item ocMILC-v1, where optimization involves both the number and the set of moments, deprojecting the set of $m_{\textrm{fgds}}$ moments that lead to the lowest denoised output variance (Equation~\ref{eq:min_ocMILC}); 
\item ocMILC-v2, where optimization extends to the number of moments, set of moments, and pivot values.
\end{itemize}
The comparison results, shown in Figure~\ref{fig:ocMILC_diffversions}, reveal that deprojecting the first $m_{\textrm{fgds}}$ moments in the natural order of the Taylor expansion (ocMILC-v0) does not improve foreground subtraction across most scales, especially when the dust model, like \texttt{d7}, deviates substantially from a MBB. In such cases, nulling constraints result in a notable penalty in residual foreground variance attributed to higher-order moments. This suggests that the mitigation of foregrounds could have been better handled if some of the low-order moments had been left for variance minimization, while the higher-order moments had been deprojected. Optimizing the choice of moments (ocMILC-v1) indeed brings improvement, further enhanced by allowing the values of pivot parameters to vary freely (ocMILC-v2), enabling a more effective accommodation of the dust model through moment expansion. Enabling the weights to also partially deproject specific moments (ocMILC) provides additional flexibility to the blind variance minimization process, allowing for the mitigation of foreground modes not adequately captured by the moment expansion. This final refinement, in conjunction with the previous optimization steps, enables ocMILC to reconstruct the cleanest CMB solution among all considered intermediate pipelines for all data sets,  achieving minimum foreground variance with limited noise penalty. This increased flexibility proves advantageous in effectively handling various thermal dust models through moment expansion.

\begin{figure}
	\centering
    \includegraphics[width=\textwidth]{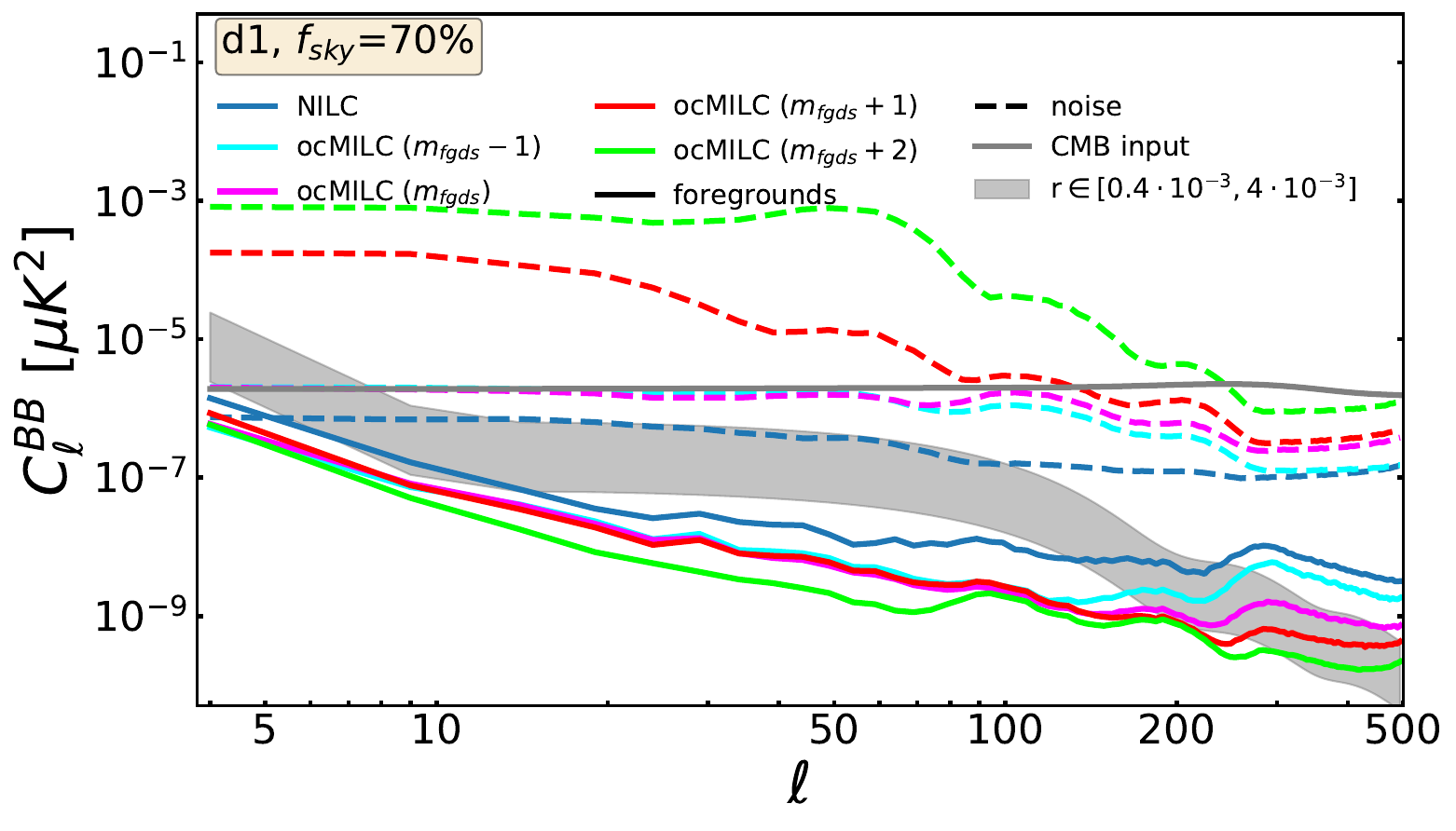}
	\caption{Angular power spectra of foreground (solid) and noise (dashed) residuals over $f_{\textrm{sky}}=70\%$ of the sky obtained from the application of NILC (blue) and ocMILC (others) to the \textit{PICO} \texttt{d1} $B$-mode data set, with variations in the number of moments deprojected compared to the number $m_{\textrm{fgds}}$ diagnosed by GNILC. The results of ocMILC are shown for four different cases, where, respectively, $m_{\textrm{fgds}}$ (magenta), $m_{\textrm{fgds}}-1$ (cyan), $m_{\textrm{fgds}}+1$ (red), and $m_{\textrm{fgds}}+2$ (green) foreground moments are deprojected across the sky and needlet scales.}
	\label{fig:ocMILC_extras}
\end{figure}

To validate the accuracy of the GNILC diagnosis in determining the optimal number $m_{\textrm{fgds}}$ of moments to deproject before incurring significant noise penalties, we apply ocMILC to the \textit{PICO} \texttt{d1} data set while imposing the deprojection of $m_{\textrm{fgds}}-1$, $m_{\textrm{fgds}}+1$, and $m_{\textrm{fgds}}+2$ moments in each point of the sky and for each needlet scale. The resulting angular power spectra are compared with those of NILC and the proper ocMILC, with $m_{\textrm{fgds}}$ moments deprojected, in Figure~\ref{fig:ocMILC_extras}. The outcomes clearly demonstrate that deprojecting one or two additional moments, beyond what is suggested by the GNILC diagnosis, leads to a significant increase in residual noise. Conversely, deprojecting $m_{\textrm{fgds}}-1$ moments results in larger Galactic contamination on intermediate and small scales. These results highlight the efficacy of the GNILC diagnosis in providing a data-driven assessment of the optimal number of moments for deprojection with ocMILC.

\begin{figure*}[h]
	\centering
	\includegraphics[width=0.53\textwidth]{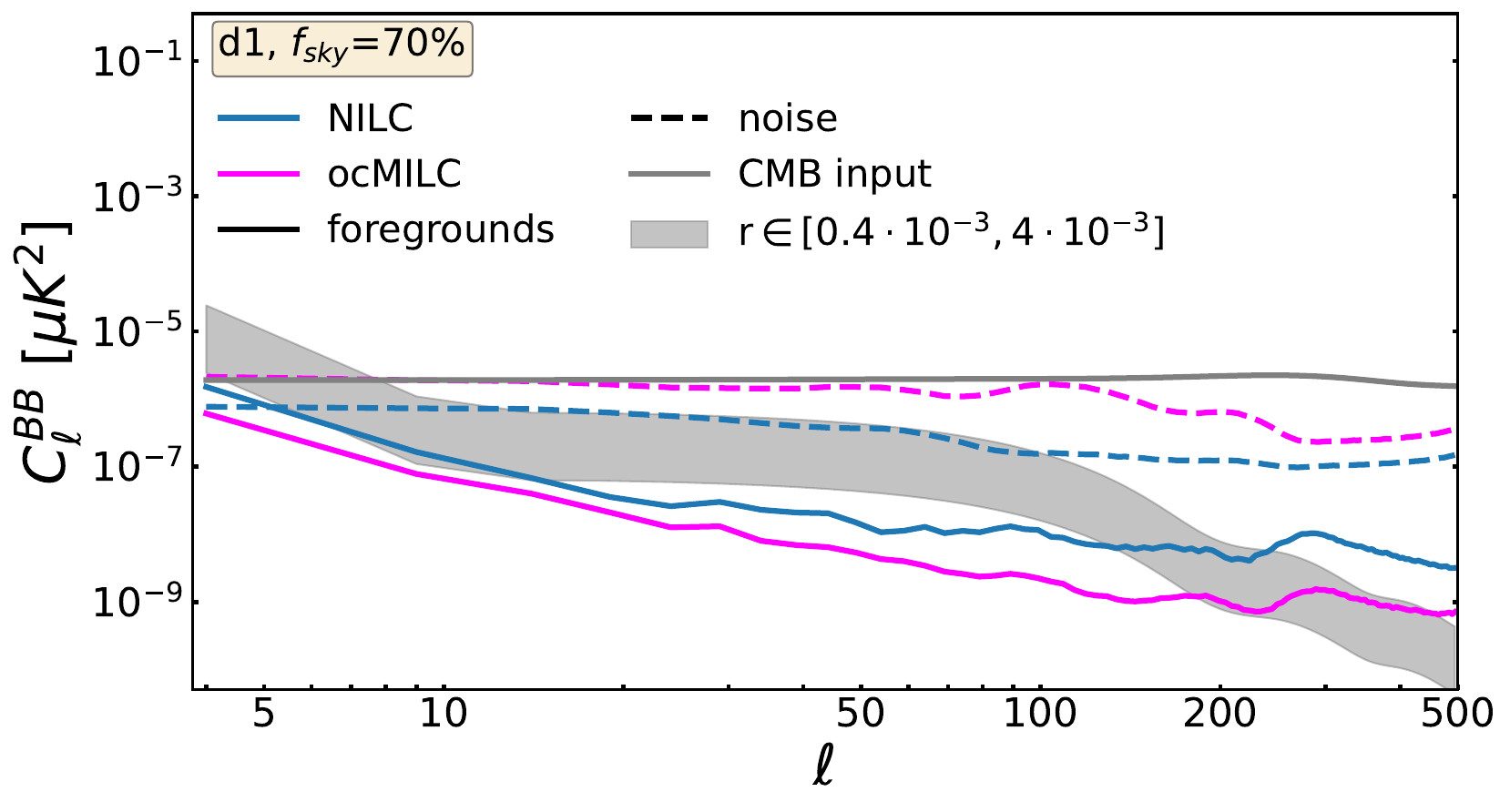}~
    \includegraphics[width=0.47\textwidth]{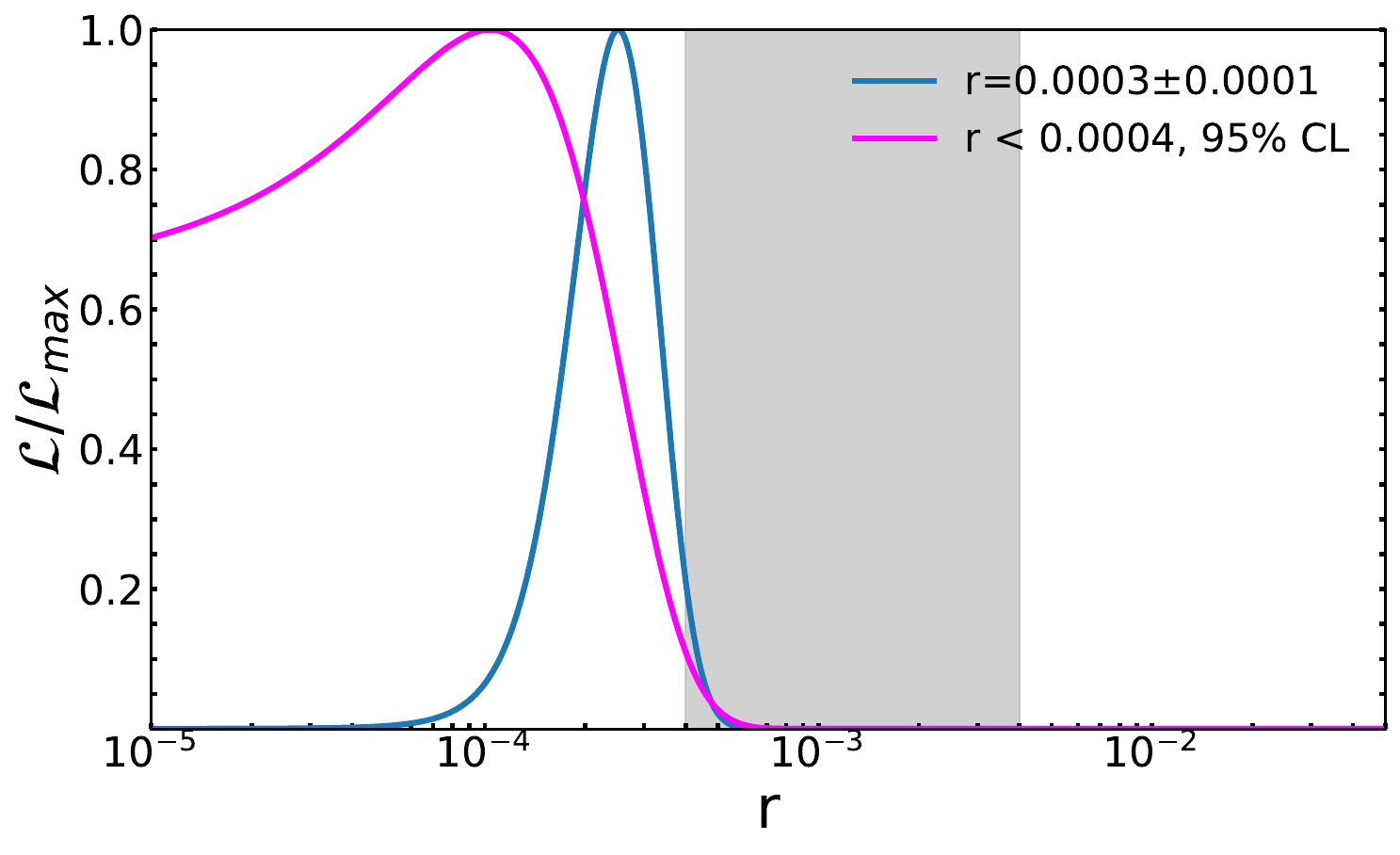}\\
    \includegraphics[width=0.53\textwidth]{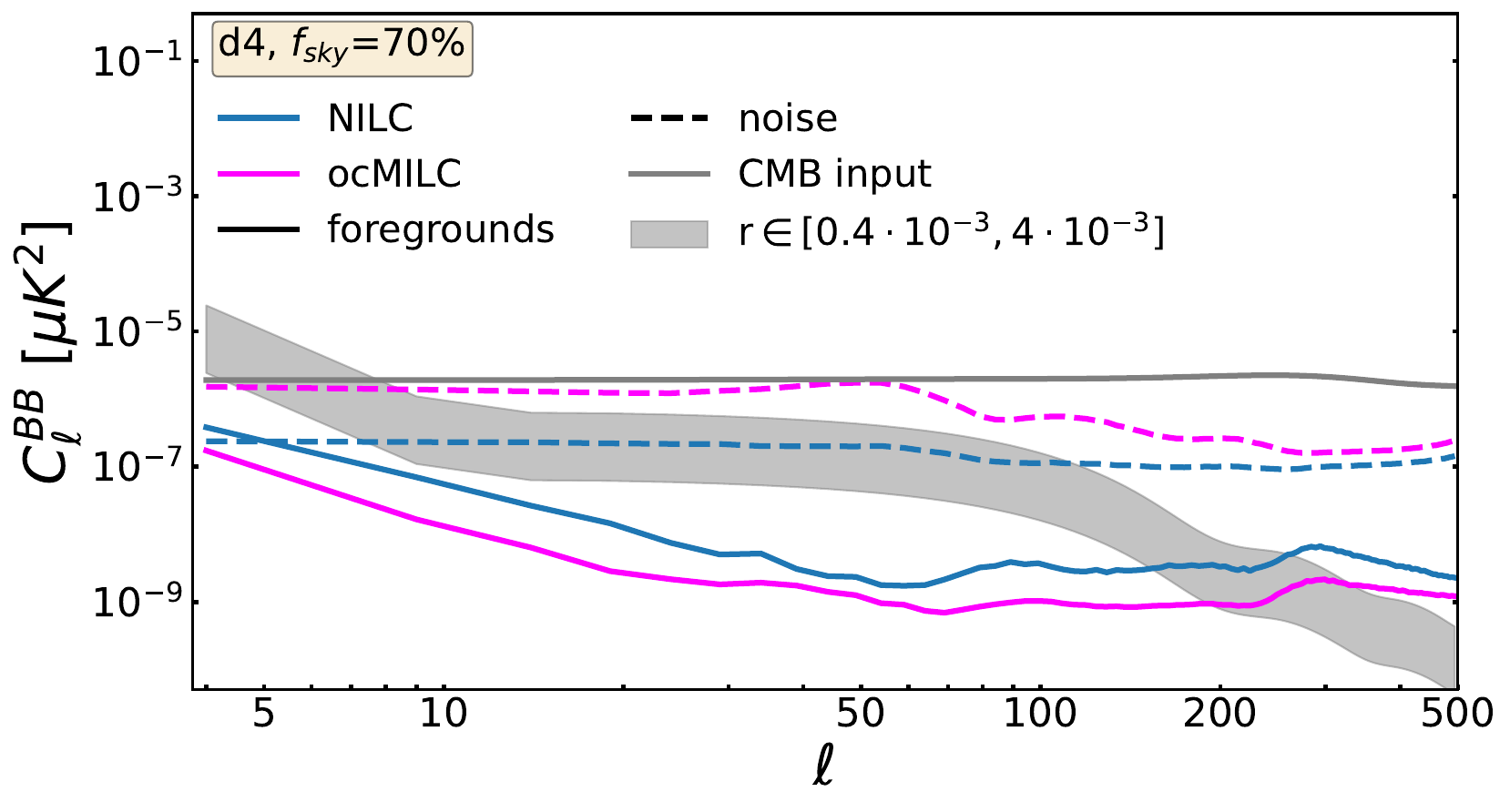}~
    \includegraphics[width=0.47\textwidth]{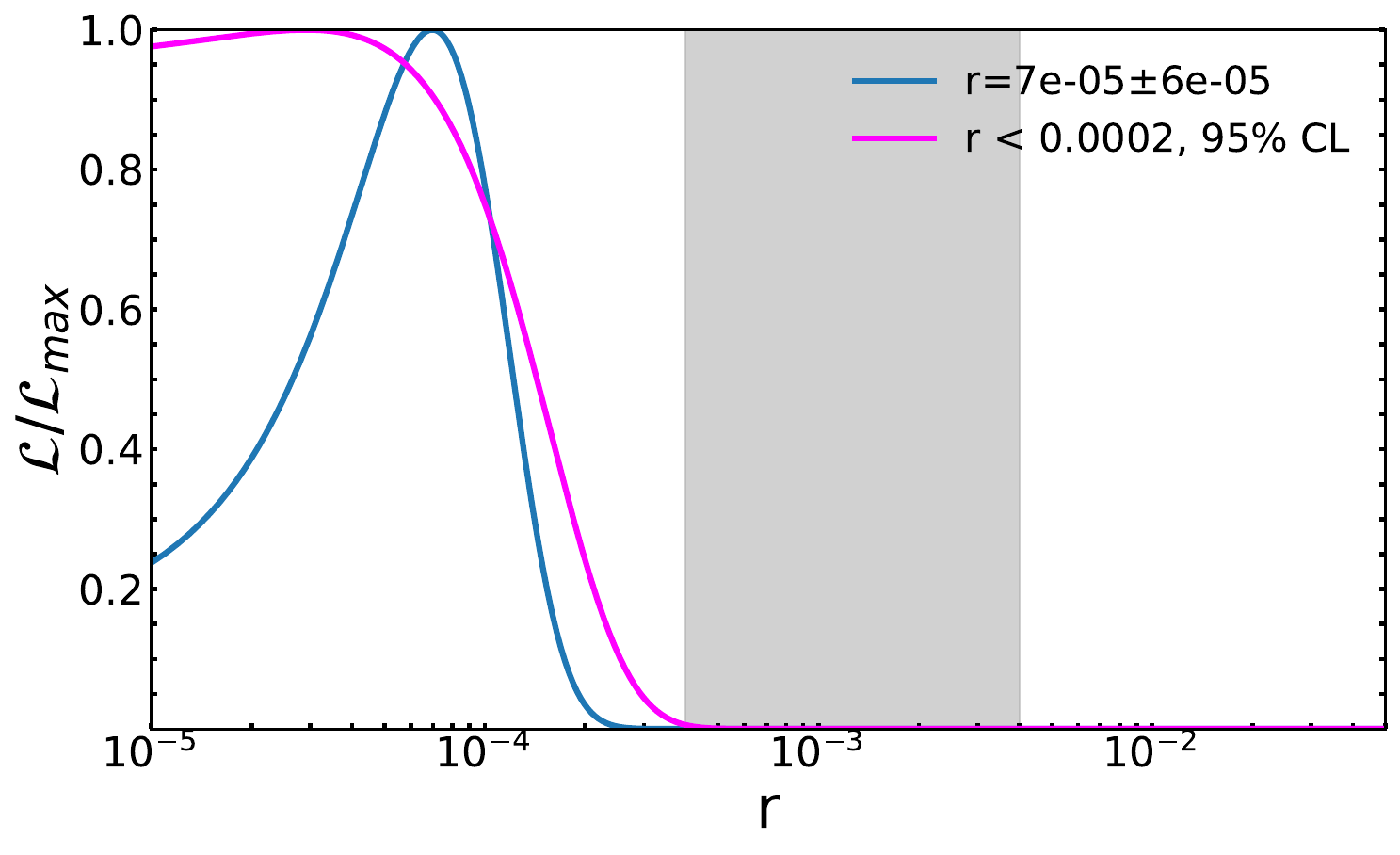}\\
    \includegraphics[width=0.53\textwidth]{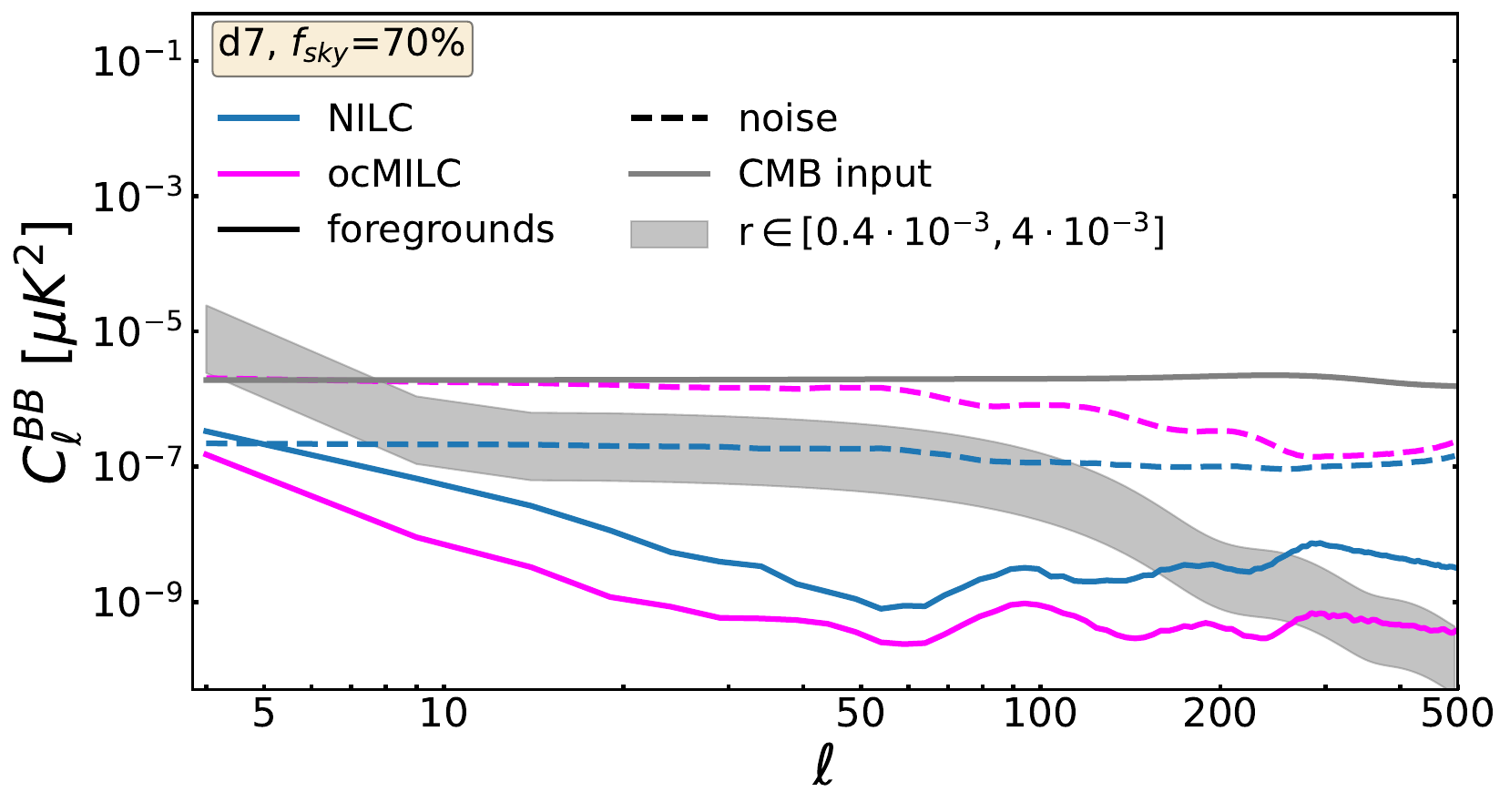}~
    \includegraphics[width=0.47\textwidth]{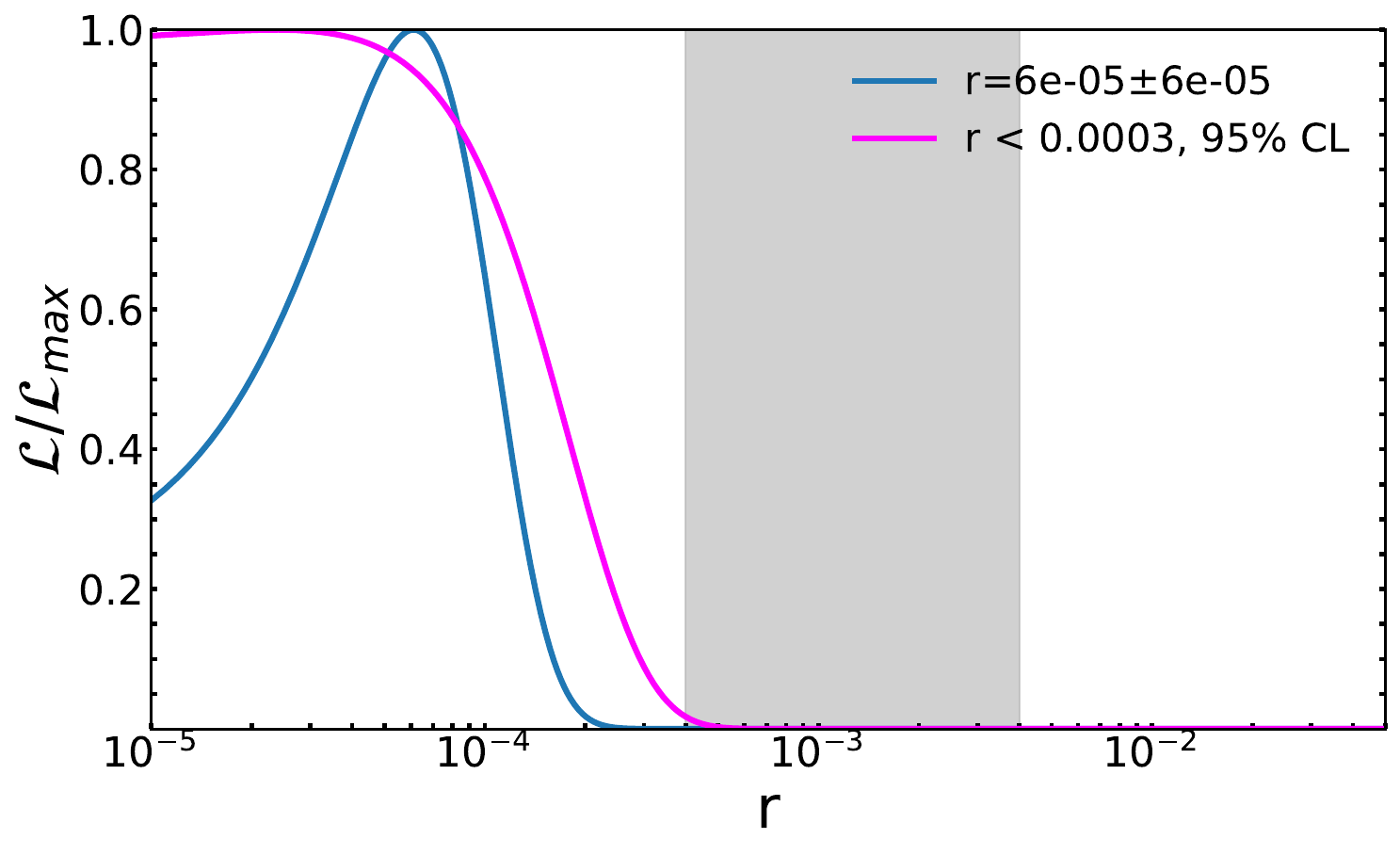}\\
    \includegraphics[width=0.53\textwidth]{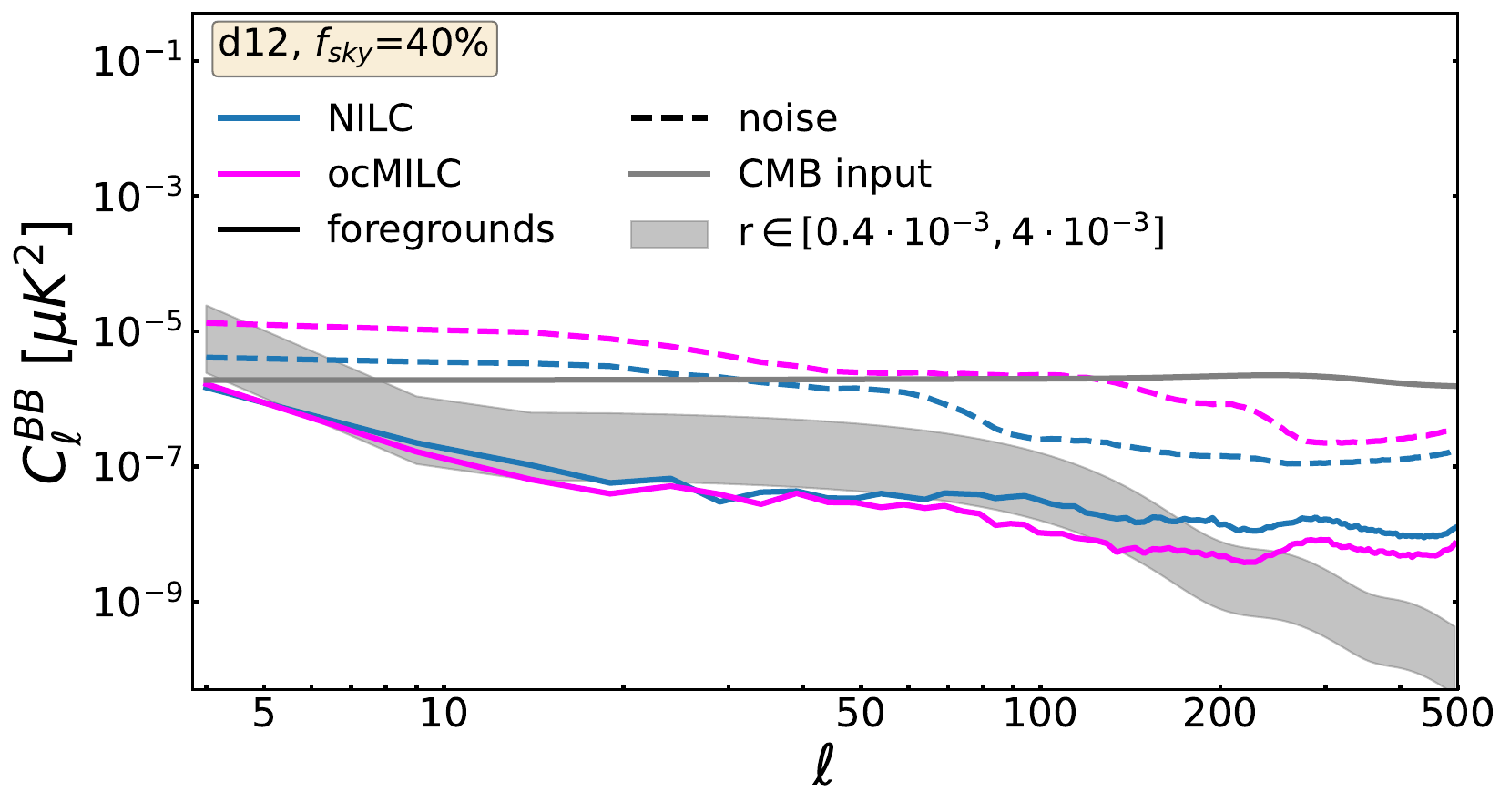}~
    \includegraphics[width=0.47\textwidth]{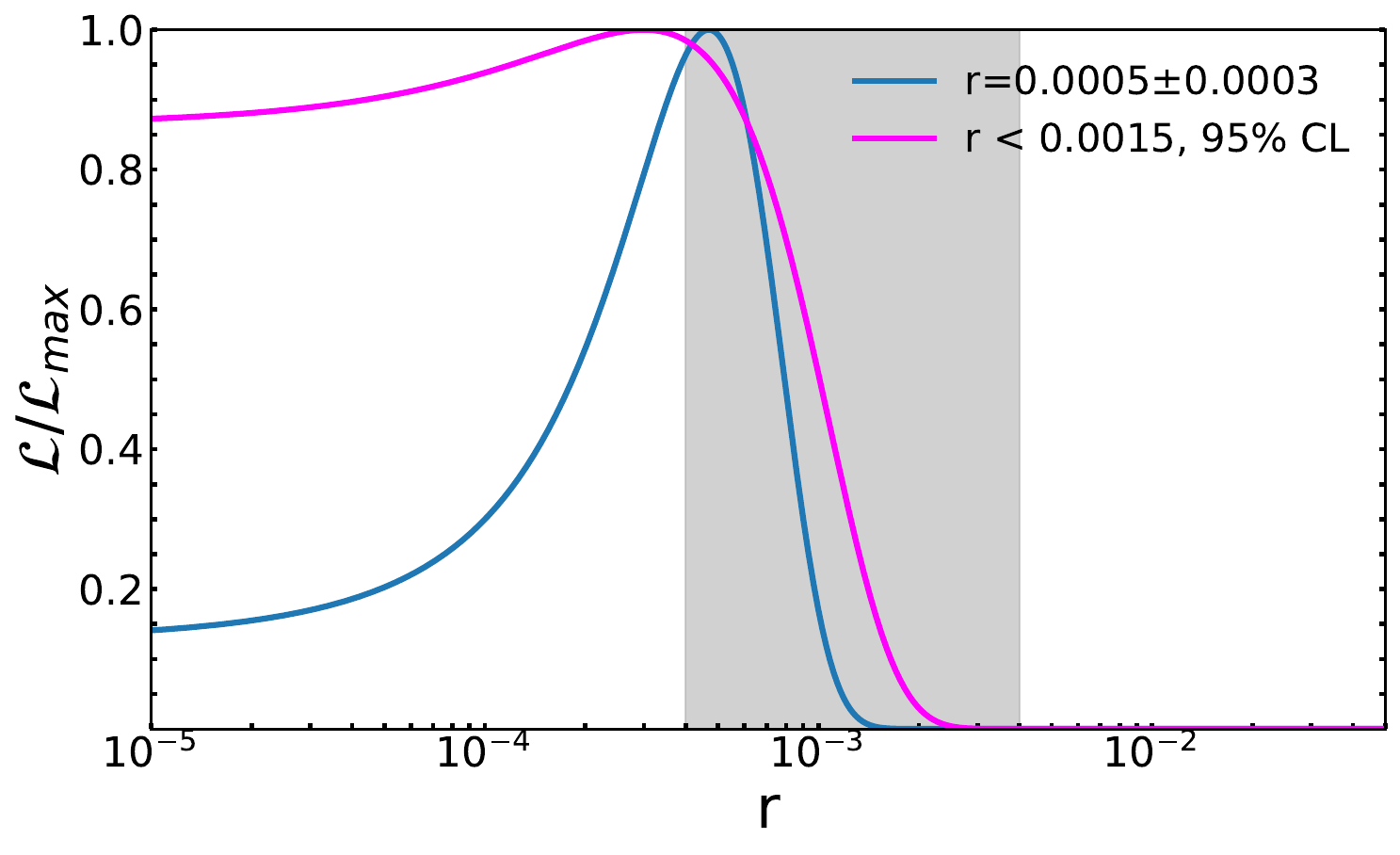}
	\caption{Left: Angular power spectra of foreground (solid) and noise (dashed) $B$-mode residuals obtained from the application of NILC (blue) and ocMILC (magenta) to \textit{PICO} \texttt{d1} (top), \texttt{d4} (second row), \texttt{d7} (third row), and \texttt{d12} (bottom) data sets. Power spectra have been computed over $f_{sky}=70\%$ of the sky for models \texttt{d1}, \texttt{d4}, and \texttt{d7}, and over $f_{sky}=40\%$ of the sky for model \texttt{d12}. 
	Right: Posterior distributions of the recovered tensor-to-scalar ratio for both pipelines. In all plots, the grey shaded area represents the range of values of the tensor-to-scalar ratio: $r\in [0.0004,0.004]$. }
	\label{fig:ocMILC_spectra}
\end{figure*}

The full ocMILC pipeline, integrating all layers of optimization, has been applied to all \textit{PICO} $B$-mode data sets, encompassing diverse foreground scenarios outlined in Section~\ref{sec:sims}. The angular power spectra of ocMILC residuals are compared with those of NILC for all foreground scenarios in Figure~\ref{fig:ocMILC_spectra}. 
For the \texttt{d1}, \texttt{d4}, and \texttt{d7} models, ocMILC yields significantly lower foreground residuals across all angular scales compared to NILC, with only a moderate increase in noise contamination relative to NILC. These improvements in terms of residual foreground contamination, regardless of the foreground model, can be attributed to the several layers of optimization in ocMILC: multiclustering from GNILC, varying number and set of deprojected moments, varying pivot parameters, and varying deprojection coefficients, across the sky and angular scales. 
For the most challenging foreground model \texttt{d12}, which includes multi-layer dust components along the line-of-sight and spectral decorrelation \cite{d12}, the reduction of residual foreground contamination is more modest than in other foreground scenarios but still outperforms NILC across angular scales overall.
For all the considered data sets, the reduced residual foreground contamination of the ocMILC CMB $B$-mode solution is achieved with only a moderate noise penalty relative to NILC, thanks to the information provided by the GNILC diagnosis. These results showcase the flexibility of ocMILC in effectively handling deviations of thermal dust emission from a simple MBB through optimized moment deprojection.

\begin{figure*}
	\centering
    \includegraphics[width=0.5\textwidth]{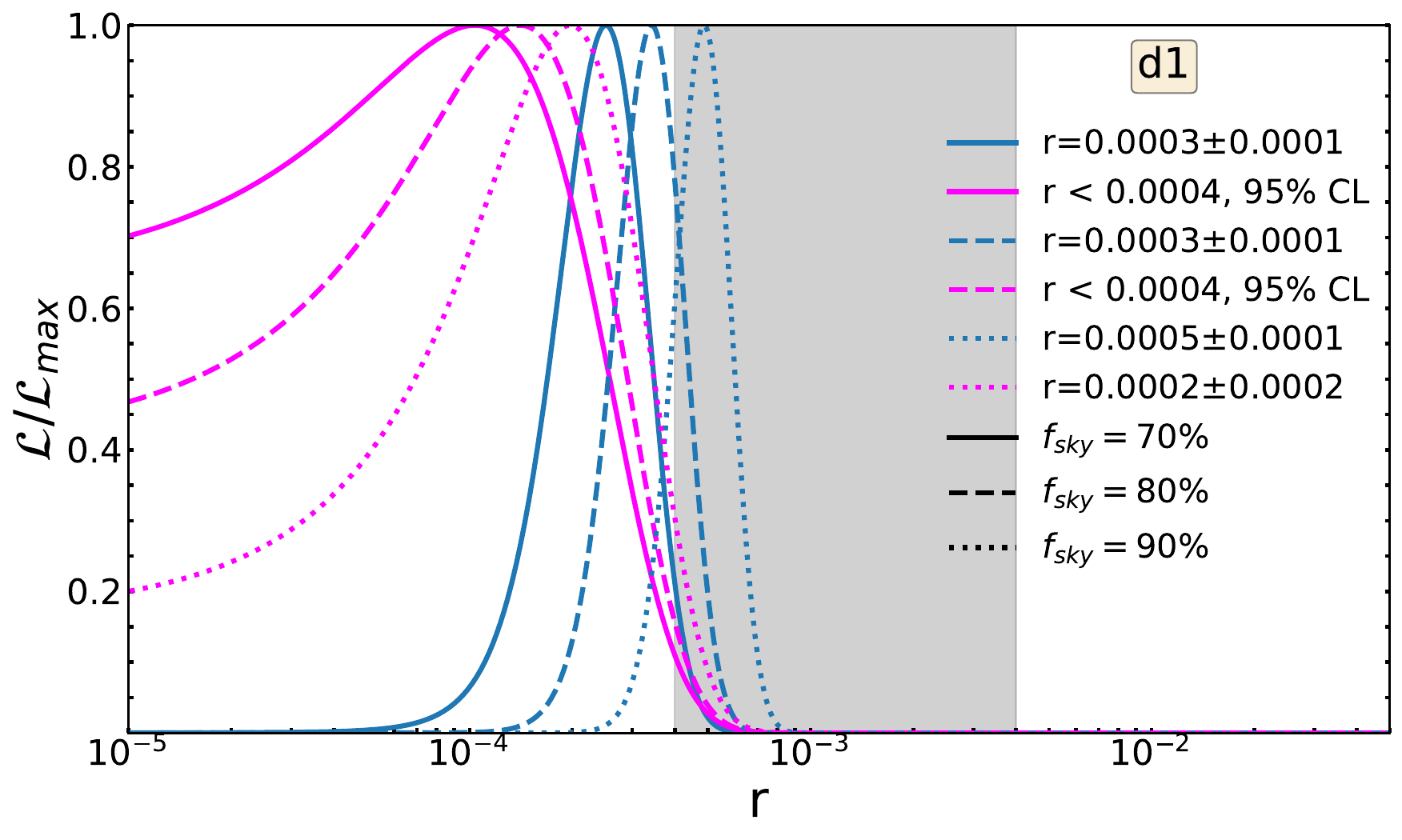}~
    \includegraphics[width=0.5\textwidth]{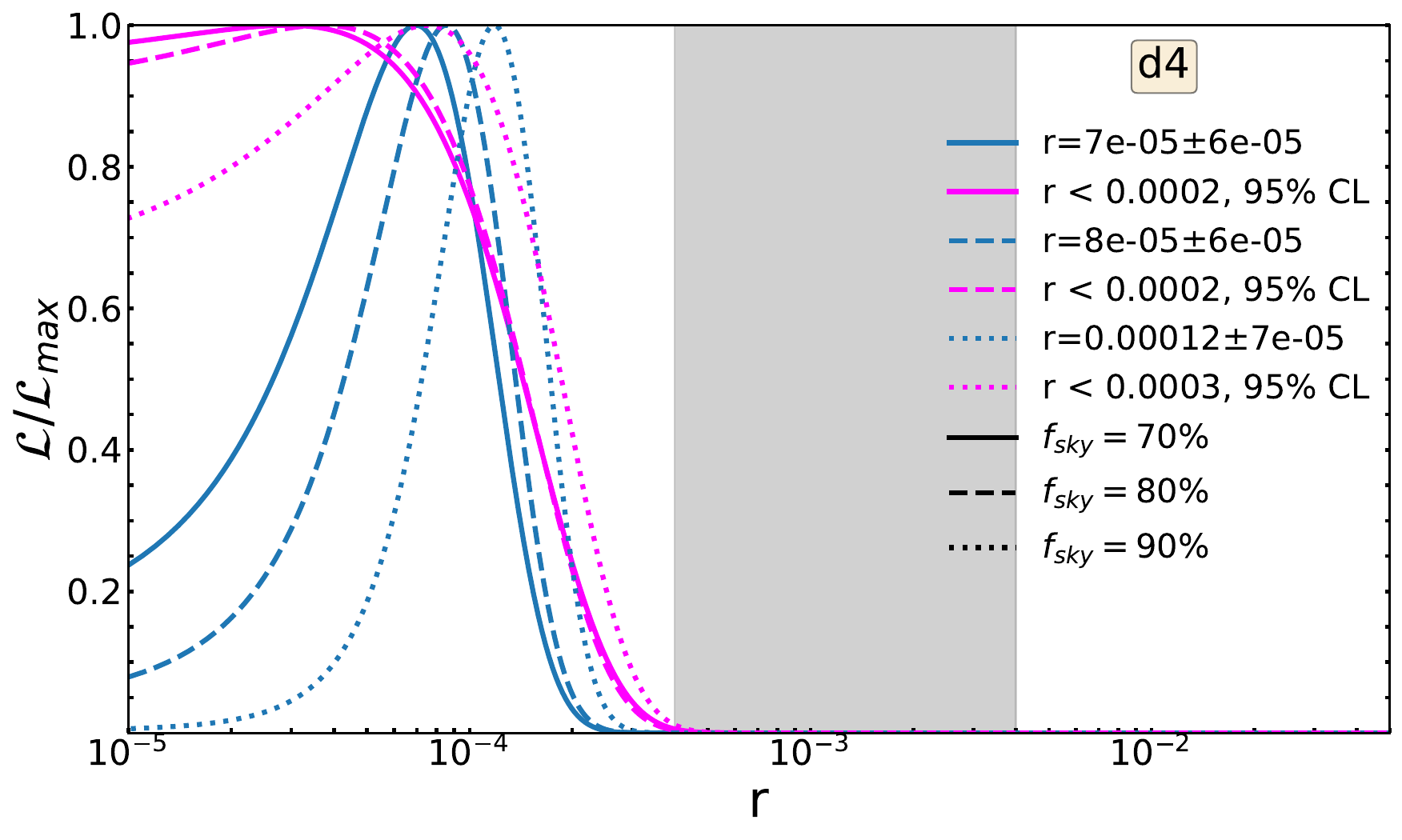}\\
    \includegraphics[width=0.5\textwidth]{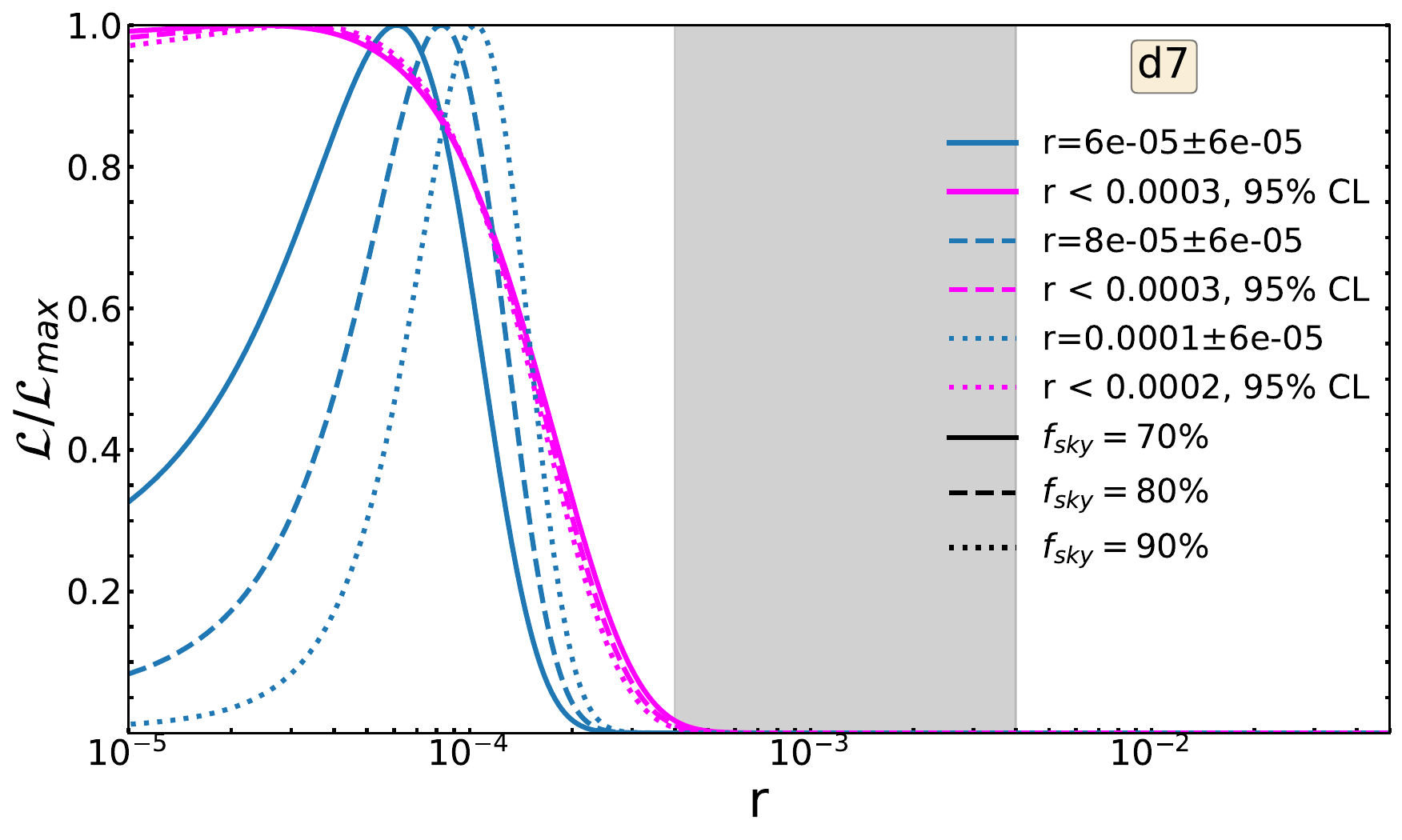}~
    \includegraphics[width=0.5\textwidth]{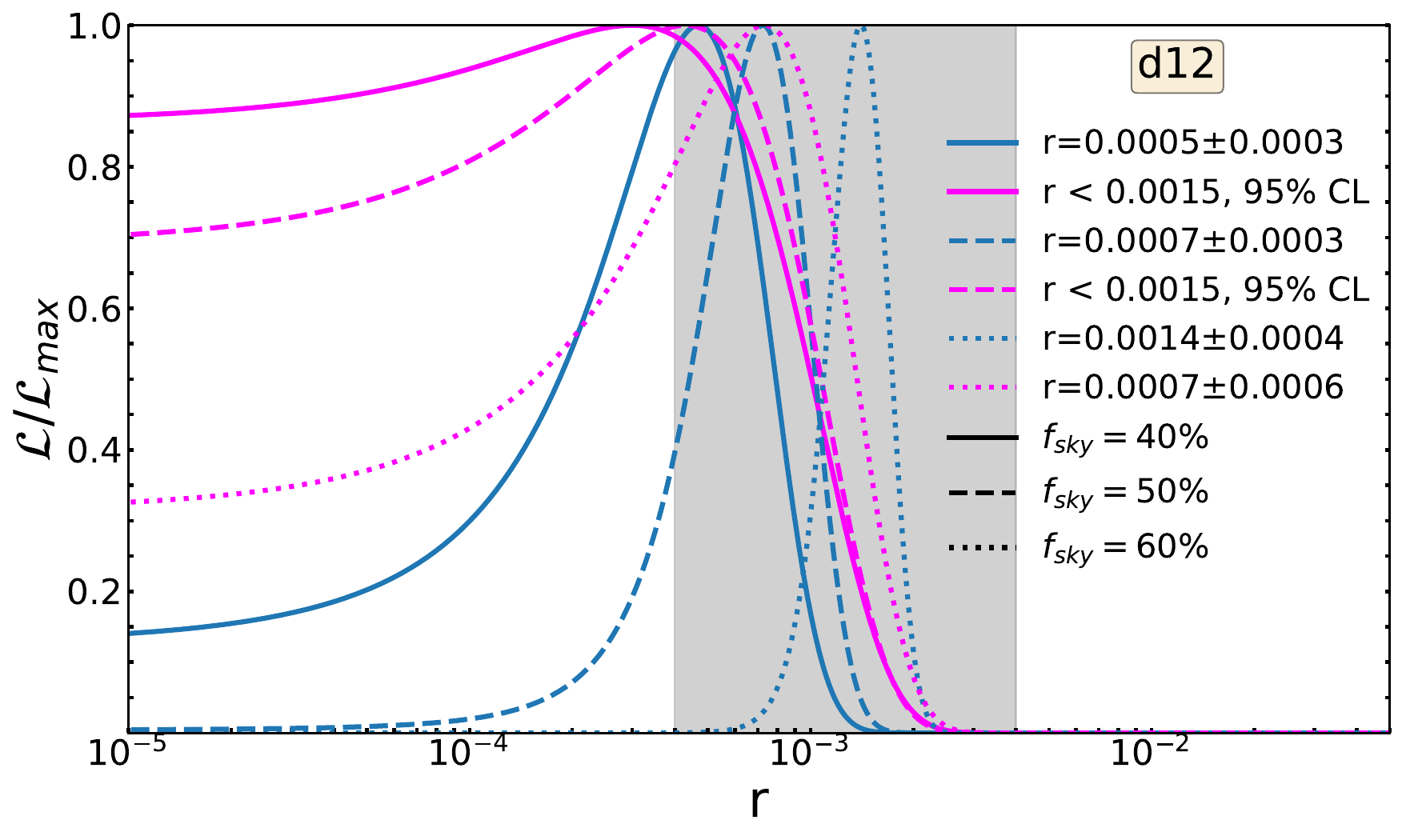}
	\caption{
	Recovered posterior distributions of the tensor-to-scalar ratio obtained from NILC (blue) and ocMILC (magenta) with increasing sky fractions. The grey shaded region corresponds to the range $r\in [0.0004, 0.004]$, where the lower bound ($r=0.0004$) aligns with the targeted sensitivity at $95\%$ confidence level for the adopted \textit{PICO} baseline instrumental configuration.
	}
	\label{fig:ocMILC_like_diffmasks}
\end{figure*}

The  resulting $r$-posteriors for NILC and ocMILC across the four distinct foreground scenarios are displayed in the right panels of Figure~\ref{fig:ocMILC_spectra}. Biased estimates of the tensor-to-scalar ratio ($r$) are reported when the value $r=0$ falls outside the $68\%$ confidence level. Notably, ocMILC, in contrast to NILC, produces unbiased posteriors for $r$ across all considered foreground scenarios even for such large fractions of sky. This bias removal is characterized by both a shift in the distribution's peak towards lower values of $r$ due to reduced foreground contamination and an increase in the distribution's width due to increased noise. The biased $r$-distribution observed in NILC results from more pronounced residual Galactic contamination across all multipoles. While previous \textit{PICO} forecasts in the literature \citep{cMILC,PICO} demonstrated that NILC and cMILC can achieve unbiased recovery of $r$ for a $50\%$ sky fraction with most foreground models, our results highlight that over larger sky fractions ($70\%$) these methods, without optimization, exhibit biases in $r$ due to lingering foreground residuals. In contrast, ocMILC remains unbiased over such extended sky fractions for all foreground models. It is interesting to note that our results, along with those from \citep{PICO}, exhibiting poorer constraints on $r$ for \texttt{d1} and \texttt{d12} compared to \texttt{d4} and \texttt{d7} foreground scenarios, align with the GNILC diagnosis in Section~\ref{sec:GNILC} which indeed revealed lower foreground complexity in the \texttt{d4} and \texttt{d7} data sets. For all considered Galactic models, except the intricate \texttt{d12} model, ocMILC achieves upper bounds on the tensor-to-scalar ratio that align with the sensitivity targeted by \textit{PICO} with the adopted baseline instrumental configuration ($r < 0.0004$ at $95\%$ CL). 
 
\begin{table*}
\caption{
Optimal configuration of moments and dust pivot parameters for deprojection, as determined by ocMILC across various sky regions and needlet scales for each foreground data set. The optimal deprojection coefficient for each moment SED is indicated in parentheses. As detailed in Section~\ref{subsec:optimization}, the optimal combination of these parameters is chosen by minimizing the denoised output variance of Equation~\ref{eq:min_ocMILC}.
\label{tab:params_ocMILC}}
\centering
\begin{adjustwidth}{-1.5cm}{}
\hspace{1.2cm}\resizebox{16.cm}{!}{
\begin{tabular}{|c|c|cccc|}
\hline 
      Band & $m_{\textrm{fgds}}$
      &
      \texttt{d1s1} & \texttt{d4s1} & \texttt{d7s1} & \texttt{d12s1} \\
      \hline
  \multirow{2}{*}{$j=1$}  & $5$ &  & \makecell{$\bar{\beta}_{\textrm{d}}=1.75,\ \bar{T}_{\textrm{d}}=19.0$; \\ $\big[\partial_{\beta_{\textrm{d}}}(-0.01)$,$\partial_{\beta_{\textrm{s}}}(0.)$,$\partial_{T_{\textrm{d}}}(0.)$,\\ $\partial^2_{\beta_{\textrm{s}}}(0.001)$,$\partial^2_{\beta_{\textrm{d}}T_{\textrm{d}}}(0.0005)\big]$} &   \makecell{$\bar{\beta}_{\textrm{d}}=1.4,\ \bar{T}_{\textrm{d}}=17.0$; \\ $\big[f_{\textrm{d}}(0.01)$,$\partial_{\beta_{\textrm{s}}}(0.)$,$\partial_{T_{\textrm{d}}}(0.)$,\\ $\partial^2_{\beta_{\textrm{d}}}(0.05)$,$\partial^2_{\beta_{\textrm{s}}}(-0.005)\big]$} & \\ [0.6cm]
   & $6$ & \makecell{ $\bar{\beta}_{\textrm{d}}=1.3,\ \bar{T}_{\textrm{d}}=17.25$; \\ $\big[f_{\textrm{d}}(0.001)$,$f_{\textrm{s}}(0.)$,$\partial_{\beta_{\textrm{s}}}(0.)$,\\ $\partial_{T_{\textrm{d}}}(0.)$,$\partial^2_{\beta_{\textrm{s}}}(0.01)$,$\partial^2_{T_{\textrm{d}}}(0.01)\big]$} & &  & \makecell{ $\bar{\beta}_{\textrm{d}}=1.25,\ \bar{T}_{\textrm{d}}=18.25$ \\ $\big[f_{\textrm{d}}(0.)$,$f_{\textrm{s}}(0.)$,$\partial_{\beta_{\textrm{d}}}(-0.005)$,\\ $\partial_{\beta_{\textrm{s}}}(0.)$,$\partial_{T_{\textrm{d}}}(0.)$,$\partial^2_{\beta_{\textrm{s}}}(0.05)\big]$} \\
 \hline
 \multirow{2}{*}{$j=2$}  & $4$ & \makecell{ $\bar{\beta}_{\textrm{d}}=1.2,\ \bar{T}_{\textrm{d}}=17.5$; \\ $\big[f_{\textrm{s}}(0.)$,$\partial_{\beta_{\textrm{s}}}(-0.0005)$,\\ $\partial^2_{\beta_{\textrm{d}}}(0.05)$,$\partial^2_{\beta_{\textrm{s}}}(0.05)\big]$} & \makecell{$\bar{\beta}_{\textrm{d}}=1.75,\ \bar{T}_{\textrm{d}}=21.5$; \\ $\big[\partial_{\beta_{\textrm{d}}}(-0.0005)$,$\partial_{\beta_{\textrm{s}}}(0.0005)$,\\ $\partial_{T_{\textrm{d}}}(-0.001)$,$\partial^2_{\beta_{\textrm{d}}}(0.01)\big]$} & \makecell{$\bar{\beta}_{\textrm{d}}=1.35,\ \bar{T}_{\textrm{d}}=21.75$; \\ $\big[f_{\textrm{d}}(0.01)$,$\partial_{\beta_{\textrm{s}}}(-0.001)$,\\ $\partial^2_{\beta_{\textrm{s}}}(0.)$,$\partial^2_{T_{\textrm{d}}}(0.)\big]$} & \\ [0.65cm]
& $5$ & \makecell{ $\bar{\beta}_{\textrm{d}}=1.7,\ \bar{T}_{\textrm{d}}=21.75$; \\ $\big[f_{\textrm{d}}(0.)$,$\partial_{\beta_{\textrm{d}}}(0.001)$,$\partial_{T_{\textrm{d}}}(0.)$,\\ $\partial_{\beta_{\textrm{s}}}(0.001)$,$\partial^2_{\beta_{\textrm{d}}}(0.01)\big]$} & \makecell{$\bar{\beta}_{\textrm{d}}=1.75,\ \bar{T}_{\textrm{d}}=19.5$; \\ $\big[\partial_{\beta_{\textrm{d}}}(-0.01)$,$\partial_{\beta_{\textrm{s}}}(0.)$,$\partial_{T_{\textrm{d}}}(0.)$,\\ $\partial^2_{\beta_{\textrm{s}}}(0.005)$,$\partial^2_{\beta_{\textrm{d}}T_{\textrm{d}}}(0.0005)\big]$} & \makecell{$\bar{\beta}_{\textrm{d}}=1.35,\ \bar{T}_{\textrm{d}}=21.5$; \\ $\big[f_{\textrm{d}}(0.01)$,$\partial_{\beta_{\textrm{s}}}(0.)$,\\ $\partial^2_{\beta_{\textrm{d}}}(0.05)$,$\partial^2_{\beta_{\textrm{s}}}(0.001)$, \\ $\partial^2_{T_{\textrm{d}}}(0.)\big]$} & \makecell{ $\bar{\beta}_{\textrm{d}}=1.7,\ \bar{T}_{\textrm{d}}=17.5$ \\ $\big[ f_{\textrm{s}}(-0.0005)$,$\partial_{\beta_{\textrm{s}}}(0.001)$, \\ $\partial_{T_{\textrm{d}}}(0.),$
$ \partial^2_{\beta_{\textrm{d}}}(-0.0005)$, \\ $\partial^2_{\beta_{\textrm{s}}}(0.05)\big]$} \\
   \hline
\multirow{2}{*}{$j=3$}  & $3$ & \makecell{ $\bar{\beta}_{\textrm{d}}=1.5,\ \bar{T}_{\textrm{d}}=19.0$; \\ $\big[f_{\textrm{s}}(-0.001)$,$\partial_{T_{\textrm{d}}}(0.)$,\\ $\partial^2_{\beta_{\textrm{s}}}(0.05)\big]$} & \makecell{$\bar{\beta}_{\textrm{d}}=1.45,\ \bar{T}_{\textrm{d}}=17.0$; \\ $\big[\partial_{\beta_{\textrm{d}}}(-0.01)$,$\partial_{\beta_{\textrm{s}}}(-0.001)$,\\ $\partial_{T_{\textrm{d}}}(0.)\big]$} & \makecell{$\bar{\beta}_{\textrm{d}}=1.45,\ \bar{T}_{\textrm{d}}=21.0$; \\ $\big[f_{\textrm{s}}(0.001)$,$\partial_{T_{\textrm{d}}}(0.)$,\\$\partial^2_{\beta_{\textrm{d}}T_{\textrm{d}}}(0.0005)\big]$} & \\ [0.65cm]
 & $4$ & \makecell{ $\bar{\beta}_{\textrm{d}}=1.65,\ \bar{T}_{\textrm{d}}=17.25$; \\ $\big[\partial_{\beta_{\textrm{d}}}(0.001)$,$\partial_{\beta_{\textrm{s}}}(0.005)$,\\ $\partial^2_{\beta_{\textrm{s}}}(0.05)$, $\partial^2_{T_{\textrm{d}}}(0.)\big]$} & \makecell{$\bar{\beta}_{\textrm{d}}=1.45,\ \bar{T}_{\textrm{d}}=21.75$; \\ $\big[\partial_{\beta_{\textrm{d}}}(-0.01)$,$\partial_{\beta_{\textrm{s}}}(0.005)$,\\ $\partial^2_{\beta_{\textrm{s}}}(0.05)$,$\partial^2_{T_{\textrm{d}}}(0.)\big]$} & \makecell{$\bar{\beta}_{\textrm{d}}=1.4,\ \bar{T}_{\textrm{d}}=20.0$; \\ $\big[\partial_{\beta_{\textrm{s}}}(0.001)$,$\partial_{T_{\textrm{d}}}(0.)$,\\ $\partial^2_{\beta_{\textrm{d}}T_{\textrm{d}}}(0.0005)$,$\partial^2_{T_{\textrm{d}}}(0.)\big]$} & \makecell{ $\bar{\beta}_{\textrm{d}}=1.75,\ \bar{T}_{\textrm{d}}=19.25$ \\ $\big[\partial_{\beta_{\textrm{s}}}(0.005)$,$ \partial^2_{\beta_{\textrm{d}}}(-0.0005)$, \\ $\partial^2_{\beta_{\textrm{s}}}(0.05)$,$\partial^2_{\beta_{\textrm{d}}T_{\textrm{d}}}(0.0005)\big]$} \\ [0.65cm]
 & $5$ & & & & \makecell{ $\bar{\beta}_{\textrm{d}}=1.35,\ \bar{T}_{\textrm{d}}=19.0$ \\ $\big[ \partial_{\beta_{\textrm{d}}}(-0.001)$,$\partial_{\beta_{\textrm{s}}}(0.01)$, \\ $\partial_{T_{\textrm{d}}}(0.)$, $\partial^2_{\beta_{\textrm{d}}T_{\textrm{d}}}(0.)$, $\partial^2_{T_{\textrm{d}}}(0.)\big]$} \\
   \hline
   \multirow{2}{*}{$j=4$}  & $2$ & \makecell{ $\bar{\beta}_{\textrm{d}}=1.2,\ \bar{T}_{\textrm{d}}=17.0$; \\ $\big[f_{\textrm{s}}(-0.005)$,$ \partial^2_{T_{\textrm{d}}}(0.)\big]$} & \makecell{$\bar{\beta}_{\textrm{d}}=1.3,\ \bar{T}_{\textrm{d}}=18.25$; \\ $\big[f_{\textrm{d}}(0.01)$,$\partial_{\beta_{\textrm{s}}}(0.005)\big]$} & \makecell{$\bar{\beta}_{\textrm{d}}=1.65,\ \bar{T}_{\textrm{d}}=17.25$; \\ $\big[f_{\textrm{d}}(-0.01)$,$f_{\textrm{s}}(-0.005)\big]$} & \makecell{$\bar{\beta}_{\textrm{d}}=1.75,\ \bar{T}_{\textrm{d}}=21.75$; \\ $\big[\partial^2_{\beta_{\textrm{d}}}(0.01)$,$\partial^2_{\beta_{\textrm{d}}T_{\textrm{d}}}(-0.001)\big]$} \\ [0.65cm]
 & $3$ & \makecell{ $\bar{\beta}_{\textrm{d}}=1.25,\ \bar{T}_{\textrm{d}}=17.75$; \\ $\big[\partial_{\beta_{\textrm{d}}}(-0.01)$,$\partial_{\beta_{\textrm{s}}}(0.01)$,\\ $\partial^2_{\beta_{\textrm{d}}T_{\textrm{d}}}(0.0005)\big]$} & \makecell{$\bar{\beta}_{\textrm{d}}=1.45,\ \bar{T}_{\textrm{d}}=17.5$; \\ $\big[\partial_{\beta_{\textrm{d}}}(-0.01)$,$\partial_{\beta_{\textrm{s}}}(0.005)$,\\ $\partial^2_{\beta_{\textrm{d}}T_{\textrm{d}}}(0.)\big]$} & \makecell{$\bar{\beta}_{\textrm{d}}=1.6,\ \bar{T}_{\textrm{d}}=20.25$; \\ $\big[\partial_{\beta_{\textrm{d}}}(-0.01)$,$ \partial_{\beta_{\textrm{s}}}(0.005)$, \\ $\partial^2_{\beta_{\textrm{d}}T_{\textrm{d}}}(0.0005)\big]$} & \makecell{ $\bar{\beta}_{\textrm{d}}=1.5,\ \bar{T}_{\textrm{d}}=21.25$ \\ $\big[f_{\textrm{s}}(-0.001)$,$\partial_{\beta_{\textrm{d}}}(-0.0005)$,  \\ $\partial^2_{T_{\textrm{d}}}(0.)\big]$} \\ [0.65cm]
 & $4$ & \makecell{ $\bar{\beta}_{\textrm{d}}=1.6,\ \bar{T}_{\textrm{d}}=17.75$; \\ $\big[\partial_{\beta_{\textrm{d}}}(0.0005)$,$\partial_{\beta_{\textrm{s}}}(0.001)$,\\ $\partial_{T_{\textrm{d}}}(0.)$, $\partial^2_{T_{\textrm{d}}}(0.)\big]$} & \makecell{$\bar{\beta}_{\textrm{d}}=1.75,\ \bar{T}_{\textrm{d}}=20.5$; \\ $\big[\partial_{\beta_{\textrm{d}}}(-0.001)$,$\partial_{\beta_{\textrm{s}}}(0.01)$,\\ $\partial^2_{\beta_{\textrm{d}}}(0.01)$,$\partial^2_{\beta_{\textrm{d}}T_{\textrm{d}}}(-0.001)\big]$} & \makecell{$\bar{\beta}_{\textrm{d}}=1.4,\ \bar{T}_{\textrm{d}}=19.5$; \\ $\big[\partial_{\beta_{\textrm{s}}}(0.001)$,$\partial_{T_{\textrm{d}}}(0.)$,\\ $\partial^2_{\beta_{\textrm{d}}T_{\textrm{d}}}(0.0005)$, $\partial^2_{T_{\textrm{d}}}(0.)\big]$} & \makecell{ $\bar{\beta}_{\textrm{d}}=1.6,\ \bar{T}_{\textrm{d}}=17.0$ \\ $\big[\partial_{\beta_{\textrm{s}}}(0.01)$, $\partial_{T_{\textrm{d}}}(0.)$, \\ $\partial^2_{\beta_{\textrm{d}}}(0.01)$,$\partial^2_{\beta_{\textrm{d}}T_{\textrm{d}}}(0.)\big]$}  \\
   \hline
\multirow{2}{*}{$j=5$}  & $1$ & \makecell{ $\bar{\beta}_{\textrm{d}}=1.2,\ \bar{T}_{\textrm{d}}=17.0$; \\ $\big[\partial^2_{\beta_{\textrm{s}}}(0.005)\big]$} & \makecell{$\bar{\beta}_{\textrm{d}}=1.2,\ \bar{T}_{\textrm{d}}=17.0$; \\ $\big[f_{\textrm{s}}(-0.005)\big]$} &  & \makecell{$\bar{\beta}_{\textrm{d}}=1.45,\ \bar{T}_{\textrm{d}}=17.0$; \\ $\big[\partial^2_{T_{\textrm{d}}}(0.)\big]$} \\ [0.65cm]
& $2$ & \makecell{ $\bar{\beta}_{\textrm{d}}=1.2,\ \bar{T}_{\textrm{d}}=17.0$; \\ $\big[f_{\textrm{s}}(0.0005), \partial^2_{T_{\textrm{d}}}(0.)\big]$} & \makecell{$\bar{\beta}_{\textrm{d}}=1.75,\ \bar{T}_{\textrm{d}}=17.0$; \\ $\big[f_{\textrm{d}}(0.001),f_{\textrm{s}}(0.0005)\big]$} & \makecell{$\bar{\beta}_{\textrm{d}}=1.4,\ \bar{T}_{\textrm{d}}=17.0$; \\ $\big[f_{\textrm{s}}(0.),\partial^2_{T_{\textrm{d}}}(0.)\big]$} & \makecell{$\bar{\beta}_{\textrm{d}}=1.75,\ \bar{T}_{\textrm{d}}=21.75$; \\ $\big[\partial_{\beta_{\textrm{d}}}(0.001),\partial^2_{\beta_{\textrm{d}}T_{\textrm{d}}}(-0.001)\big]$} \\ [0.65cm]
 & $3$ & \makecell{ $\bar{\beta}_{\textrm{d}}=1.75,\ \bar{T}_{\textrm{d}}=21.75$; \\ $\big[f_{\textrm{s}}(0.001)$,$\partial_{\beta_{\textrm{d}}}(0.001)$,\\ $\partial^2_{T_{\textrm{d}}}(0.)\big]$} & \makecell{$\bar{\beta}_{\textrm{d}}=1.4,\ \bar{T}_{\textrm{d}}=21.75$; \\ $\big[\partial_{\beta_{\textrm{d}}}(-0.01)$,$\partial_{\beta_{\textrm{s}}}(0.01)$,\\ $\partial^2_{T_{\textrm{d}}}(0.)\big]$} & \makecell{$\bar{\beta}_{\textrm{d}}=1.35,\ \bar{T}_{\textrm{d}}=17.0$; \\ $\big[f_{\textrm{s}}(0.)$,$\partial_{\beta_{\textrm{s}}}(0.001)$,\\ $\partial^2_{T_{\textrm{d}}}(0.)\big]$} & \makecell{ $\bar{\beta}_{\textrm{d}}=1.3,\ \bar{T}_{\textrm{d}}=21.5$ \\ $\big[f_{\textrm{s}}(0.0005)$,$\partial_{\beta_{\textrm{d}}}(-0.005)$, \\ $\partial^2_{T_{\textrm{d}}}(0.)\big]$} \\ [0.65cm]
 & $4$ & \makecell{ $\bar{\beta}_{\textrm{d}}=1.55,\ \bar{T}_{\textrm{d}}=21.75$; \\ $\big[f_{\textrm{s}}(-0.0005)$,$\partial_{\beta_{\textrm{d}}}(-0.001)$,\\ $\partial_{\beta_{\textrm{s}}}(0.001)$,$\partial^2_{T_{\textrm{d}}}(0.)\big]$} & \makecell{$\bar{\beta}_{\textrm{d}}=1.3,\ \bar{T}_{\textrm{d}}=17.25$; \\ $\big[f_{\textrm{s}}(0.0005)$,$\partial_{\beta_{\textrm{d}}}(-0.0005)$,\\ $\partial_{T_{\textrm{d}}}(0.0005)$, $\partial^2_{\beta_{\textrm{d}}T_{\textrm{d}}}(0.)\big]$} & \makecell{$\bar{\beta}_{\textrm{d}}=1.6,\ \bar{T}_{\textrm{d}}=21.75$; \\ $\big[f_{\textrm{s}}(-0.0005)$,$\partial_{\beta_{\textrm{d}}}(-0.005)$,\\ $\partial_{\beta_{\textrm{s}}}(0.005)$, $\partial^2_{T_{\textrm{d}}}(0.)\big]$} & \makecell{ $\bar{\beta}_{\textrm{d}}=1.6,\ \bar{T}_{\textrm{d}}=17.0$ \\ $\big[\partial_{\beta_{\textrm{s}}}(0.01)$,$\partial_{T_{\textrm{d}}}(0.)$,\\ $\partial^2_{\beta_{\textrm{d}}}(0.01)$,$\partial^2_{\beta_{\textrm{d}}T_{\textrm{d}}}(0.)\big]$}  \\
   \hline
\end{tabular} }
\end{adjustwidth}
\end{table*}

Thanks to optimized moment deprojection, ocMILC enables the observation of larger regions of uncontaminated sky compared to NILC and cMILC, without significant bias introduced by residual foregrounds. This is demonstrated in Figure~\ref{fig:ocMILC_like_diffmasks}, showing the variation of the recovered posterior on $r$ with increasing sky fractions, ranging from $f_{\rm sky}=70\%$ to $90\%$ for \texttt{d1}, \texttt{d4}, and \texttt{d7}, and from $f_{\rm sky}=40\%$ to $60\%$  for \texttt{d12}. Across all foreground models, ocMILC consistently yields lower unbiased upper bounds on the tensor-to-scalar ratio compared to NILC, whose posteriors exhibit consistent bias at high significance across all data sets over such large sky fractions.
Lastly, it is important to highlight that ocMILC successfully handles the intricate multi-layer dust model \texttt{d12} \citep{d12}, avoiding the residual bias reported for NILC in previous PICO forecasts \citep{PICO} concerning this specific foreground model.

Table~\ref{tab:params_ocMILC} presents the optimal moment configurations determined by ocMILC across different sky regions and angular scales for each foreground data set. For each needlet band and GNILC cluster (sky region with uniform foreground complexity), the table includes information on the optimal number and combination of moments for deprojection, along with the corresponding optimal values of deprojection coefficients and dust pivot parameters.
Examining these configurations allows us to draw conclusions consistent with the discussions in previous sections:
\begin{itemize}
 \item The optimal pivot values of the spectral parameters vary for different foreground models, sky regions, and ranges of angular scales. 
 \item Higher-order moments are sometimes preferred for deprojection (e.g. \texttt{d4s1}), while the subtraction of lower-order components can be effectively handled by minimum variance.
 \item Most moments require only partial deprojection. 
 \end{itemize}

The ranges of $m_{\textrm{fgds}}$ values derived from the GNILC diagnosis for different Galactic models and angular scales, as reported in Table~\ref{tab:params_ocMILC} and used by ocMILC, are similar to those shown in Table~\ref{tab:ms_nl}. However, minor differences exist in a few cases, particularly at large and intermediate scales and for small clusters. This discrepancy arises because ocMILC and Table~\ref{tab:params_ocMILC} adopt the more conservative estimate $m_{\textrm{fgds}}=m_{\textrm{GNILC}}-1$, where $m_{\textrm{GNILC}}$ is computed considering only instrumental noise as a nuisance component, and one mode is subtracted to account for the expected CMB contribution. In contrast, Table~\ref{tab:ms_nl} reports $m_{\textrm{fgds}}=m_{\textrm{GNILC}}$, including CMB in the nuisance term. These minor variations result from the fact that, on large and intermediate angular scales, the CMB lensing $B$-modes are not consistently detected across the entire sky as an independent component above the \textit{PICO} noise level. This is expected, as lensing modes primarily contribute at small scales. However, these differences are negligible and do not impact the performance of the ocMILC technique.

\begin{figure*}
	\centering
    \includegraphics[width=0.53\textwidth]{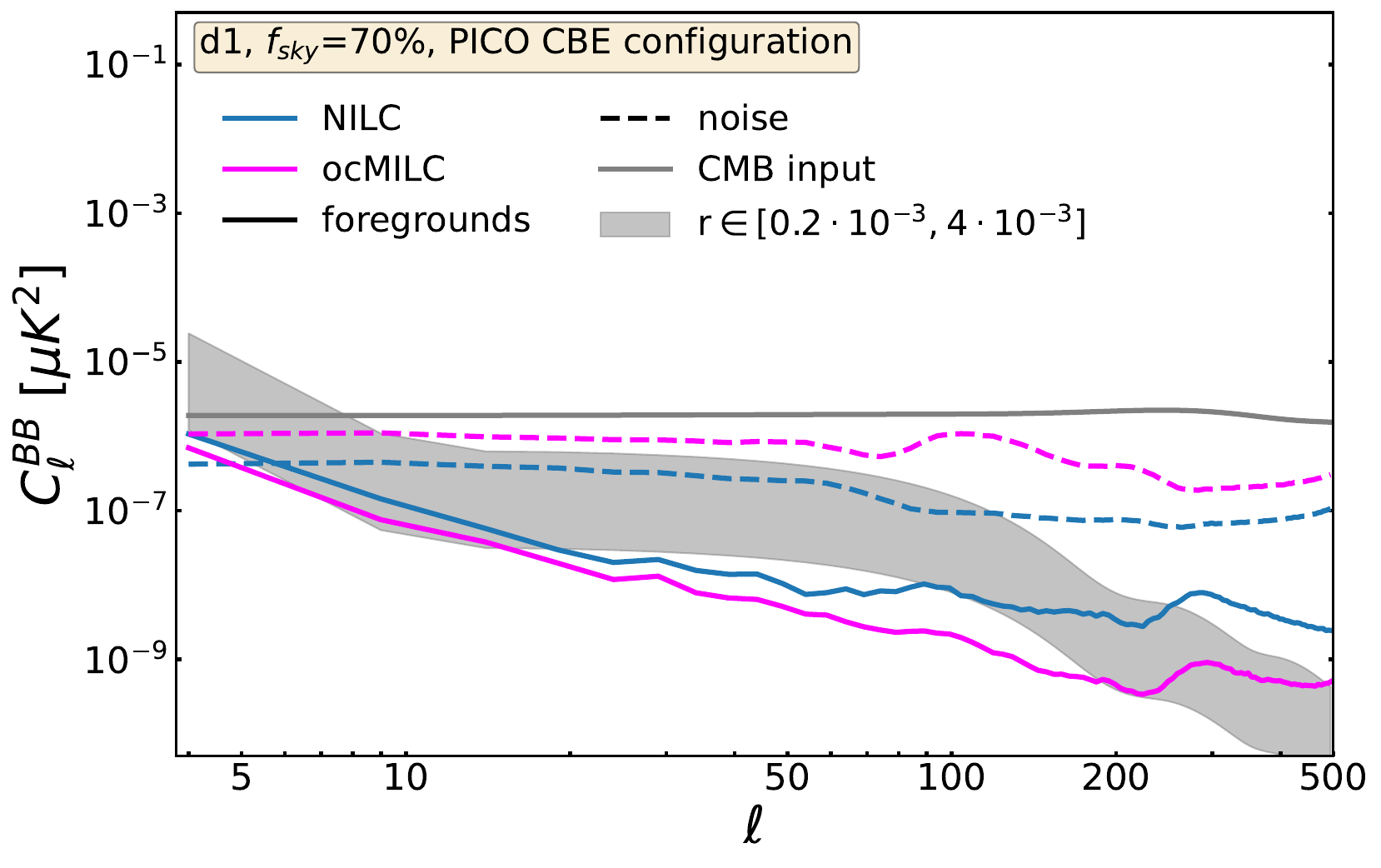}~
    \includegraphics[width=0.47\textwidth]{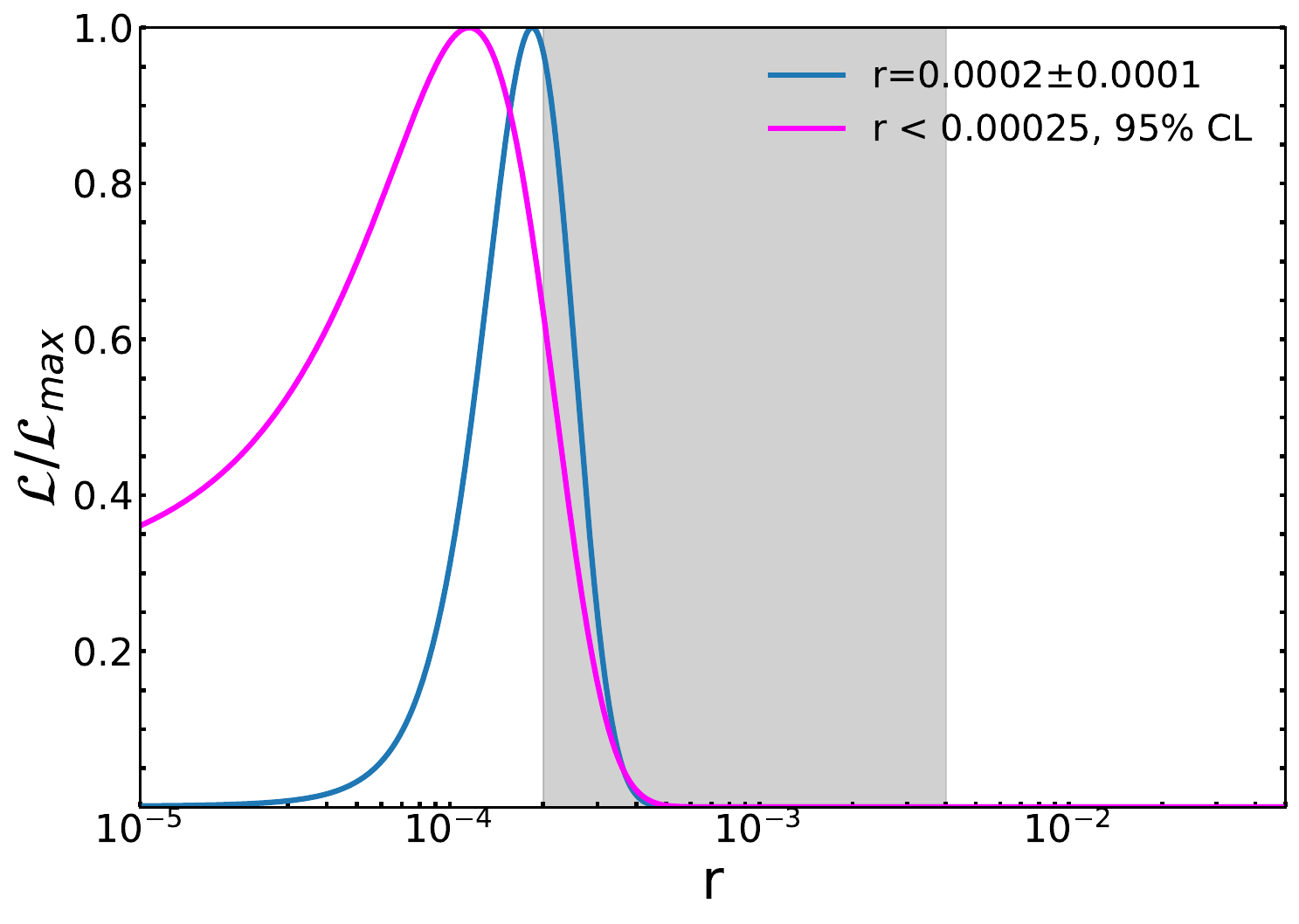}
	\caption{
	Same as the top panels of Figure~\ref{fig:ocMILC_spectra}, but for \textit{PICO} CBE sensitivities. In both panels, the grey shaded area represents the range of values of the tensor-to-scalar ratio: $r\in [0.0002,0.004]$. The lower bound ($r=0.0002$) corresponds to the targeted sensitivity at $95\%$ CL for \textit{PICO} CBE configuration, while the upper bound ($r=0.004$) roughly corresponds to the energy scale of Starobinski's inflation model.
	}
	\label{fig:ocMILC_PICO_BCE}
\end{figure*}

Throughout this paper, we adopted the baseline instrumental configuration of \textit{PICO}, consistent with required specifications detailed in the \textit{PICO} report \citep{PICO_inst}, which provides an overall sensitivity of $0.87$ $\mu K\cdot\text{arcmin}$. 
To complement our results, we present NILC and ocMILC outcomes for the \textit{PICO} \texttt{d1} data set, employing the alternative \textit{PICO} configuration known as CBE (Current Best Estimate) \citep{PICO_inst,PICO}, featuring an enhanced overall sensitivity of $0.61$ $\mu K\cdot\text{arcmin}$. The results are shown in Figure~\ref{fig:ocMILC_PICO_BCE}, showcasing the angular power spectra of residuals averaged over all realizations and corresponding posteriors of the estimated tensor-to-scalar ratio. Even with the increased sensitivity of the \textit{PICO} CBE configuration, ocMILC still outperforms NILC, exhibiting lower foreground residuals across all angular scales with only a moderate rise in noise contamination.  As anticipated, the noise residuals in both NILC and ocMILC CMB $B$-mode reconstructions are lower compared to those obtained with the baseline \textit{PICO} configuration (displayed in the top panel of Figure~\ref{fig:ocMILC_spectra}). Consequently, this leads to lower upper bounds on the estimated tensor-to-scalar ratio ($r<0.00025$ at $95\%$ CL), as shown in the right panel of Figure~\ref{fig:ocMILC_PICO_BCE}. Furthermore, ocMILC improves upon the biased upper bound obtained with NILC for such a large sky fraction ($70\%$). The ocMILC constraint aligns with the quoted forecast in \citep{PICO} for \textit{PICO} CBE sensitivities, albeit for a significantly larger sky fraction compared to \citep{PICO}.

\subsection{Sensitivity to AME and synchrotron emission}
\label{sec:discuss}

The results in Section~\ref{subsec:optimization_res} demonstrated the capability of ocMILC to handle diverse dust models efficiently, thanks to the optimization of moment deprojection. This was achieved by integrating information from the GNILC diagnosis and optimizing the selection of moments, pivot values and deprojection coefficients across various sky regions and angular scales. In this section, our focus shifts to assessing the pipeline's robustness against varied models of Galactic synchrotron emission, while maintaining a fixed thermal dust emission model (\texttt{d1}) to isolate the impact of synchrotron model variations.  Additionally, we investigate the pipeline's response to potential polarization of the anomalous microwave emission (AME). 

Specifically, in addition to \texttt{s1}, two alternative PySM synchrotron models are considered:
\begin{enumerate}
 \item \texttt{s5}: This model features a power-law SED as defined in Equation~\ref{eq:sync}, but the spectral index map from \texttt{s1} has been rescaled to match the larger variability observed in S-PASS data \citep{krach_2018}. In addition, small-scale non-gaussian anisotropies have been incorporated into the templates using the Logarithm of the Polarization Fraction Tensor (\texttt{logpoltens}) formalism. 
\item \texttt{s7}: This model features a power-law SED with some curvature, 
 \begin{equation}
        X_{\textrm{s}}(\hat{n},\nu) = X_{\textrm{s}}(\hat{n},\nu_{\textrm{s}}) \left(\frac{\nu}{\nu_{\textrm{s}}} \right)^{\beta_{\textrm{s}}(\hat{n})+C_{\textrm{s}}(\hat{n})\log{\frac{\nu}{\nu_s}}}\,,
        \label{eq:sync_with_curv}
    \end{equation}
where $\nu_{\textrm{s}} = 23$\,GHz, and the $X_{\textrm{s}}(\hat{n},\nu_{\textrm{s}})$ and $\beta_{\textrm{s}}$ maps remain consistent with those in \texttt{s5}, while a spatially varying curvature term $C_{\textrm{s}}$ is introduced based on the analysis of intensity data observed in a patch of the ARCADE experiment \citep{Kogut_2012}.
\end{enumerate}

As of now, there is no reported detection of AME polarization, and the most stringent upper bounds on the AME polarization fraction are at the level of $1\%$ or lower \citep{2011ame,2015ame,AME_commander,2017MNRAS.464.4107G}, so that the actual contribution of AME to polarized emission still remains uncertain. Therefore, we explore three AME scenarios with different constant degrees of polarization across the sky: $\Pi_{\textrm{AME}}=0.5,\ 1,$ and $2\%$.  The AME model from PySM corresponds to the sum of two spinning dust components based on the Commander fit of \textit{Planck} temperature data \citep{2016A&A...594A..10P}. The SED of the first component features a spatially anisotropic peak frequency, while the SED of the other component maintains a constant peak frequency across the sky. The polarization angles of AME are set to match those of thermal dust emission.

\begin{table*}
\caption{Number $m_{\textrm{fgds}}$ of independent modes characterizing Galactic foreground emission, as detected by the GNILC in each needlet band for the baseline \textit{PICO} configuration, as a function of the synchrotron (top panel) and AME (bottom panel) scenarios. In cases where multiple values are observed, parentheses indicate the corresponding sky fraction in percentage associated with each value of $m_{\textrm{fgds}}$. \label{tab:ms_nl_diffmodels}}
\centering
\begin{tabular}{c}
\textbf{$B$-modes with different synchrotron models}
\end{tabular} \\
\centering
\begin{adjustwidth}{-1.5cm}{}
\hspace{1.2cm}\resizebox{16.2cm}{!}{\begin{tabular}{|c|ccccc|}
\hline 
      \multirow{2}{*}{Model}
      &
      \multicolumn{5}{c|}{$m_{\textrm{fgds}}$\ ($f_{\textrm{sky}}(\%)$)} \\
      & $j=1$ & $j=2$ & $j=3$ & $j=4$ & $j=5$ \\
      \hline
  \texttt{d1s1} & $6$ & $4\ (13),5\ (87)$ & $3\ (4),4\ (95),5\ (1)$ & $2\ (11.5),3\ (28),4\ (60.5)$ & $1\ (0.7),2\ (68),3\ (16.3),4\ (15)$ \\ [0.2cm]
  \texttt{d1s5}& $6$ & $5$ & $3\ (4),4\ (52),5\ (44)$ & $2\ (1), 3\ (38),4\ (50),5\ (11)$ & $1\ (0.7),2\ (58.5),3\ (25.5),4\ (15.3)$ \\ [0.2cm]
  \texttt{d1s7} & $6$ & $5$ & $3\ (4),4\ (52),5\ (44)$ & $2\ (1), 3\ (38),4\ (50),5\ (11)$ & $1\ (0.7),2\ (59),3\ (25),4\ (15.3)$ \\
 \hline
\end{tabular}} \vspace{0.2 cm}\\
\end{adjustwidth}
\begin{tabular}{c}
\textbf{$B$-modes with AME with different polarization fractions}
\end{tabular} \\
\begin{adjustwidth}{-2.7cm}{}
\hspace{2.35cm}\resizebox{16.25cm}{!}{\begin{tabular}{|c|ccccc|}
\hline 
      \multirow{2}{*}{Model}
      & \multicolumn{5}{c|}{$m_{\textrm{fgds}}$\ ($f_{\textrm{sky}}(\%)$)} \\
      & $j=1$ & $j=2$ & $j=3$ & $j=4$ & $j=5$ \\
      \hline 
  $\Pi_{\textrm{AME}} =0.5\%$ & $7$ & $4\ (6.4),5\ (64), 6\ (29.6)$ & $3\ (4),4\ (89.5),5\ (6.5)$ & $2\ (11),3\ (28.5),4\ (60.5)$ & \makecell{ $1\ (3),2\ (67),$ \\ $3\ (25.5),4\ (4.5)$} \\[0.5cm]
  $\Pi_{\textrm{AME}} =1\%$ & $7$ & $5\ (27.5),6\ (72.5)$ & \makecell{$3\ (4),4\ (74),$ \\ $5\ (10.5), 6\ (11.5)$} & \makecell{ $2\ (10),3\ (29.6),4\ (55.2),$ \\ $ 5\ (5), 6\ (0.2)$} & $2\ (71.5),3\ (27),4\ (1.5)$ \\[0.5cm]
  $\Pi_{\textrm{AME}} =2\%$ & $8$ & $6\ (81.5),7\ (18.5)$ & $3\ (4),4\ (45),5\ (24), 6\ (27)$ & \makecell{ $1\ (0.8), 2\ (62.8),3\ (19.6)$ \\ $4\ (14.8),5\ (2)$} & \makecell{$1\ (2),2\ (51),$ \\ $3\ (20.4),4\ (26.6)$}\\
 \hline
\end{tabular}}
\end{adjustwidth}
\end{table*}

We begin by examining the outcomes of the GNILC diagnosis for the various synchrotron and AME models in Table~\ref{tab:ms_nl_diffmodels}, contrasting them with those of the \texttt{d1s1} data set, where the synchrotron was represented by the \texttt{s1} model and AME was assumed to be unpolarized. Clearly, the number of independent foreground $B$-modes detected by GNILC at each needlet scale and the associated sky areas tend to increase with more complex synchrotron models and higher degrees of AME polarization. The larger variability of the synchrotron spectral index in the \texttt{s5} model compared to \texttt{s1} is noticeable to GNILC, particularly at intermediate and small scales ($j=2,\ 3$ and $4$), where larger regions of the sky exhibit higher values of $m_{\textrm{fgds}}$. In contrast, GNILC is not particularly sensitive to the spatially varying curvature of the \texttt{s7} model (equation~\ref{eq:sync_with_curv}) for the frequency range and sensitivity of \textit{PICO}, as the number of independent foreground modes detected remains similar for \texttt{s5} and \texttt{s7}. 
Comparing the first lines of the top and bottom tables highlights that even a modest AME polarization fraction of $\Pi_{\textrm{AME}}=0.5\%$ is detected by GNILC in all needlet bands for the frequency range and sensitivity of \textit{PICO}. This trend persists with increasing polarization fractions ($\Pi_{\textrm{AME}}=1$ and $2\%$), suggesting that higher degrees of AME polarization further enhance the effective foreground complexity across the sky and angular scales.

\begin{figure*}
	\centering
	\includegraphics[width=0.5\textwidth]{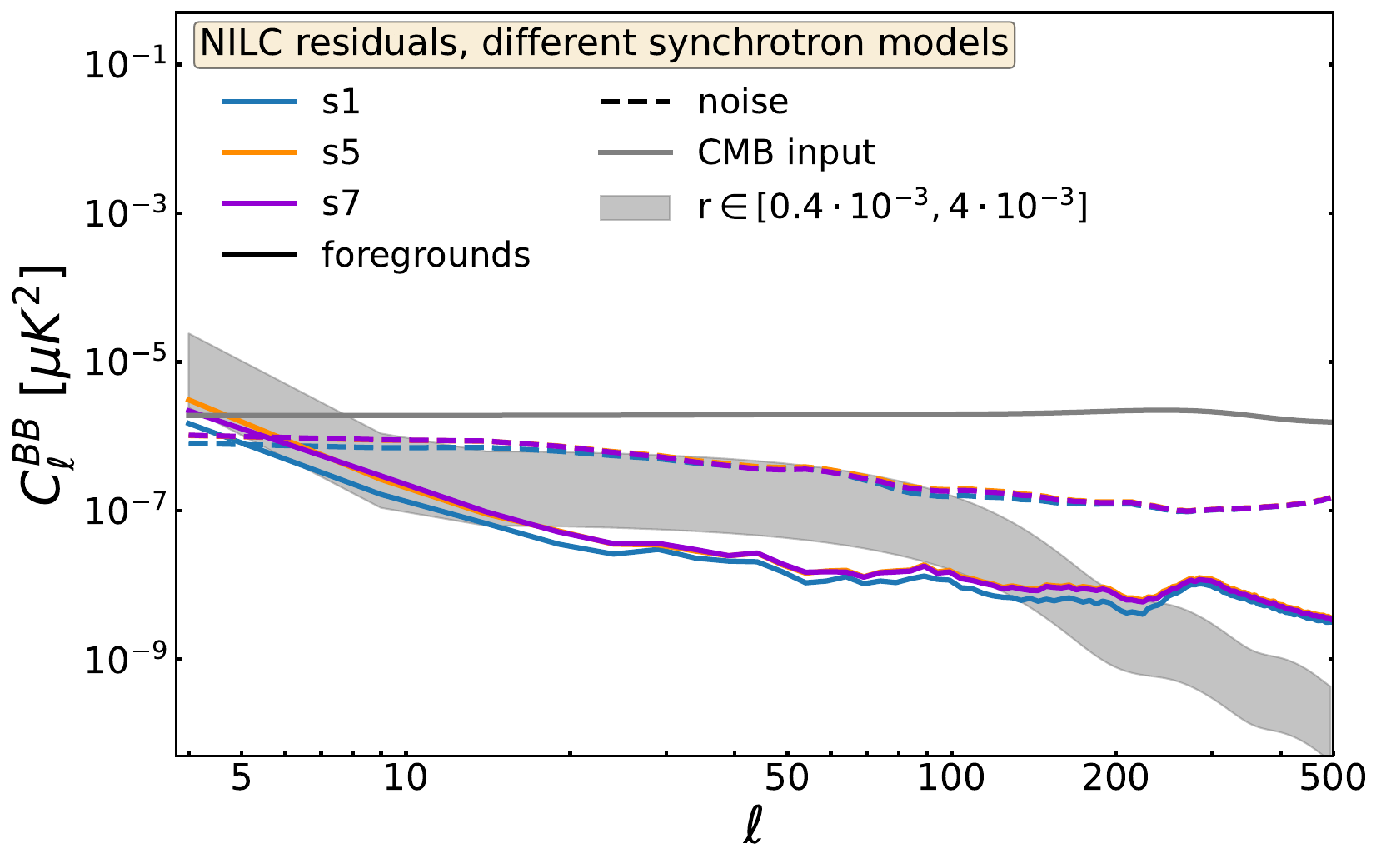}~
    \includegraphics[width=0.5\textwidth]{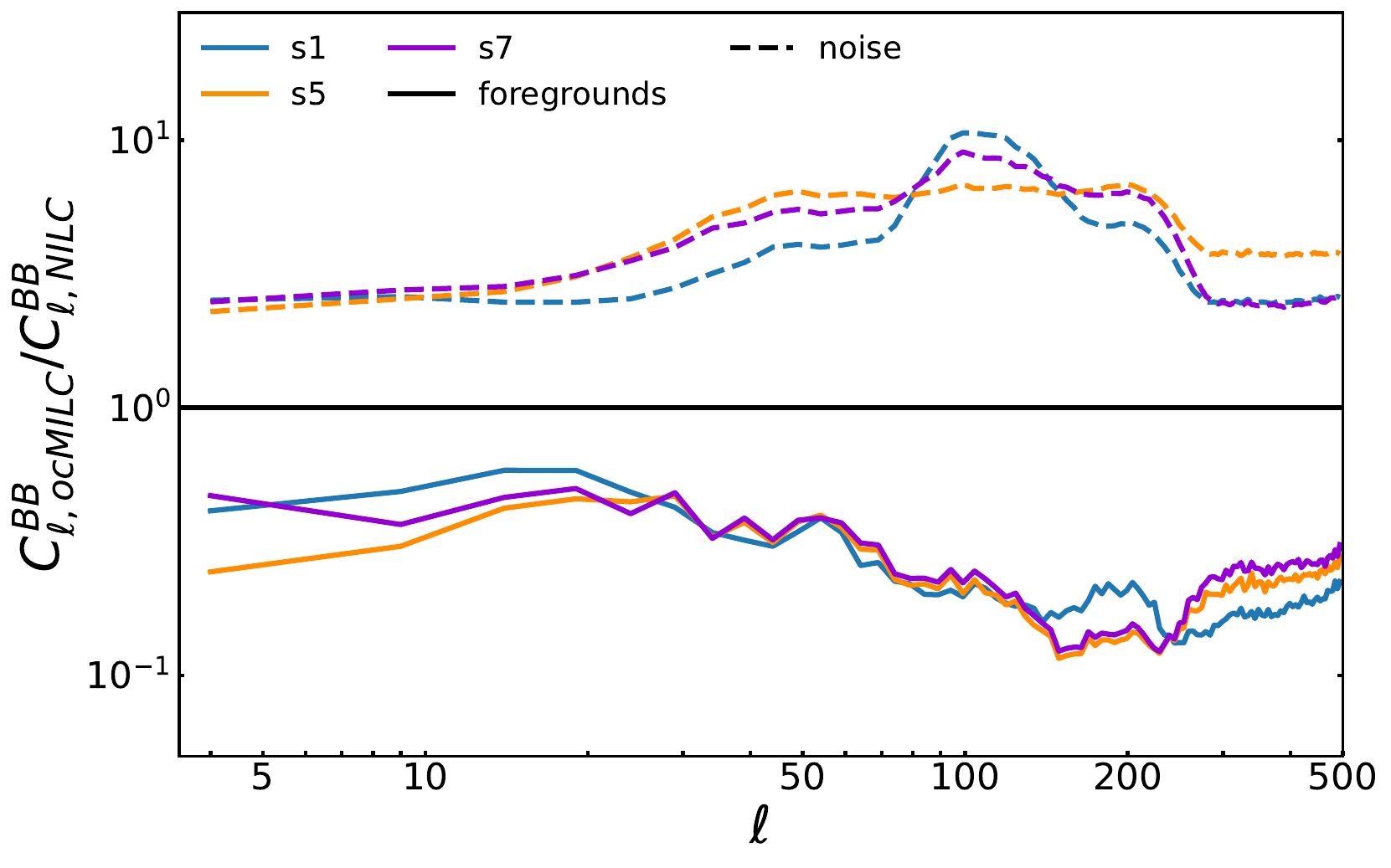} \\
    \includegraphics[width=0.5\textwidth]{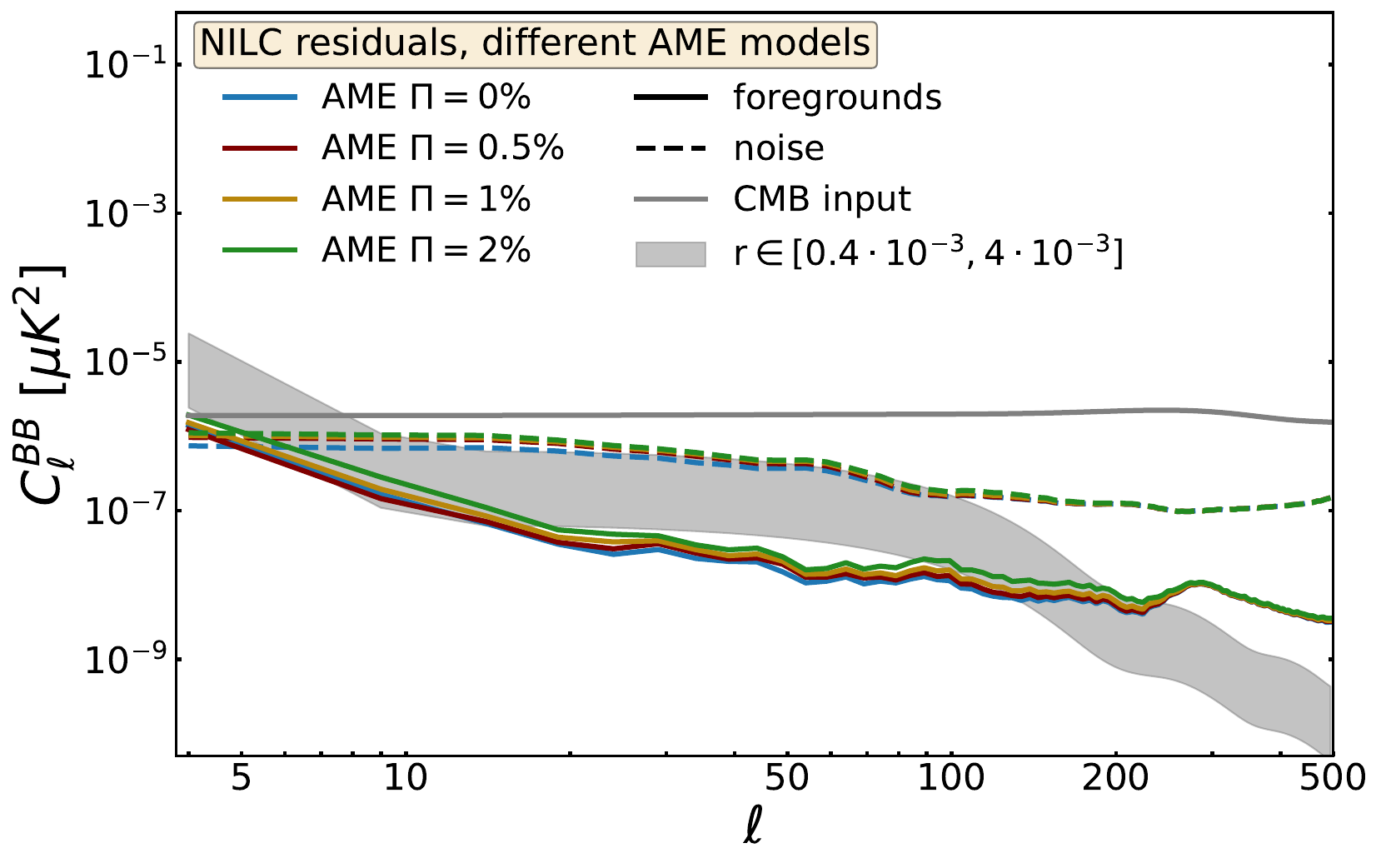}~
    \includegraphics[width=0.5\textwidth]{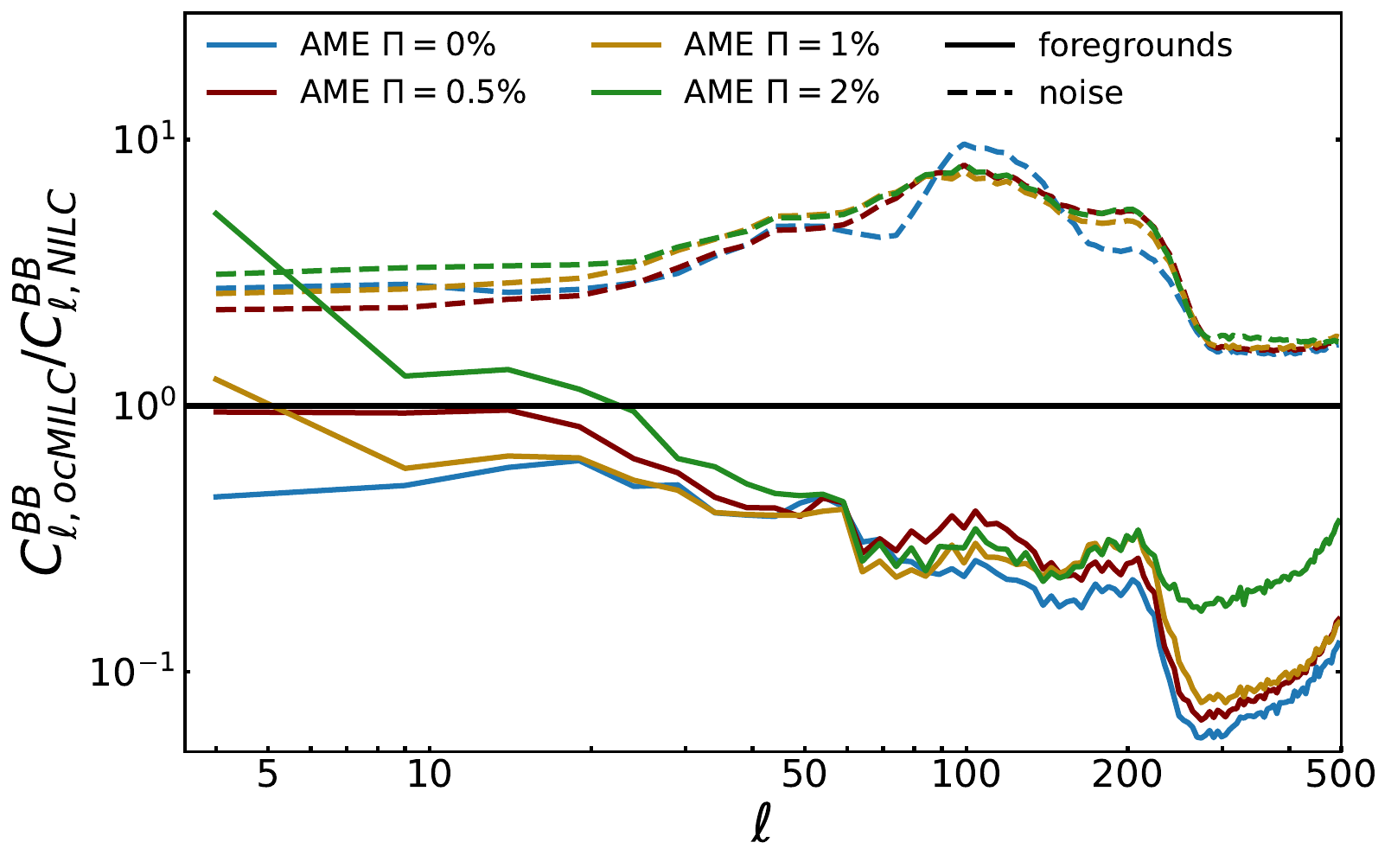}
	\caption{Left: Angular power spectra of foreground (solid) and noise (dashed) $B$-mode residuals over $70\%$ of the sky resulting from NILC applied to the \textit{PICO} \texttt{d1} data set with varying synchrotron models (top) or varying AME polarization models (bottom). Right: Ratio of the power spectra between ocMILC and NILC for foreground (solid) and noise (dashed) $B$-mode residuals across different synchrotron and AME scenarios considered.}
	\label{fig:ocMILC_diffmodels}
\end{figure*}

After evaluating the effects of more complex synchrotron and AME models in the GNILC diagnosis, we proceed to assess their impact on NILC and ocMILC CMB $B$-mode reconstruction in Figure~\ref{fig:ocMILC_diffmodels}. This is to determine whether there is a need to extend the optimization procedure of ocMILC to accommodate the higher complexity of synchrotron and AME polarized emissions. Indeed, despite ocMILC's optimization including synchrotron moments, the current pipeline maintains a fixed pivot value of $\bar{\beta}_{\textrm{s}} = -3$ for the synchrotron spectral index (Section~\ref{subsec:optimization}). However, a higher variability of this parameter across the sky may necessitate optimizing the corresponding pivot values, similar to what was done for dust spectral parameters. Furthermore, the framework of moment expansion has not yet been extended to AME \citep{moments}, so the current ocMILC pipeline does not attempt to deproject any AME moments, leaving them for blind variance minimization.

The angular power spectra of foreground and noise residuals in the NILC CMB $B$-mode reconstruction under various synchrotron and AME scenarios are displayed in the left panels of Figure~\ref{fig:ocMILC_diffmodels}. Having a more complex synchrotron emission (\texttt{s5} and \texttt{s7}) results in a slight increase in foreground residuals across all angular scales up to $\ell \sim 300$. However, the impact of varying synchrotron complexity on NILC residuals is notably less pronounced compared to the impact of varying dust models, as discussed in previous sections. Consequently, further optimization of ocMILC on the pivot values of the synchrotron spectral index is not expected to significantly enhance ocMILC results. Across all extensions of synchrotron emission considered, the ocMILC's CMB solution with a fixed pivot of $\bar{\beta}_{\textrm{s}} = -3$ indeed exhibits lower foreground contamination compared to NILC (top right panel), with percentage improvements similar to those obtained with the least complex model \texttt{s1}. In the case of the \texttt{d1s5} data set, we even observe a $10\%$ greater improvement of ocMILC over NILC at large angular scales compared to \texttt{d1s1}. Additionally, the increase in noise residuals in ocMILC remains consistent across all synchrotron scenarios. The negligible impact of synchrotron curvature is anticipated since, as demonstrated in \cite{moments,cMILC}, curvature is accounted for by the second-order term in the moment expansion (Equation~\ref{eq:sync_exp}), which is included in our optimized deprojection. These results show that, currently, with the available and foreseeable more complex synchrotron models, employing a fixed pivot for the synchrotron spectral index ($\bar{\beta}_{\textrm{s}} = -3$) still yields highly effective moment deprojection with ocMILC.

Concerning AME scenarios, we observe a slight increase in NILC residuals for $\ell \lesssim 250$ as the AME polarization fraction increases (bottom left panel of Figure~\ref{fig:ocMILC_diffmodels}). This indicates that the presence of a polarized AME component affects the variance minimization, given the sensitivity and frequency coverage of \textit{PICO}. Examining the ratio between ocMILC and NILC residuals (bottom right panel), we find that, while the improvement achieved by ocMILC at intermediate and small angular scales ($\ell \gtrsim 70$) slightly degrades with increasing AME polarization compared to the case without polarized AME, it remains highly significant. However, for very low multipoles ($\ell \lesssim 15$), ocMILC foreground residuals become comparable to or larger than NILC foreground residuals depending on the degree of AME polarization. This highlights the necessity, in such scenarios and considering the frequency coverage of \textit{PICO}, of incorporating AME moments into the ocMILC pipeline in the first needlet band. We leave this extension for future investigation, noting that there is currently no clear observational evidence for such levels of polarized AME emission across the sky.

\section{Conclusions}
\label{sec:concl}

The success of upcoming CMB experiments hinges on the ability to measure the primordial CMB $B$-mode polarization down to $r\lesssim 10^{-3}$ without bias to validate the inflationary scenario. This requires exquisite control of instrumental systematics and Galactic foreground subtraction. While the blind NILC method enables the reconstruction of the CMB $B$-mode polarization field without encountering biases from foreground mismodelling that parametric methods may face, fully blind overall-variance minimization might not always be sufficient to reduce large-scale Galactic contamination below the primordial signal, especially for the most challenging foreground scenarios, depending on the available sensitivity, frequency channels, and the targeted $r$ value. 

Deprojecting moments of Galactic emission with the semi-blind cMILC method theoretically allows for minimizing foreground variance rather than overall variance. However, the associated noise penalty, including the increased variance of unconstrained foreground moments, does not always guarantee optimal outcomes for the most complex foreground scenarios. Specifically, deprojecting foreground moments in the natural order of the Taylor expansion may not be optimal, particularly for the moments that dominate the overall variance, which may be better suited for blind variance minimization. To address this trade-off, we introduced an optimized variant of cMILC, named ocMILC (Section~\ref{sec:ocMILC}), which refines the deprojection process through a data-driven approach by optimizing the number and selection of moments to deproject, as well as pivot values and deprojection coefficients. 

In ocMILC, the optimization of the number of moments to deproject, depending on sky areas and angular scales, relies on the GNILC diagnosis of data complexity (Section~\ref{sec:GNILC}), which maps out the number of independent observable modes of Galactic emission, $m_{\textrm{fgds}}$, in the given data set across the sky and at different angular scales (Figure~\ref{fig:ms_PICO_diffmodels_nl4} and Table~\ref{tab:ms_nl}). This diagnosis informs ocMILC on the optimal number of moments to deproject locally before paying significant noise penalties. This strategy is corroborated by the results obtained in Figure~\ref{fig:ocMILC_extras}. The specific subset of moments to deproject locally is also optimized to retain some dominant moments for blind variance minimization, while selectively deprojecting higher-order moments that are not predominantly captured by covariance estimation. The optimization of local pivot values of the dust spectral parameters across different sky regions and ranges of angular scales allows for accommodating local moment expansions associated with local averaging processes. Finally, allowing the deprojection coefficients $\epsilon_{k}$ to be non-zero for partial deprojection of certain moments offers additional flexibility for limiting noise penalties. In addition, it allows to better handle foreground modes not properly described by the moment expansion of a single MBB SED.
The optimal pivot values, subset of moments, and deprojection coefficients are determined to minimize the denoised output variance (Equation~\ref{eq:min_ocMILC}),  and hence the residual foreground variance, in each GNILC cluster. 

Testing on simulated \textit{PICO} data demonstrated ocMILC's superiority over NILC in reducing Galactic contamination across all angular scales and all the considered thermal dust models over large sky fractions, while maintaining noise levels within the target sensitivity for the tensor-to-scalar ratio (Figures~\ref{fig:ocMILC_spectra}-\ref{fig:ocMILC_like_diffmasks}). Furthermore, ocMILC outperformed the standard cMILC, showcasing its efficacy in handling diverse dust emission models, including the multi-layer dust model \texttt{d12} with decorrelation, while controlling noise penalties. While current synchrotron emission models do not necessitate pivot value optimization, the inclusion of mildly polarized AME emissions may require the incorporation of AME moments into ocMILC, a potential avenue for future exploration given the lack of observational evidence. Overall, ocMILC stands as a robust optimization for semi-blind CMB polarization reconstruction, offering promise for application to other satellite missions like \textit{LiteBIRD} \citep{PTEP}, with potential extensions to ground-based experiments \citep{SO_2019,CMBS4} through appropriate $E$-to-$B$ leakage corrections \cite{Ghosh2021,2022JCAP...07..044Z,NILC_cutsky}.

\acknowledgments

We thank Marina Migliaccio and Domenico Marinucci for useful discussions. This research used resources of the National Energy Research Scientific Computing Center (NERSC), a U.S. Department of Energy Office of Science User Facility located at Lawrence Berkeley National Laboratory. AC acknowledges funding support by AASS PhD Program, ASI/LiteBIRD grant n. 2020-9-HH.0 and ASI/LiteBIRD grant CUP: F84I20000230005. AC was also supported by the InDark and LiteBIRD INFN projects. MR thanks the Spanish Agencia Estatal de Investigación (AEI, MICIU) for the financial support provided under the projects with references PID2019-110610RB-C21 and PID2022-139223OB-C2, and acknowledges support from the CSIC programme ‘Ayuda a la Incorporaci\'on de Cient\'ificos Titulares’ provided under the project 202250I159.

\bibliographystyle{JHEP}
\bibliography{biblio}

\providecommand{\href}[2]{#2}\begingroup\raggedright\begin{thebibliography}{10}

\bibitem{Planck_cosmopars}
{Planck Collaboration}, N.~{Aghanim}, Y.~{Akrami}, M.~{Ashdown}, J.~{Aumont}, C.~{Baccigalupi} et~al., \emph{{Planck 2018 results. VI. Cosmological parameters}}, \href{https://doi.org/10.1051/0004-6361/201833910}{\emph{\aap} {\bfseries 641} (2020) A6} [\href{https://arxiv.org/abs/1807.06209}{{\ttfamily 1807.06209}}].

\bibitem{PhysRevD.107.023510}
A.~La~Posta, U.~Natale, E.~Calabrese, X.~Garrido and T.~Louis, \emph{Assessing the consistency between cmb temperature and polarization measurements with application to planck, act, and spt data}, \href{https://doi.org/10.1103/PhysRevD.107.023510}{\emph{\prd} {\bfseries 107} (2023) 023510}.

\bibitem{SO_2019}
P.~{Ade}, J.~{Aguirre}, Z.~{Ahmed}, S.~{Aiola}, A.~{Ali}, D.~{Alonso} et~al., \emph{{The Simons Observatory: science goals and forecasts}}, \href{https://doi.org/10.1088/1475-7516/2019/02/056}{\emph{\jcap} {\bfseries 2019} (2019) 056} [\href{https://arxiv.org/abs/1808.07445}{{\ttfamily 1808.07445}}].

\bibitem{CMBS4}
K.~{Abazajian}, G.~{Addison}, P.~{Adshead}, Z.~{Ahmed}, S.W.~{Allen}, D.~{Alonso} et~al., \emph{{CMB-S4 Science Case, Reference Design, and Project Plan}}, \href{https://doi.org/10.48550/arXiv.1907.04473}{\emph{arXiv e-prints} (2019) arXiv:1907.04473} [\href{https://arxiv.org/abs/1907.04473}{{\ttfamily 1907.04473}}].

\bibitem{PTEP}
{LiteBIRD Collaboration}, E.~{Allys}, K.~{Arnold}, J.~{Aumont}, R.~{Aurlien}, S.~{Azzoni} et~al., \emph{{Probing cosmic inflation with the LiteBIRD cosmic microwave background polarization survey}}, \href{https://doi.org/10.1093/ptep/ptac150}{\emph{Progress of Theoretical and Experimental Physics} {\bfseries 2023} (2023) 042F01} [\href{https://arxiv.org/abs/2202.02773}{{\ttfamily 2202.02773}}].

\bibitem{PICO_inst}
S.~{Hanany}, M.~{Alvarez}, E.~{Artis}, P.~{Ashton}, J.~{Aumont}, R.~{Aurlien} et~al., \emph{{PICO: Probe of Inflation and Cosmic Origins}}, \href{https://doi.org/10.48550/arXiv.1902.10541}{\emph{arXiv e-prints} (2019) arXiv:1902.10541} [\href{https://arxiv.org/abs/1902.10541}{{\ttfamily 1902.10541}}].

\bibitem{2020MNRAS.499..550Q}
Y.~{Qin}, V.~{Poulin}, A.~{Mesinger}, B.~{Greig}, S.~{Murray} and J.~{Park}, \emph{{Reionization inference from the CMB optical depth and E-mode polarization power spectra}}, \href{https://doi.org/10.1093/mnras/staa2797}{\emph{\mnras} {\bfseries 499} (2020) 550} [\href{https://arxiv.org/abs/2006.16828}{{\ttfamily 2006.16828}}].

\bibitem{Planck_reio}
{Planck Collaboration}, R.~{Adam}, N.~{Aghanim}, M.~{Ashdown}, J.~{Aumont}, C.~{Baccigalupi} et~al., \emph{{Planck intermediate results. XLVII. Planck constraints on reionization history}}, \href{https://doi.org/10.1051/0004-6361/201628897}{\emph{\aap} {\bfseries 596} (2016) A108} [\href{https://arxiv.org/abs/1605.03507}{{\ttfamily 1605.03507}}].

\bibitem{2015PhRvD..92l3535A}
R.~{Allison}, P.~{Caucal}, E.~{Calabrese}, J.~{Dunkley} and T.~{Louis}, \emph{{Towards a cosmological neutrino mass detection}}, \href{https://doi.org/10.1103/PhysRevD.92.123535}{\emph{\prd} {\bfseries 92} (2015) 123535} [\href{https://arxiv.org/abs/1509.07471}{{\ttfamily 1509.07471}}].

\bibitem{2016PhRvD..94h3522G}
E.~{Giusarma}, M.~{Gerbino}, O.~{Mena}, S.~{Vagnozzi}, S.~{Ho} and K.~{Freese}, \emph{{Improvement of cosmological neutrino mass bounds}}, \href{https://doi.org/10.1103/PhysRevD.94.083522}{\emph{\prd} {\bfseries 94} (2016) 083522} [\href{https://arxiv.org/abs/1605.04320}{{\ttfamily 1605.04320}}].

\bibitem{1997PhRvL..78.2058K}
M.~{Kamionkowski}, A.~{Kosowsky} and A.~{Stebbins}, \emph{{A Probe of Primordial Gravity Waves and Vorticity}}, \href{https://doi.org/10.1103/PhysRevLett.78.2058}{\emph{\prl} {\bfseries 78} (1997) 2058} [\href{https://arxiv.org/abs/astro-ph/9609132}{{\ttfamily astro-ph/9609132}}].

\bibitem{Commander}
H.K.~{Eriksen}, J.B.~{Jewell}, C.~{Dickinson}, A.J.~{Banday}, K.M.~{G{\'o}rski} and C.R.~{Lawrence}, \emph{{Joint Bayesian Component Separation and CMB Power Spectrum Estimation}}, \href{https://doi.org/10.1086/525277}{\emph{\apj} {\bfseries 676} (2008) 10} [\href{https://arxiv.org/abs/0709.1058}{{\ttfamily 0709.1058}}].

\bibitem{FGBuster}
R.~{Stompor}, S.~{Leach}, F.~{Stivoli} and C.~{Baccigalupi}, \emph{{Maximum likelihood algorithm for parametric component separation in cosmic microwave background experiments}}, \href{https://doi.org/10.1111/j.1365-2966.2008.14023.x}{\emph{\mnras} {\bfseries 392} (2009) 216} [\href{https://arxiv.org/abs/0804.2645}{{\ttfamily 0804.2645}}].

\bibitem{Bsecret}
E.~{de la Hoz}, P.~{Vielva}, R.B.~{Barreiro} and E.~{Mart{\'\i}nez-Gonz{\'a}lez}, \emph{{On the detection of CMB B-modes from ground at low frequency}}, \href{https://doi.org/10.1088/1475-7516/2020/06/006}{\emph{\jcap} {\bfseries 2020} (2020) 006} [\href{https://arxiv.org/abs/2002.12206}{{\ttfamily 2002.12206}}].

\bibitem{Azzoni2021}
S.~{Azzoni}, M.H.~{Abitbol}, D.~{Alonso}, A.~{Gough}, N.~{Katayama} and T.~{Matsumura}, \emph{{A minimal power-spectrum-based moment expansion for CMB B-mode searches}}, \href{https://doi.org/10.1088/1475-7516/2021/05/047}{\emph{\jcap} {\bfseries 2021} (2021) 047} [\href{https://arxiv.org/abs/2011.11575}{{\ttfamily 2011.11575}}].

\bibitem{Vacher2022}
L.~{Vacher}, J.~{Aumont}, L.~{Montier}, S.~{Azzoni}, F.~{Boulanger} and M.~{Remazeilles}, \emph{{Moment expansion of polarized dust SED: A new path towards capturing the CMB B-modes with LiteBIRD}}, \href{https://doi.org/10.1051/0004-6361/202142664}{\emph{\aap} {\bfseries 660} (2022) A111} [\href{https://arxiv.org/abs/2111.07742}{{\ttfamily 2111.07742}}].

\bibitem{Commander3}
M.~{Galloway}, K.J.~{Andersen}, R.~{Aurlien}, R.~{Banerji}, M.~{Bersanelli}, S.~{Bertocco} et~al., \emph{{BEYONDPLANCK. III. Commander3}}, \href{https://doi.org/10.1051/0004-6361/202243137}{\emph{\aap} {\bfseries 675} (2023) A3} [\href{https://arxiv.org/abs/2201.03509}{{\ttfamily 2201.03509}}].

\bibitem{ILC}
C.L.~{Bennett}, R.S.~{Hill}, G.~{Hinshaw}, M.R.~{Nolta}, N.~{Odegard}, L.~{Page} et~al., \emph{{First-Year Wilkinson Microwave Anisotropy Probe (WMAP) Observations: Foreground Emission}}, \href{https://doi.org/10.1086/377252}{\emph{\apjs} {\bfseries 148} (2003) 97} [\href{https://arxiv.org/abs/astro-ph/0302208}{{\ttfamily astro-ph/0302208}}].

\bibitem{SMICA}
J.~{Delabrouille}, J.F.~{Cardoso} and G.~{Patanchon}, \emph{{Multidetector multicomponent spectral matching and applications for cosmic microwave background data analysis}}, \href{https://doi.org/10.1111/j.1365-2966.2003.07069.x}{\emph{\mnras} {\bfseries 346} (2003) 1089} [\href{https://arxiv.org/abs/astro-ph/0211504}{{\ttfamily astro-ph/0211504}}].

\bibitem{SEVEM}
R.~{Fern{\'a}ndez-Cobos}, P.~{Vielva}, R.B.~{Barreiro} and E.~{Mart{\'\i}nez-Gonz{\'a}lez}, \emph{{Multiresolution internal template cleaning: an application to the Wilkinson Microwave Anisotropy Probe 7-yr polarization data}}, \href{https://doi.org/10.1111/j.1365-2966.2011.20182.x}{\emph{\mnras} {\bfseries 420} (2012) 2162} [\href{https://arxiv.org/abs/1106.2016}{{\ttfamily 1106.2016}}].

\bibitem{NILC}
J.~{Delabrouille}, J.-F.~{Cardoso}, M.~{Le Jeune}, M.~{Betoule}, G.~{Fay} and F.~{Guilloux}, \emph{{A full sky, low foreground, high resolution CMB map from WMAP}}, \href{https://doi.org/10.1051/0004-6361:200810514}{\emph{\aap} {\bfseries 493} (2009) 835} [\href{https://arxiv.org/abs/0807.0773}{{\ttfamily 0807.0773}}].

\bibitem{Delta-map}
K.~{Ichiki}, H.~{Kanai}, N.~{Katayama} and E.~{Komatsu}, \emph{{Delta-map method of removing CMB foregrounds with spatially varying spectra}}, \href{https://doi.org/10.1093/ptep/ptz009}{\emph{Progress of Theoretical and Experimental Physics} {\bfseries 2019} (2019) 033E01} [\href{https://arxiv.org/abs/1811.03886}{{\ttfamily 1811.03886}}].

\bibitem{cMILC}
M.~{Remazeilles}, A.~{Rotti} and J.~{Chluba}, \emph{{Peeling off foregrounds with the constrained moment ILC method to unveil primordial CMB B modes}}, \href{https://doi.org/10.1093/mnras/stab648}{\emph{\mnras} {\bfseries 503} (2021) 2478} [\href{https://arxiv.org/abs/2006.08628}{{\ttfamily 2006.08628}}].

\bibitem{MCNILC}
A.~{Carones}, M.~{Migliaccio}, G.~{Puglisi}, C.~{Baccigalupi}, D.~{Marinucci}, N.~{Vittorio} et~al., \emph{{Multiclustering needlet ILC for CMB B-mode component separation}}, \href{https://doi.org/10.1093/mnras/stad2423}{\emph{\mnras} {\bfseries 525} (2023) 3117} [\href{https://arxiv.org/abs/2212.04456}{{\ttfamily 2212.04456}}].

\bibitem{Tassis2015}
K.~{Tassis} and V.~{Pavlidou}, \emph{{Searching for inflationary B modes: can dust emission properties be extrapolated from 350 GHz to 150 GHz?}}, \href{https://doi.org/10.1093/mnrasl/slv077}{\emph{\mnras} {\bfseries 451} (2015) L90} [\href{https://arxiv.org/abs/1410.8136}{{\ttfamily 1410.8136}}].

\bibitem{moments}
J.~{Chluba}, J.C.~{Hill} and M.H.~{Abitbol}, \emph{{Rethinking CMB foregrounds: systematic extension of foreground parametrizations}}, \href{https://doi.org/10.1093/mnras/stx1982}{\emph{\mnras} {\bfseries 472} (2017) 1195} [\href{https://arxiv.org/abs/1701.00274}{{\ttfamily 1701.00274}}].

\bibitem{Pelgrims2021}
V.~{Pelgrims}, S.E.~{Clark}, B.S.~{Hensley}, G.V.~{Panopoulou}, V.~{Pavlidou}, K.~{Tassis} et~al., \emph{{Evidence for line-of-sight frequency decorrelation of polarized dust emission in Planck data}}, \href{https://doi.org/10.1051/0004-6361/202040218}{\emph{\aap} {\bfseries 647} (2021) A16} [\href{https://arxiv.org/abs/2101.09291}{{\ttfamily 2101.09291}}].

\bibitem{Remazeilles2016}
M.~{Remazeilles}, C.~{Dickinson}, H.K.K.~{Eriksen} and I.K.~{Wehus}, \emph{{Sensitivity and foreground modelling for large-scale cosmic microwave background B-mode polarization satellite missions}}, \href{https://doi.org/10.1093/mnras/stw441}{\emph{\mnras} {\bfseries 458} (2016) 2032} [\href{https://arxiv.org/abs/1509.04714}{{\ttfamily 1509.04714}}].

\bibitem{d12}
G.~{Mart{\'\i}nez-Solaeche}, A.~{Karakci} and J.~{Delabrouille}, \emph{{A 3D model of polarized dust emission in the Milky Way}}, \href{https://doi.org/10.1093/mnras/sty204}{\emph{\mnras} {\bfseries 476} (2018) 1310} [\href{https://arxiv.org/abs/1706.04162}{{\ttfamily 1706.04162}}].

\bibitem{PICO}
R.~{Aurlien}, M.~{Remazeilles}, S.~{Belkner}, J.~{Carron}, J.~{Delabrouille}, H.K.~{Eriksen} et~al., \emph{{Foreground separation and constraints on primordial gravitational waves with the PICO space mission}}, \href{https://doi.org/10.1088/1475-7516/2023/06/034}{\emph{\jcap} {\bfseries 2023} (2023) 034} [\href{https://arxiv.org/abs/2211.14342}{{\ttfamily 2211.14342}}].

\bibitem{Fuskeland2023}
U.~{Fuskeland}, J.~{Aumont}, R.~{Aurlien}, C.~{Baccigalupi}, A.J.~{Banday}, H.K.~{Eriksen} et~al., \emph{{Tensor-to-scalar ratio forecasts for extended LiteBIRD frequency configurations}}, \href{https://doi.org/10.1051/0004-6361/202346155}{\emph{\aap} {\bfseries 676} (2023) A42} [\href{https://arxiv.org/abs/2302.05228}{{\ttfamily 2302.05228}}].

\bibitem{GNILC}
{Planck Collaboration}, N.~{Aghanim}, M.~{Ashdown}, J.~{Aumont}, C.~{Baccigalupi}, M.~{Ballardini} et~al., \emph{{Planck intermediate results. XLVIII. Disentangling Galactic dust emission and cosmic infrared background anisotropies}}, \href{https://doi.org/10.1051/0004-6361/201629022}{\emph{\aap} {\bfseries 596} (2016) A109} [\href{https://arxiv.org/abs/1605.09387}{{\ttfamily 1605.09387}}].

\bibitem{NPIPE}
{Planck Collaboration}, Y.~{Akrami}, K.J.~{Andersen}, M.~{Ashdown}, C.~{Baccigalupi}, M.~{Ballardini} et~al., \emph{{Planck intermediate results. LVII. Joint Planck LFI and HFI data processing}}, \href{https://doi.org/10.1051/0004-6361/202038073}{\emph{\aap} {\bfseries 643} (2020) A42} [\href{https://arxiv.org/abs/2007.04997}{{\ttfamily 2007.04997}}].

\bibitem{sroll}
{Planck Collaboration}, N.~{Aghanim}, M.~{Ashdown}, J.~{Aumont}, C.~{Baccigalupi}, M.~{Ballardini} et~al., \emph{{Planck intermediate results. XLVI. Reduction of large-scale systematic effects in HFI polarization maps and estimation of the reionization optical depth}}, \href{https://doi.org/10.1051/0004-6361/201628890}{\emph{\aap} {\bfseries 596} (2016) A107} [\href{https://arxiv.org/abs/1605.02985}{{\ttfamily 1605.02985}}].

\bibitem{sroll2}
J.M.~{Delouis}, L.~{Pagano}, S.~{Mottet}, J.L.~{Puget} and L.~{Vibert}, \emph{{SRoll2: an improved mapmaking approach to reduce large-scale systematic effects in the Planck High Frequency Instrument legacy maps}}, \href{https://doi.org/10.1051/0004-6361/201834882}{\emph{\aap} {\bfseries 629} (2019) A38} [\href{https://arxiv.org/abs/1901.11386}{{\ttfamily 1901.11386}}].

\bibitem{cosmoglobe}
D.J.~{Watts}, A.~{Basyrov}, J.R.~{Eskilt}, M.~{Galloway}, E.~{Gjerl{\o}w}, L.T.~{Hergt} et~al., \emph{{COSMOGLOBE DR1 results. I. Improved Wilkinson Microwave Anisotropy Probe maps through Bayesian end-to-end analysis}}, \href{https://doi.org/10.1051/0004-6361/202346414}{\emph{\aap} {\bfseries 679} (2023) A143} [\href{https://arxiv.org/abs/2303.08095}{{\ttfamily 2303.08095}}].

\bibitem{healpix}
K.M.~{G{\'o}rski}, E.~{Hivon}, A.J.~{Banday}, B.D.~{Wandelt}, F.K.~{Hansen}, M.~{Reinecke} et~al., \emph{{HEALPix: A Framework for High-Resolution Discretization and Fast Analysis of Data Distributed on the Sphere}}, \href{https://doi.org/10.1086/427976}{\emph{\apj} {\bfseries 622} (2005) 759} [\href{https://arxiv.org/abs/astro-ph/0409513}{{\ttfamily astro-ph/0409513}}].

\bibitem{healpy}
A.~{Zonca}, L.~{Singer}, D.~{Lenz}, M.~{Reinecke}, C.~{Rosset}, E.~{Hivon} et~al., \emph{{healpy: equal area pixelization and spherical harmonics transforms for data on the sphere in Python}}, \href{https://doi.org/10.21105/joss.01298}{\emph{The Journal of Open Source Software} {\bfseries 4} (2019) 1298}.

\bibitem{2020A&A...641A...6P}
{Planck Collaboration}, N.~{Aghanim}, Y.~{Akrami}, M.~{Ashdown}, J.~{Aumont}, C.~{Baccigalupi} et~al., \emph{{Planck 2018 results. VI. Cosmological parameters}}, \href{https://doi.org/10.1051/0004-6361/201833910}{\emph{\aap} {\bfseries 641} (2020) A6} [\href{https://arxiv.org/abs/1807.06209}{{\ttfamily 1807.06209}}].

\bibitem{pysm}
B.~{Thorne}, J.~{Dunkley}, D.~{Alonso} and S.~{N{\ae}ss}, \emph{{The Python Sky Model: software for simulating the Galactic microwave sky}}, \href{https://doi.org/10.1093/mnras/stx949}{\emph{\mnras} {\bfseries 469} (2017) 2821} [\href{https://arxiv.org/abs/1608.02841}{{\ttfamily 1608.02841}}].

\bibitem{pysm3}
A.~{Zonca}, B.~{Thorne}, N.~{Krachmalnicoff} and J.~{Borrill}, \emph{{The Python Sky Model 3 software}}, \href{https://doi.org/10.21105/joss.03783}{\emph{The Journal of Open Source Software} {\bfseries 6} (2021) 3783} [\href{https://arxiv.org/abs/2108.01444}{{\ttfamily 2108.01444}}].

\bibitem{2016A&A...594A..10P}
{Planck Collaboration}, R.~{Adam}, P.A.R.~{Ade}, N.~{Aghanim}, M.I.R.~{Alves}, M.~{Arnaud} et~al., \emph{{Planck 2015 results. X. Diffuse component separation: Foreground maps}}, \href{https://doi.org/10.1051/0004-6361/201525967}{\emph{\aap} {\bfseries 594} (2016) A10} [\href{https://arxiv.org/abs/1502.01588}{{\ttfamily 1502.01588}}].

\bibitem{d4}
A.M.~{Meisner} and D.P.~{Finkbeiner}, \emph{{Modeling Thermal Dust Emission with Two Components: Application to the Planck High Frequency Instrument Maps}}, \href{https://doi.org/10.1088/0004-637X/798/2/88}{\emph{\apj} {\bfseries 798} (2015) 88} [\href{https://arxiv.org/abs/1410.7523}{{\ttfamily 1410.7523}}].

\bibitem{d7_magnet}
B.T.~{Draine} and B.~{Hensley}, \emph{{Magnetic Nanoparticles in the Interstellar Medium: Emission Spectrum and Polarization}}, \href{https://doi.org/10.1088/0004-637X/765/2/159}{\emph{\apj} {\bfseries 765} (2013) 159} [\href{https://arxiv.org/abs/1205.7021}{{\ttfamily 1205.7021}}].

\bibitem{Planck2018_compsep}
{Planck Collaboration}, Y.~{Akrami}, M.~{Ashdown}, J.~{Aumont}, C.~{Baccigalupi}, M.~{Ballardini} et~al., \emph{{Planck 2018 results. IV. Diffuse component separation}}, \href{https://doi.org/10.1051/0004-6361/201833881}{\emph{\aap} {\bfseries 641} (2020) A4} [\href{https://arxiv.org/abs/1807.06208}{{\ttfamily 1807.06208}}].

\bibitem{Rybicki}
D.B.~{Rybicki} and A.P.~{Lightman}, \emph{Radiative Processes in Astrophysics} (1979).

\bibitem{sync_temp}
C.L.~{Bennett}, D.~{Larson}, J.L.~{Weiland}, N.~{Jarosik}, G.~{Hinshaw}, N.~{Odegard} et~al., \emph{{Nine-year Wilkinson Microwave Anisotropy Probe (WMAP) Observations: Final Maps and Results}}, \href{https://doi.org/10.1088/0067-0049/208/2/20}{\emph{\apjs} {\bfseries 208} (2013) 20} [\href{https://arxiv.org/abs/1212.5225}{{\ttfamily 1212.5225}}].

\bibitem{sync_index}
M.A.~{Miville-Desch{\^e}nes}, N.~{Ysard}, A.~{Lavabre}, N.~{Ponthieu}, J.F.~{Mac{\'\i}as-P{\'e}rez}, J.~{Aumont} et~al., \emph{{Separation of anomalous and synchrotron emissions using WMAP polarization data}}, \href{https://doi.org/10.1051/0004-6361:200809484}{\emph{\aap} {\bfseries 490} (2008) 1093} [\href{https://arxiv.org/abs/0802.3345}{{\ttfamily 0802.3345}}].

\bibitem{page_2007}
L.~{Page}, G.~{Hinshaw}, E.~{Komatsu}, M.R.~{Nolta}, D.N.~{Spergel}, C.L.~{Bennett} et~al., \emph{{Three-Year Wilkinson Microwave Anisotropy Probe (WMAP) Observations: Polarization Analysis}}, \href{https://doi.org/10.1086/513699}{\emph{\apjs} {\bfseries 170} (2007) 335} [\href{https://arxiv.org/abs/astro-ph/0603450}{{\ttfamily astro-ph/0603450}}].

\bibitem{krach_2018}
N.~{Krachmalnicoff}, E.~{Carretti}, C.~{Baccigalupi}, G.~{Bernardi}, S.~{Brown}, B.M.~{Gaensler} et~al., \emph{{S-PASS view of polarized Galactic synchrotron at 2.3 GHz as a contaminant to CMB observations}}, \href{https://doi.org/10.1051/0004-6361/201832768}{\emph{\aap} {\bfseries 618} (2018) A166} [\href{https://arxiv.org/abs/1802.01145}{{\ttfamily 1802.01145}}].

\bibitem{Vacher2023}
L.~{Vacher}, J.~{Chluba}, J.~{Aumont}, A.~{Rotti} and L.~{Montier}, \emph{{High precision modeling of polarized signals: Moment expansion method generalized to spin-2 fields}}, \href{https://doi.org/10.1051/0004-6361/202243913}{\emph{\aap} {\bfseries 669} (2023) A5} [\href{https://arxiv.org/abs/2205.01049}{{\ttfamily 2205.01049}}].

\bibitem{2021MNRAS.500..976R}
A.~{Rotti} and J.~{Chluba}, \emph{{Combining ILC and moment expansion techniques for extracting average-sky signals and CMB anisotropies}}, \href{https://doi.org/10.1093/mnras/staa3292}{\emph{\mnras} {\bfseries 500} (2021) 976} [\href{https://arxiv.org/abs/2006.02458}{{\ttfamily 2006.02458}}].

\bibitem{Adak2021}
D.~{Adak}, \emph{{A new approach of estimating the galactic thermal dust and synchrotron polarized emission template in the microwave bands}}, \href{https://doi.org/10.1093/mnras/stab2392}{\emph{\mnras} {\bfseries 507} (2021) 4618} [\href{https://arxiv.org/abs/2104.13778}{{\ttfamily 2104.13778}}].

\bibitem{2022JCAP...10..063G}
S.~{Ghosh}, Y.~{Liu}, L.~{Zhang}, S.~{Li}, J.~{Zhang}, J.~{Wang} et~al., \emph{{Performance forecasts for the primordial gravitational wave detection pipelines for AliCPT-1}}, \href{https://doi.org/10.1088/1475-7516/2022/10/063}{\emph{\jcap} {\bfseries 2022} (2022) 063} [\href{https://arxiv.org/abs/2205.14804}{{\ttfamily 2205.14804}}].

\bibitem{2023arXiv230908170Z}
Z.~{Zhang}, Y.~{Liu}, S.-Y.~{Li}, H.~{Li} and H.~{Li}, \emph{{A Constrained NILC method for CMB B mode observations}}, \href{https://doi.org/10.48550/arXiv.2309.08170}{\emph{arXiv e-prints} (2023) arXiv:2309.08170} [\href{https://arxiv.org/abs/2309.08170}{{\ttfamily 2309.08170}}].

\bibitem{Dou2023}
J.~{Dou}, S.~{Ghosh}, L.~{Santos} and W.~{Zhao}, \emph{{Foreground removal with ILC methods for AliCPT-1}}, \href{https://doi.org/10.48550/arXiv.2310.19627}{\emph{arXiv e-prints} (2023) arXiv:2310.19627} [\href{https://arxiv.org/abs/2310.19627}{{\ttfamily 2310.19627}}].

\bibitem{depro_paper}
Y.S.~{Abylkairov}, O.~{Darwish}, J.C.~{Hill} and B.D.~{Sherwin}, \emph{{Partially constrained internal linear combination: A method for low-noise CMB foreground mitigation}}, \href{https://doi.org/10.1103/PhysRevD.103.103510}{\emph{\prd} {\bfseries 103} (2021) 103510} [\href{https://arxiv.org/abs/2012.04032}{{\ttfamily 2012.04032}}].

\bibitem{GNILC_intro}
M.~{Remazeilles}, J.~{Delabrouille} and J.-F.~{Cardoso}, \emph{{Foreground component separation with generalized Internal Linear Combination}}, \href{https://doi.org/10.1111/j.1365-2966.2011.19497.x}{\emph{\mnras} {\bfseries 418} (2011) 467} [\href{https://arxiv.org/abs/1103.1166}{{\ttfamily 1103.1166}}].

\bibitem{Stolyarov2005}
V.~{Stolyarov}, M.P.~{Hobson}, A.N.~{Lasenby} and R.B.~{Barreiro}, \emph{{All-sky component separation in the presence of anisotropic noise and dust temperature variations}}, \href{https://doi.org/10.1111/j.1365-2966.2005.08610.x}{\emph{\mnras} {\bfseries 357} (2005) 145} [\href{https://arxiv.org/abs/astro-ph/0405494}{{\ttfamily astro-ph/0405494}}].

\bibitem{Narcowich2006}
F.J.~Narcowich, P.~Petrushev and J.D.~Ward, \emph{Localized tight frames on spheres}, \href{https://doi.org/10.1137/040614359}{\emph{SIAM Journal on Mathematical Analysis} {\bfseries 38} (2006) 574} [\href{https://arxiv.org/abs/https://doi.org/10.1137/040614359}{{\ttfamily https://doi.org/10.1137/040614359}}].

\bibitem{Marinucci2008}
D.~{Marinucci}, D.~{Pietrobon}, A.~{Balbi}, P.~{Baldi}, P.~{Cabella}, G.~{Kerkyacharian} et~al., \emph{{Spherical needlets for cosmic microwave background data analysis}}, \href{https://doi.org/10.1111/j.1365-2966.2007.12550.x}{\emph{\mnras} {\bfseries 383} (2008) 539} [\href{https://arxiv.org/abs/0707.0844}{{\ttfamily 0707.0844}}].

\bibitem{MASTER}
E.~{Hivon}, K.M.~{G{\'o}rski}, C.B.~{Netterfield}, B.P.~{Crill}, S.~{Prunet} and F.~{Hansen}, \emph{{MASTER of the Cosmic Microwave Background Anisotropy Power Spectrum: A Fast Method for Statistical Analysis of Large and Complex Cosmic Microwave Background Data Sets}}, \href{https://doi.org/10.1086/338126}{\emph{\apj} {\bfseries 567} (2002) 2} [\href{https://arxiv.org/abs/astro-ph/0105302}{{\ttfamily astro-ph/0105302}}].

\bibitem{MASTER2}
G.~{Polenta}, D.~{Marinucci}, A.~{Balbi}, P.~{de Bernardis}, E.~{Hivon}, S.~{Masi} et~al., \emph{{Unbiased estimation of an angular power spectrum}}, \href{https://doi.org/10.1088/1475-7516/2005/11/001}{\emph{\jcap} {\bfseries 2005} (2005) 001} [\href{https://arxiv.org/abs/astro-ph/0402428}{{\ttfamily astro-ph/0402428}}].

\bibitem{pymaster}
D.~{Alonso}, J.~{Sanchez}, A.~{Slosar} and {LSST Dark Energy Science Collaboration}, \emph{{A unified pseudo-C$_{{\ensuremath{\ell}}}$ framework}}, \href{https://doi.org/10.1093/mnras/stz093}{\emph{\mnras} {\bfseries 484} (2019) 4127} [\href{https://arxiv.org/abs/1809.09603}{{\ttfamily 1809.09603}}].

\bibitem{Hamimeche2008}
S.~{Hamimeche} and A.~{Lewis}, \emph{{Likelihood analysis of CMB temperature and polarization power spectra}}, \href{https://doi.org/10.1103/PhysRevD.77.103013}{\emph{\prd} {\bfseries 77} (2008) 103013} [\href{https://arxiv.org/abs/0801.0554}{{\ttfamily 0801.0554}}].

\bibitem{Kogut_2012}
A.~{Kogut}, \emph{{Synchrotron Spectral Curvature from 22 MHz to 23 GHz}}, \href{https://doi.org/10.1088/0004-637X/753/2/110}{\emph{\apj} {\bfseries 753} (2012) 110} [\href{https://arxiv.org/abs/1205.4041}{{\ttfamily 1205.4041}}].

\bibitem{2011ame}
C.~{Dickinson}, M.~{Peel} and M.~{Vidal}, \emph{{New constraints on the polarization of anomalous microwave emission in nearby molecular clouds}}, \href{https://doi.org/10.1111/j.1745-3933.2011.01138.x}{\emph{\mnras} {\bfseries 418} (2011) L35} [\href{https://arxiv.org/abs/1108.0308}{{\ttfamily 1108.0308}}].

\bibitem{2015ame}
R.~{G{\'e}nova-Santos}, J.A.~{Rubi{\~n}o-Mart{\'\i}n}, R.~{Rebolo}, A.~{Pel{\'a}ez-Santos}, C.H.~{L{\'o}pez-Caraballo}, S.~{Harper} et~al., \emph{{QUIJOTE scientific results - I. Measurements of the intensity and polarisation of the anomalous microwave emission in the Perseus molecular complex}}, \href{https://doi.org/10.1093/mnras/stv1405}{\emph{\mnras} {\bfseries 452} (2015) 4169} [\href{https://arxiv.org/abs/1501.04491}{{\ttfamily 1501.04491}}].

\bibitem{AME_commander}
D.~{Herman}, B.~{Hensley}, K.J.~{Andersen}, R.~{Aurlien}, R.~{Banerji}, M.~{Bersanelli} et~al., \emph{{BEYONDPLANCK. XV. Limits on large-scale polarized anomalous microwave emission from Planck LFI and WMAP}}, \href{https://doi.org/10.1051/0004-6361/202243081}{\emph{\aap} {\bfseries 675} (2023) A15} [\href{https://arxiv.org/abs/2201.03530}{{\ttfamily 2201.03530}}].

\bibitem{2017MNRAS.464.4107G}
R.~{G{\'e}nova-Santos}, J.A.~{Rubi{\~n}o-Mart{\'\i}n}, A.~{Pel{\'a}ez-Santos}, F.~{Poidevin}, R.~{Rebolo}, R.~{Vignaga} et~al., \emph{{QUIJOTE scientific results - II. Polarisation measurements of the microwave emission in the Galactic molecular complexes W43 and W47 and supernova remnant W44}}, \href{https://doi.org/10.1093/mnras/stw2503}{\emph{\mnras} {\bfseries 464} (2017) 4107} [\href{https://arxiv.org/abs/1605.04741}{{\ttfamily 1605.04741}}].

\bibitem{Ghosh2021}
S.~{Ghosh}, J.~{Delabrouille}, W.~{Zhao} and L.~{Santos}, \emph{{Towards ending the partial sky E-B ambiguity in CMB observations}}, \href{https://doi.org/10.1088/1475-7516/2021/02/036}{\emph{\jcap} {\bfseries 2021} (2021) 036} [\href{https://arxiv.org/abs/2007.09928}{{\ttfamily 2007.09928}}].

\bibitem{2022JCAP...07..044Z}
Z.~{Zhang}, Y.~{Liu}, S.-Y.~{Li}, D.-L.~{Wu}, H.~{Li} and H.~{Li}, \emph{{Efficient ILC analysis on polarization maps after EB leakage correction}}, \href{https://doi.org/10.1088/1475-7516/2022/07/044}{\emph{\jcap} {\bfseries 2022} (2022) 044} [\href{https://arxiv.org/abs/2109.12619}{{\ttfamily 2109.12619}}].

\bibitem{NILC_cutsky}
A.~{Carones}, M.~{Migliaccio}, D.~{Marinucci} and N.~{Vittorio}, \emph{{Analysis of Needlet Internal Linear Combination performance on B-mode data from sub-orbital experiments}}, \href{https://doi.org/10.1051/0004-6361/202244824}{\emph{\aap} {\bfseries 677} (2023) A147} [\href{https://arxiv.org/abs/2208.12059}{{\ttfamily 2208.12059}}].

\end{thebibliography}\endgroup

\end{document}